\documentclass[reprint,amsmath,amssymb,aps,rmp,floatfix]{revtex4-2}

\pdfoutput=1

\usepackage{graphicx}% Include figure files
\usepackage{dcolumn}% Align table columns on decimal point
\usepackage{bm}% bold math
\usepackage{epsfig} %%%
\usepackage{color}
\begin{document}

\preprint{APS/123-QED}

\title{Positronium Physics and 
Biomedical Applications}

\author{Steven D. Bass}
\email{Steven.Bass@cern.ch}
\affiliation{\mbox{Kitzb\"uhel Centre for Physics,
Kitzb\"uhel, Austria}}
\affiliation{\mbox{Marian Smoluchowski Institute of Physics, Jagiellonian University, 
PL 30-348 Krakow, Poland}}

\author{Sebastiano Mariazzi}
\email{Sebastiano.Mariazzi@unitn.it}
\affiliation{\mbox{Department of Physics, University of Trento via Sommarive 14, 38123 Povo, Trento, Italy}, \mbox{TIFPA/INFN, via Sommarive 14, 38123 Povo, Trento, Italy}}

\author{Pawel Moskal}
\email{P.Moskal@uj.edu.pl}
\affiliation{\mbox{Marian Smoluchowski Institute of Physics, Jagiellonian University, 
PL 30-348 Krakow, Poland}}
\affiliation{\mbox{Center for Theranostics, Jagiellonian University, Krakow, Poland}}

\author{Ewa Stepien}
\email{E.Stepien@uj.edu.pl}
\affiliation{\mbox{Marian Smoluchowski Institute of Physics, Jagiellonian University, 
PL 30-348 Krakow, Poland}}
\affiliation{\mbox{Center for Theranostics, Jagiellonian University, Krakow, Poland}}

\date{\today}% It is always \today, today,

\begin{abstract}
Positronium is the simplest bound state, built of an electron and a positron. 
Studies of positronium in vacuum and its decays in medium tell us about Quantum Electrodynamics, QED, and about the structure of matter and biological 
processes of living organisms at the nanoscale, respectively. 
Spectroscopic measurements constrain our understanding of QED bound state theory. 
{
Searches for 
rare decays 
and 
measurements of the effect of gravitation on positronium
are used to look for new physics phenomena.
In biological materials
positronium decays}
are sensitive 
to the inter- and intra-molecular structure and 
to the metabolism of living organisms ranging from single cells to human beings. 
This leads to new ideas of positronium imaging in medicine
using the fact that during positron emission tomography (PET) as much as 40\% of positron annihilation occurs through the production of positronium atoms inside the patient's body. A new generation of the high sensitivity and multi-photon total-body PET systems opens perspectives for clinical applications of positronium as a biomarker of tissue pathology and the degree of tissue oxidation.
\end{abstract}

\maketitle

%\tableofcontents

\section{Introduction}
\label{sec:introduction}

Positronium ``atoms'' are special as short lived 
bound states of an electron $e^-$ and its antiparticle, the
positron $e^+$.
They are at same time their own ``anti-atoms".
Positronium is topical in both fundamental physics research as well as in applications in biology and medicine, 
with prime focus here on its role in new positron emission tomography, PET, technologies.

The physics of positronium is expected to be described
by Quantum Electrodynamics, QED,
which is our most accurately tested theory,
up to one part in $10^{12}$, with tiny radiative corrections from the
strong and weak interactions.
Recent experiments have revealed some surprises
pushing the boundaries of QED
bound state theory 
\cite{Adkins:2022omi}
with the observation of anomalies up 
to 4.5 standard deviations at the precision of 
$10^{-4}$ between measurements and theory in hyperfine splittings of positronium energy levels.
Possible couplings of positronium to new interactions 
are being probed through precision symmetry tests and
rare decay measurements.
These experiments promise to yield new understanding of charged lepton bound states.

While there are uncertainties at this level, positronium is sufficiently well  understood to enable its role in applications from fundamental physics experiments involving the study of gravitation on antimatter 
%through 
to diagnostic tests in medicine.
About 40\% of the positrons in PET scans go through positronium formation and decay in the body. Building on this result, 
positronium is being explored  as a vital ingredient in next generation total-body 
PET devices  where two or more
photons are detected simultaneously from individual positronium decays using the new technique of  
multi-photon tomography. 
Quantum entanglement of the emitted photons may further enhance the diagnostic power.

In this Colloquium we explore this physics first with an introduction to 
positronium
and then covering the present status of precision positronium measurements and current anomalies between data and bound state theory. We explain the mechanisms of positronium  formation and decays in materials and then, with key focus on biological substances, the application in next generation PET devices. 
New positronium imaging technologies have the promise to  revolutionise total body PET imaging with benefit to medical diagnostics.

Positronium comes in two ground states,
$^1 S_0$ para-positronium, denoted p-Ps, 
where the spins of the electron and positron 
add up to zero 
and 
$^3 S_1$ ortho-positronium, 
denoted o-Ps, 
where the spins of the electron and positron add up to one.
The binding energy, 
\begin{equation}
E_B \approx
- m_e \alpha^2 / 4
=
-6.8 \ {\rm eV},
\label{eq:1}
\end{equation}
is much less than the electron mass $m_e = 0.51$ MeV 
with
$\alpha \approx \frac{1}{137}$ 
the fine structure constant.
\hbox{p-Ps} is slightly lighter by 0.84 meV due to 
the interaction between the electron and positron spins 
and also the existence of virtual annihilation processes~\cite{Cassidy2018tgq}.

Spin-zero p-Ps 
decays in vacuum to two photons
with 
a lifetime of 125 picoseconds 
and spin-one o-Ps 
decays to three photons
with
a lifetime of 142 nanoseconds.
The factor of more than a thousand times longer lifetime of o-Ps
enables efficient distinction of these two states.
The main reason for the difference in  lifetimes comes from
an extra factor of 
the fine structure
constant
$\alpha$
that enters with the three photon decay compared to two photon decays.

Positronium was first discovered by
\textcite{Deutsch1951zza}
following the initial prediction of positron antimatter
by 
\textcite{Dirac1931kp},
the discovery of the
positron by
\textcite{Anderson1933mb} and
prediction of 
the $e^+ e^-$ bound state in
\textcite{Mohorovicic1934aa}.

The two positronium ground states o-Ps and p-Ps, 
being bound states of
$e^-$ and $e^+$,
are both odd under parity transformations P.
Under charge conjugation C
o-Ps is odd and p-Ps is even.
C symmetry conservation
determines the decays 
of o-Ps and p-Ps 
into an odd and even number of photons respectively, with photons being 
C symmetry odd~\cite{Berko1980,Cassidy2018tgq}.
Since positronium is unstable 
with leading decay to two or three massless photons 
(for p-Ps and o-Ps),
it is not an eigenstate of time reversal transformations T.
This property has the consequence that
final state interactions involving 
photon-photon rescattering 
interactions
at higher order in
$\alpha$
can mimic a tiny
CP and CPT violation in
positronium decays.

{
Positronium spectroscopy research is presently focused on precision measurements of hyperfine transitions, HFS, and also Rydberg states,  the latter with the aim to determine the Rydberg constant based on positronium 
\cite{Cassidy2018tgq}.
}
Several few standard deviations discrepancies
have been reported
between the precision HFS measurements and QED bound state calculations
performed using 
the simplifications of
non-relativistic QED
effective theory,
with the differences
entering at
precision of a
few parts in 10,000
or less
\cite{Gurung2020hms,Heiss2018jbl,Karshenboim2003vs}.
Positronium decay measurements are so far in agreement with QED bound state theory at  similar accuracy.
An important ingredient in modelling is that positronium should
satisfy the fundamental symmetries of its constituents.
Rare decays are
strongly constrained by precision measurements of the electron anomalous magnetic moment
and electric dipole moment
with a prime topic being 
the search for
invisible decays
in connection with
possible dark matter candidates called
``mirror matter'' particles \cite{Raaijmakers2019hqj}.
In connection with gravitation, 
positronium is also playing an important role in precision tests of gravity on antimatter planned at CERN, 
the experiments 
AEgIS~\cite{Doser2018zfc}
and GBAR~\cite{Perez2012, Dufour2015}.

In materials, positronium formation 
and decay is sensitive to the immediate chemical environment.
This has  interesting medical applications
with sensitivity to
the healthiness of biological tissue
where the positronium is produced, 
and may serve as a hallmark telling one about the size of inter- and intra-molecular voids and the concentration in them of bio-molecules such as, e.g., oxygen O$_2$~\cite{NRP2019}. These properties of  positronium suggest its role as a biomarker,
a characteristic that is objectively measured and evaluated as an indicator of normal biological  (healthy) or pathogenic (cancerous) processes.
This result has inspired new ideas for positronium imaging - 
a new technique in medical diagnosis 
that enables imaging of positronium properties inside the bodies of living 
organisms~\cite{Moskal2021science}.
{
Electromagnetic
decays of positronium should exhibit 
quantum entanglement of the final state photons
\cite{Acin2000,Hiesmayr2017xgx}, 
with ideas how this may be exploited in 
positronium imaging 
and next generation PET devices discussed in
\textcite{McNamara2014, Hiesmayr2018rcm}.
}

The plan of this article is as follows.
In Section II we
discuss the status of precision QED measurements and theory, which constrains detailed modelling of the positronium system.
In Section III we 
turn our attention to materials systems where positronium production and decays
depend on the chemical environment.
This leads to discussion of positronium in
fundamental physics experiments and in medical applications.
Positronium spectroscopy, its role in 
the AEgIS and GBAR antimatter experiments at CERN and
Bose-Einstein condensates as well as quantum entanglement in positronium decays are summarized in Section IV.
Biological and
medical applications are discussed in Section V and VI
including new developments with positronium imaging 
and the emerging application of positronium as a biomarker 
for assessing the tissue pathology {\it in  vivo}.
Conclusions and an outlook to future opportunities are given in Section VII.

Complementary reviews
of positronium physics, each with a different emphasis are given in
\textcite{Adkins:2022omi},
\textcite{Cassidy2018tgq} as well as
\textcite{Bass2019ibo,
Nagashima2014,
Goworek2014,
Karshenboim2005iy,
Karshenboim2003vs,
Gninenko2002jn,
Berko1980}.
An introduction to
applications in 
medicine and biology is given in
\textcite{NRP2019} and \textcite{Harpen2003zz}.

\section{Positronium in the Standard Model}
\label{sec:positronium-SM}

As a bound state of an electron and positron with dynamics determined by QED, 
the physics of positronium is strongly constrained
by precision QED observables.
QED is a gauge theory invariant 
under local U(1)
transformations of the phase of the electron and other charged fermions.
The QED Lagrangian
reads
\begin{equation}
{\cal L} =
\bar{\psi} i \gamma^{\mu} 
(\partial_{\mu} 
 + i e A_{\mu})
\psi 
- m_e {\bar \psi} \psi
- {\frac{1}{4}} F_{\mu \nu} F^{\mu \nu} .
\label{eq:2}
\end{equation}
Here $\psi$ represents the electron field, $A_{\mu}$ is the photon,
$e$ is the electric charge
and
$F_{\mu \nu} = \partial_{\mu} A_{\nu} - \partial_{\nu} A_{\mu}$
is the electromagnetic field tensor;
$\alpha = e^2/4 \pi$
is the fine structure constant.
Electrons and positrons interact through massless photon exchange.

Measurements of the 
electron's anomalous magnetic moment
$a_e = (g-2)/2$
and
atomic physics
measurements of the
fine structure constant 
using atom interferometry
with Caesium, Cs, and Rubidium, Rb, atoms
are consistent with each other and with QED theory to one part in
$10^{12}$.

The electron's anomalous magnetic moment
$a_e$
is non vanishing, differing
from the Born term level Dirac value 
$a_e=0$
by a
perturbative QED expansion in $\alpha$ which is known to ${\cal O}(\alpha^5)$
\cite{Aoyama2017uqe}.
Precision measurement of 
$a_e$ thus
allows determination of the fine structure constant. 
The atom interferometry measurements give a direct measurement of $\alpha$.
Any ``beyond the Standard Model'' effects involving 
new particles active in radiative corrections would give an
extra correction to $a_e$ 
but not the direct
Cs and Rb 
interferometry measurements.
Thus, 
comparing these different determinations of 
$\alpha$ 
gives a precision test of QED
as well as constraining possible new physics scenarios.

QED radiative corrections
involving heavy muons and tau leptons as well as  hadronic corrections 
from Quantum Chromodynamics, QCD, 
each enter $a_e$
at the level of 
$2 \times 10^{-12}$
and weak interactions 
at the level of 
$3 \times 10^{-14}$
so the
anomalous magnetic moment 
is a very precise test of electron photon interactions.

The most accurate measurement of $a_e$ is~\cite{Hanneke2008tm}
\begin{equation}
a_e^{\rm exp}
=
0.001 159 652 180 73 (28).
\label{eq:3}
\end{equation}
If one substitutes
the most recent 
direct $\alpha$
measurements from atom interferometry measurements using both 
Cs 
\cite{Parker2018vye}
\begin{equation}
1 / \alpha |_{\rm Cs} = 137.035 999 046 (27) 
\label{eq:8}
\end{equation}
and Rb \cite{Morel2020dww}
\begin{equation}
1 / \alpha |_{\rm Rb} = 137.035 999 206 (11) .
\label{eq:9}
\end{equation}
into the perturbative QED expansion for $a_e$ one finds
agreement
to one part in $10^{12}$
when comparing with $a_e^{\rm exp}$ in Eq.~(\ref{eq:3}), viz.
\begin{equation}
a_e^{\rm exp} - a_e^{\rm th}|_{\rm Cs}
=
(-88 \pm 36 ) \times 10^{-14}
\label{eq:10}
\end{equation}
and
\begin{equation}
a_e^{\rm exp} - a_e^{\rm th}|_{\rm Rb}
=
(+44 \pm 30 ) \times 10^{-14}
\label{eq:11}
\end{equation}
when we substitute the $\alpha$ values in Eqs.~(\ref{eq:8},\ref{eq:9})
into 
the QED perturbative expansion for $a_e$
to obtain the value $a_e^{\rm th}|_{\rm atom}$.

QED is working very well! 
For practical calculations of positronium 
spectroscopy and decays one needs an extra step of QED bound state theory.

Bound state calculations are hard, even in QED.
Some model simplifications are needed to make the calculations tractable.
The non-relativistic
Schr\"odinger equation for the 
$e^- e^+$ system
gives the correct leading order 
expression for the positronium binding energy,
Eq.~(\ref{eq:1}).
With this in mind a 
rigorous effective theory formalism has been developed for calculating 
positronium 
spectroscopy and 
decays to 
multiple-photon final states.
This is called non-relativistic QED, NRQED 
\cite{Caswell1985ui};
for reviews see
\textcite{Karshenboim2003vs,Kinoshita1990ai,Labelle1992hd}.

NRQED involves a perturbation expansion in 
$v/c \sim \alpha$ 
where $v$ is the 
electron and positron velocities in the positronium,
$c$ is the speed of light and $\alpha$ the fine structure constant.
This approximation 
allows for doable calculations.
One introduces a cut-off on relativistic effects 
from
the fundamental QED Lagrangian, Eq.~(\ref{eq:2}).
These 
are then implemented through adding extra ``correction terms'' in the NRQED Lagrangian.
The parameters are
adjusted to fit the results of experiments, and 
then the NRQED Lagrangian is used to calculate new observables.
One assumes that the incident electron positron pair is non-relativistic with relativistic terms in the interactions taken care of by the NRQED interactions.
The fundamental discrete symmetries of QED should carry over to the truncated NRQED.

Positronium energy levels have been calculated 
to order 
$m_e \alpha^6$ 
\cite{Czarnecki1998zv,Pachucki1997vm},
plus
some contributions
calculated to order 
$m_e \alpha^7$
-- see \textcite{Cassidy2018tgq} and references therein.
Experiments in positronium spectroscopy  including anomalies 
at order $10^{-4}$ 
between 
precision measurements and NRQED predictions
are
discussed in Section IV below.

For QED decays of positronium to photon final states,
radiative corrections
to the tree level processes have been evaluated in NRQED calculations 
up to two loop level 
\cite{Adkins2002fg}.
The Born term 
level decay rates
$
\Gamma( \hbox{o-Ps} \to 3 \gamma)
=
2 (\pi^2 - 9) \alpha^6 m_e /{9 \pi}
$
and
$
\Gamma( \hbox{p-Ps} \to 2 \gamma)
=
\alpha^5 m_e / 2
$
are multiplied by 
radiative correction terms
of the form 
$\{1 + c_{nm} \alpha^n \ln^m \alpha \}$
where 
$n>m$
and $c_{nm}$ are coefficients evaluated from the Feynman diagrams,
presently up to $n=3$.
Branching ratios 
$\sim 10^{-6}$ 
for subleading decays of o-Ps to 5 photons and p-Ps to 4 photons are suppressed relative to the leading 3 and 2 photon decays by factors of $(\alpha/\pi)^2$.
Radiative corrections from
QCD and weak interactions 
as well as QED radiative corrections involving heavy leptons are tiny and
presently beyond experimental accuracy.

The most accurate measurements of o-Ps decays are consistent
with each other and with NRQED theory.
Working in vacuum, 
\textcite{Vallery2003iz} 
found
\begin{equation}
\Gamma = 
(7.0404 \pm 0.0010 \pm 0.0008) \times 10^6 \ {\rm s}^{-1}.
\label{eq:G-oPs-1}
\end{equation}
\textcite{Kataoka2008hj} found
\begin{equation}
\Gamma = 
(7.0401 \pm 0.0007) 
\times 10^6 \ {\rm s}^{-1}
\label{eq:G-oPs-2}
\end{equation} 
with the o-Ps produced in SiO$_2$ powder.
Including both the 3 and 5 photon contributions NRQED gives 
the QED decay rate prediction
$
\Gamma = 
(7.039 979 
\pm 0.000011) 
\times 10^6 \ {\rm s}^{-1}
$
\cite{Adkins2002fg}.
The measurements
are consistent with the NRQED theory prediction with the caveat that the present experimental uncertainties 
on the decay rate 
are about 100 times larger than the NRQED
theoretical error.
The leading ${\cal O}(\alpha)$ correction to the decay rates
is needed to agree with the data.
The ${\cal O}(\alpha^2)$ 
terms are of order the same size as
the experimental error; 
${\cal O}(\alpha^3)$ terms are well within the experimental uncertainties, as are 
QCD radiative corrections.

For the p-Ps decay rate
one finds
$
\Gamma_p = (7989.6178 \pm 0.0002) \times 10^6 \ {\rm s}^{-1} 
$
from NRQED theory with the 4 photon decay included
\cite{Adkins2003eh}, which compares 
with the experimental result
\cite{AlRamadhan1994zz}
\begin{equation}
\Gamma_p = 
(7990.9 \pm 1.7) \times 10^6 \ {\rm s}^{-1}
\label{eq:G-pPs}
\end{equation}
with the experimental error 10,000 times the
size of the theoretical error within NRQED.

Going beyond QED decays to photon final states, branching ratios for possible decays involving new particles 
beyond the Standard Model 
are strongly constrained by precision measurements of the electron's anomalous magnetic moment %
with limits on couplings of any new particles to the electron.
If a new interaction were to couple to the electron with coupling $\alpha_{\rm eff}$
it would give a leading contribution to $a_e$ of size
$\alpha_{\rm eff}/2 \pi$.
Taken alone,
the $a_e$ measurements
imply 
constraints on the branching ratios of o-Ps decays to two photons plus a new light vector particle and to a photon plus new light pseudoscalar of
less than $10^{-9}$ and $10^{-6}$ respectively
\cite{Bass2019ibo,Gninenko2002jn}.
Possible invisible decays of o-Ps have been looked for in the context of mirror matter models of dark matter with the branching ratio constraint from o-Ps decays in vacuum measured to be 
$< 3 \times 10^{-5}$ at 90\% 
confidence level~\cite{Raaijmakers2019hqj}.

As a bound state, positronium should inherit the symmetries of its constituents.
Fundamental QED interactions 
encoded in the Lagrangian, Eq.(\ref{eq:2}),
respect the discrete symmetries of 
P, C, T, 
and their combinations: 
CP and CPT, 
with CPT a 
fundamental property of 
relativistic quantum field theories.

{The
precision confirmation of QED through the electron's anomalous magnetic moment $a_e$
implicitly 
implies CPT as a good symmetry 
for electrons, positrons and photons.
More directly,
the symmetries of CPT and C have been shown to work 
to the level of 
$2 \times 10^{-12}$
through measurement of $g-2$ 
for both electrons and positrons \cite{VanDyck1987ay}, 
\begin{equation}
g(e^-) / g(e^+) 
= 
1 + (0.5 \pm 2.1) \times 10^{-12} .
\label{eq:12}
\end{equation}
For spin-one o-Ps,
CPT is tested 
in the 3 photon decay
through measurement of a CPT odd correlation}
$
A_{CPT} =
\langle \vec{S} . (\vec{k}_1 \times \vec{k}_2) \rangle
$
which 
measures the T-odd integrated moments between the
polarization vector $\vec{S}$ of the o-Ps and the momenta of the emitted photons with magnitude $k_1 \geq k_2 \geq k_3$.
The most precise and recent measurement is consistent with zero with the reached precision of $\pm 0.00095$ 
\cite{Moskal2021nature}.

For CP symmetry the electron electric dipole moment, eEDM, puts strong constraints on any new CP violating interactions coupling to the electron. 
The eEDM has been precisely measured
showing that any eEDM is tiny
\cite{Andreev2018ayy},
\begin{equation}
| d_e | < 1.1 \times 10^{-29} e {\rm cm} .
\label{eq:13}
\end{equation}
This measurement
puts strong limits on new CP violating interactions coupling to the electron.
If interpreted in terms of possible CP violating
new heavy particle exchanges with couplings similar order of magnitude to Standard Model ones, then one finds a constraint on the new physics scale of similar size to constraints from the LHC high-energy experiments at CERN \cite{Andreev2018ayy}.
If instead the eEDM is
taken as due to the exchange of near massless particles, then one finds a bound on their coupling to the electron of $\alpha_{\rm eff} \sim 5 \times 10^{-9}$ \cite{Bass2019ibo}.
Measurements of CP violating correlations involving the polarisation of the o-Ps 
with the momentum vectors of the emitted photons are consistent with zero at ${\cal O}(10^{-3})$ 
\cite{Yamazaki2009hp}.
In future experiments 
up to ${\cal O}(10^{-5})$ precision 
in CP and CPT violating correlations
is expected from measurements with the J-PET tomograph in Krakow where new correlations involving polarisation of the final state photons are also measured \cite{Moskal2016moj}.

While the underlying QED conserves CP and CPT,
finite values 
for CP and CPT violating correlations in o-Ps decays
at the level of 
${\cal O}(10^{-9}) - {\cal O}(10^{-10})$ 
can occur associated with the fact that unstable Ps is not an eigenstate of T symmetry.
These non zero correlations
are found
in detailed calculations of the 
final state interactions with the leading contribution 
coming from 
light by light scattering of two of 
the three photons in 
the final state~\cite{Bernreuther1988tt}.

Summarizing,
precision studies of positronium decays 
are consistent with QED theoretical 
predictions 
within the accuracy of present experiments.
Spectroscopic measurements discussed 
in Section IV.A reveal discrepancies compared to theory that need to be understood.
Positronium structure remains an exciting topic of investigation.
For these measurements 
one needs development 
of precision
methods and phenomenology of positronium production and decay in materials discussed in the next Section. 
The present experimental precision on the
positronium decay rates in vacuum is however sufficient for applications, e.g., 
to studies in biological materials relevant to medicine discussed in Sections~\ref{sec:biology} and \ref{sec:medical-applications} below.

\section{Positronium production and decay in materials}
\label{sec:positronium-in-materials}
For experiments involving tests of fundamental physics with positronium  as well as
in applications, e.g., to medicine, one first needs to produce the positronium.
How is it made?
Positronium is produced via positron interactions within materials {and then emitted into the vacuum and employed in} fundamental physics experiments or its decays are directly studied in medium as a function of the medium properties.
In medicine one finds sensitivity to the 
healthiness of the tissue 
{that the positronium is produced in.}

\begin{figure*}[t!]
\vspace{-0.8cm}
\centering
\includegraphics[width=0.78\textwidth]{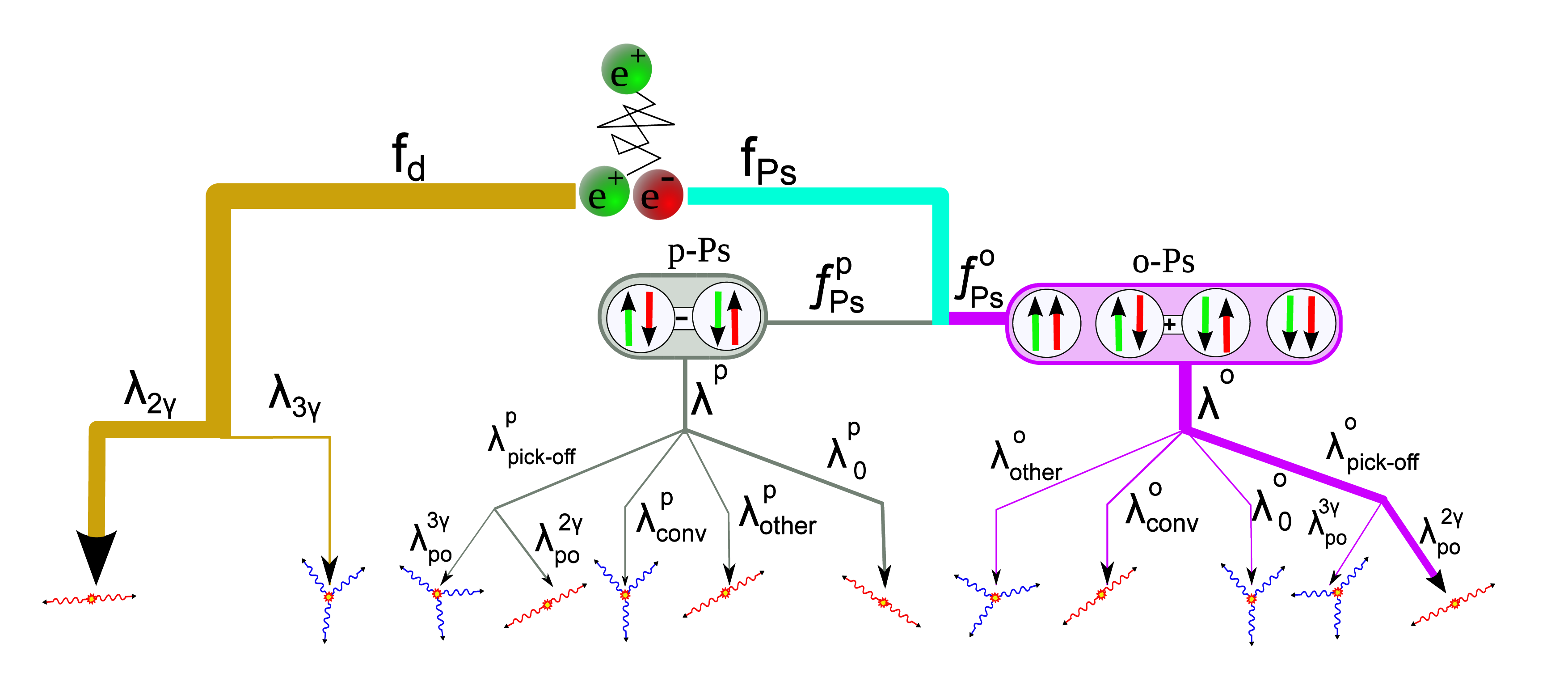}
\caption{
\label{fig:Ps-decay-scheme}
Diagram of the $e^+ e^-$ annihilation processes in matter. 
The scheme indicates annihilations to 2$\gamma$ and 3$\gamma$ only;
${\rm f_d}$ and ${\rm f_{Ps}}$ indicate fraction of direct annihilations and annihilations via positronium formation, respectively. 
In biological materials, a ${\rm f_d}$/${\rm f_{Ps}}$ ratio ranging between about 3/2 ~\cite{Harpen2003zz,Kotera2005} and about 1/4 ~\cite{Blanco2016} is reported. ${\rm f_d}$/${\rm f_{Ps}}$ ratios smaller than 1 are common in mesoporous materials (see for example \textcite{Goworek2014}). 
The total decay rate is due to the positronium self-annihilation~($\lambda_0$), pick-off processes~($\lambda_{\rm pick-off}$), ortho-para conversion reactions~($\lambda_{\rm conv}$) and other chemical processes as e.g. oxidation~($\lambda_{\rm other}$):  $\lambda$~=~$\lambda_0~+~\lambda_{\rm pick-off}~+~\lambda_{\rm conv}~+~\lambda_{\rm other}$. \
In water $\lambda^p_0=7990.9
{\rm \mu s}^{-1} \gg \lambda_{\rm pick-off}=
512.8 {\rm \mu s}^{-1} \gg
\lambda_{\rm conv} + \lambda_{\rm other}\approx 27
{\rm \mu s}^{-1} (\mbox{O}_2 \, {\mbox{saturated}}) >
\lambda^0_0=7.0401
{\rm \mu s}^{-1}$.
In mesoporous silica
$
\lambda^p_0=7990.9
{\rm \mu s}^{-1} \gg 
\lambda_{\rm conv}\approx 25
{\rm \mu s}^{-1}$ 
       (O$_2$ at 1 atm.) $>
\lambda^0_0=7.0401
{\rm \mu s}^{-1} >  
\lambda_{\rm pick-off}\approx 1
{\rm \mu s}^{-1} 
$.
Explanation is given in the text.
}
\end{figure*}

As a first step one needs
$e^+$
production which proceeds
either 
via pair creation from 
$\gamma \to e^- e^+$
or through $\beta^+$ decays.
The positrons then interact with a 
material and 
annihilate directly with electrons ($e^+ e^- \to {\rm  photons}$) 
or make positronium 
which is either {emitted and used in} experiments or its decay properties are  studied directly in medium with various processes at play, 
depicted in Fig.~\ref{fig:Ps-decay-scheme} 
and discussed in this Section.
When positronium is formed it may self-annihilate or, alternatively,  
decay through annihilation of the positron with an electron in the medium or via intermediate reactions with molecules in the system.
One only considers the
leading 2~photon decays of p-Ps and 3~photon decays of o-Ps for applications, with the branching ratios 
for two extra photons production
suppressed by a factor of 
$(\alpha/ \pi)^2$ and safely taken as negligible.

In this Section we first outline positron production and then positronium 
formation and decay processes in medium with 
a focus on its use in
fundamental physics experiments (Section~IV)
and medical applications
(Section~\ref{sec:medical-applications}) 
where key media are water and mesoporous materials, 
materials with intermediate
pores (inter-atomic voids) size in the range of 2–50~nm.
Water constitutes the largest percentage of cells and in general biological materials. Mesoporous silica targets are used as efficient positron to positronium converters for the production and emission into vacuum of positronium for physics experiments.

\subsection{Positron sources and positron thermalization}

Positrons are routinely produced in physics and biomedical laboratories around the world via two processes: 
$e^- e^+$ pair production 
in the electric field of the nucleus
and 
through use of 
$\beta^+$ radioactive sources~\cite{Coleman2003,Hugenschmidt2016}. 
With $e^- e^+$ pair production, 
photons with energy larger than around 1.2~MeV 
are implanted in materials with high atomic number Z such as tungsten and platinum, 
and their energy is converted to the mass of $e^- e^+$ pairs. 
The high-energy photons can be generated via bremsstrahlung from decelerating electrons previously accelerated to relativistic energy by employing an electron linear accelerator (LINACs),
see for example \textcite{Howell1982,Wada2012,Charlton2021}. 
As an alternative, $\gamma$ rays can be released from nuclear processes (see for example \textcite{Schut2004,Hugenschmidt2008,Hawari2011,Sato2015}).
With $\beta^+$ decays the starting nuclei transform into daughter nuclei 
    (with atomic number Z reduced by one) through emission of a positron and a neutrino. A large variety of $\beta^+$ radio nuclides with a half-life ranging from less than a second up to several years and maximum positron energy ranging between several hundreds of keV and a few MeV are available. The most commonly used in physical laboratories is $^{22}$Na (half-life of 2.6 years with maximum positron energy 0.54~MeV \cite{Hugenschmidt2016}) while for biomedical applications there is a growing interest in  $^{44}$Sc (half-life of 4~hours with maximum positron energy of 1.47~MeV \cite{Matulewicz2021,Choinski2021,Hernandez2014,Rosar2020}).
    {
    $^{44}$Sc radioisotope can be obtained from the $^{44}$Ti/$^{44}$Sc generator~\cite{Filosofov2010,Pruszynski2010}, and also by irradiation with protons or deuterons
    of an enriched $^{44}$Ca
    target
    \cite{Choinski2021,Mikolajczak2021}. $^{44}$Ti transforms to $^{44}$Sc with the half-lifetime of 60 years via electron capture
    \cite{Roesch2012}. Long lifetime of Titanium-44 makes the  $^{44}$Ti/$^{44}$Sc generator convenient for the laboratory and clinical applications.
   
    In the decay of $^{22}$Na and $^{44}$Sc radionuclides, %
    } excited daughter nuclei are produced that then de-excites through the emission of a prompt photon via the following reaction chains
\begin{equation}
\nonumber
^{22}\mbox{Na}~\to~^{22}\!\mbox{Ne}^*+e^+~+~\nu~+~\gamma(1.27~\mbox{MeV},3.6~\mbox{ps}), 
\end{equation}
\begin{equation}
^{44}\mbox{Sc}~\to~^{44}\!\mbox{Ca}^*~+~e^+~+~\nu~+~\gamma(1.16~\mbox{MeV},2.61~\mbox{ps}). 
\label{eq:sodium-scandium}
\end{equation}
Here the energies of prompt photon and mean de-excitation times 
\cite{Matulewicz2021,Choinski2021,Kaminska2016fsn,Thirolf2015} 
are given in brackets.

When produced positrons are implanted in materials, they rapidly lose kinetic energy \cite{Kubica1975} in a variety of interactions (bremsstrahlung, ionization, electron excitation, phonon excitation, vibrational and rotational excitation, positronium formation etc...) approaching the thermal energy \cite{Schultz1988,Puska1994}. The efficiency of the positron stopping process depends both on the positron energy range and on the type of material, where the positron is implanted. 
{For energies of a few tens of MeV, the dominant energy loss mechanism for positrons (as well as for electrons) is bremsstrahlung in which the positron interacts with the Coulomb field of the nuclei and the atomic orbital electrons emitting photons \cite{Schultz1988,Pages1972}. At implantation energies lower than a few MeV, this energy loss mechanism becomes less efficient for positrons than for electrons, due to the different sign of the electric charge,
e.g., with positrons attracted and electrons repulsed by the electric
charge of the nucleus (and vice versa by the atomic electrons)
\cite{Feng1981,Kim1986}.} 

{Below several hundreds of keV \cite{Hansen1983,Schultz1988} down to few electronvolts or few tenths of electronvolt, in the case of metals, the most important energy loss processes are ionization and electron excitation \cite{Valkealahti1983,Valkealahti1984,Schultz1985,Champion2005}.}
In this energy range the rate of energy transfer is very high (up to $10^{17}-10^{18}$ eV/s) and the positron energy can be reduced to a few tens of electronvolt within
some picoseconds 
\cite{Perkins1970,Schultz1988}. At lower energies, electron excitation processes vanish and other mechanisms involving
phonon scattering \cite{Perkins1970,Schultz1988,Nieminen1980,Gullikson1986} and vibrational and rotational excitation processes become dominant \cite{Blanco2013,Blanco2016}. These last mechanisms are less efficient than the electron excitation \cite{Dupasquier1985,Schultz1988}. 
However, the total complete thermalization time is estimated to be roughly 3~ps for positrons implanted with 1 keV in aluminum at the temperature of 600 K \cite{Nieminen1980}.
In semiconductors and insulators, where the electron excitations stop when the positron energy decreases under the band gap and a wider region of energy must be lost through phonon excitation, the thermalization results longer than in metals \cite{Gullikson1986,Nielsen1986,Mills1986,Schultz1988}. In the case of positron implantation in water, the contribution of ionization vanishes below around 50~eV, while the contribution
given by electronic excitations becomes negligible below around 7~eV \cite{Blanco2013,Blanco2016}. Vibrational and rotational excitations are expected to overcome the contribution given by other energy loss processes below around 10~eV \cite{Blanco2016}. In water the entire process of thermalization takes about 5~ps to 10~ps \cite{Stepanov2021} while the mean diffusion range is 1.5 mm and 2.1 mm for positrons from $^{22}$Na and $^{44}$Sc, respectively \cite{Thirolf2015}. 
In rare-gas solids, the absence of an optical-phonon branch further reduces the energy loss efficiency \cite{Schultz1988}. As a result, positrons can diffuse for lengths of several $\mu$m retaining several eV of kinetic energy \cite{Mills1994} or, in other words, positrons retain an eV energy for a few tens of picoseconds.

\subsection{Positronium formation mechanisms}
\label{positronium-formation}

In materials with a wide energy band gap, a positron with kinetic energy less than the band gap can also lose energy in positronium formation \cite{Schultz1988}. Ps formation is energetically possible if the positron energy is within the 
so-called Ore gap \cite{Ore1949}, i.e.,  between the ionization threshold ($I$) of the material and $I$-6.8 eV (where 6.8 eV is the Ps binding energy in vacuum). This process has been extensively investigated in the case of ice at the beginning of the 1980s \cite{Eldrup1983,VanHouse1984,Eldrup1985}.

\begin{figure}[t!]
\vspace{-0.3cm}
\centering
\includegraphics[width=0.156\textwidth]{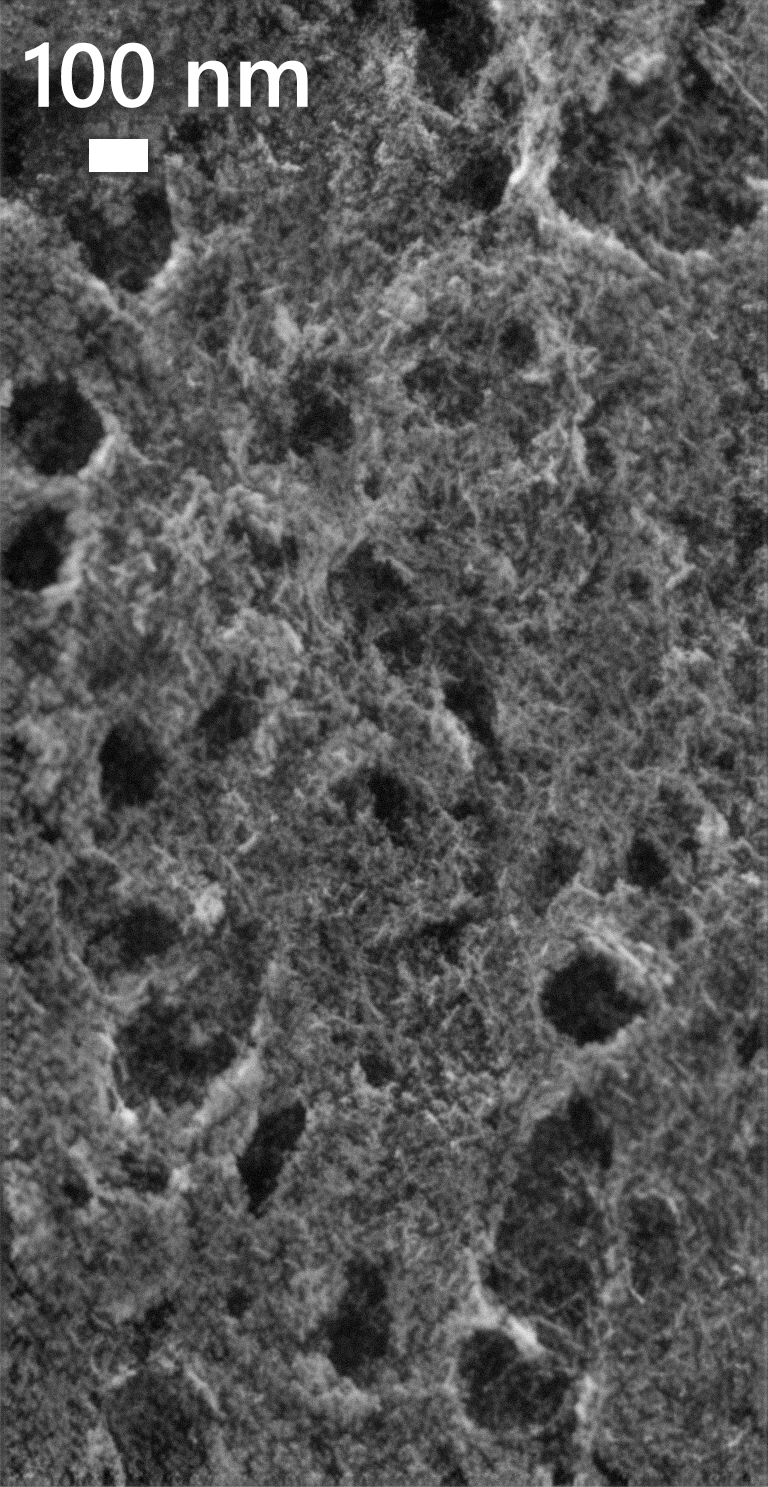}
\includegraphics[width=0.265\textwidth]{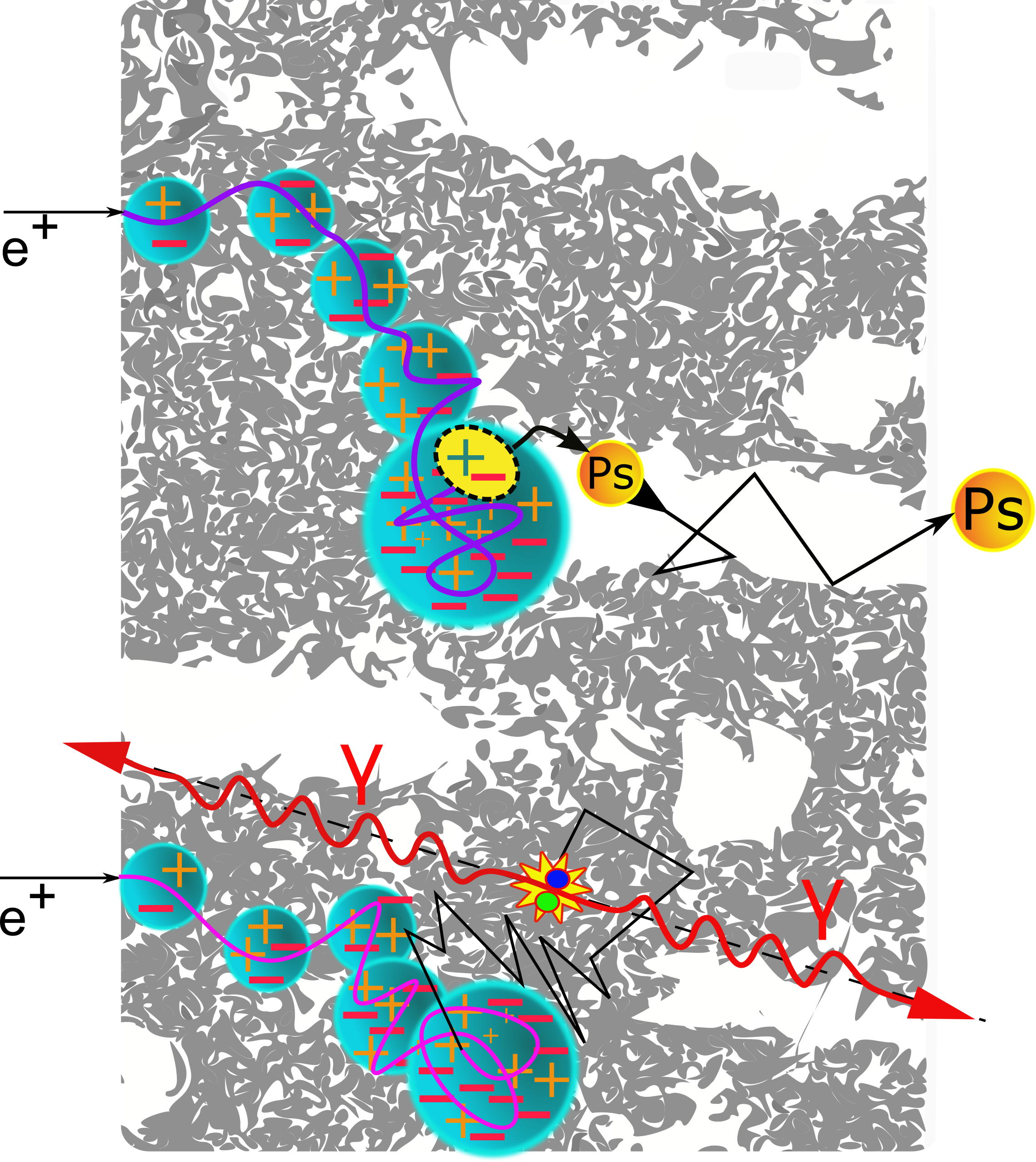}
\caption{
\label{fig:porous-material}
Left: Scanning electron microscope picture 
of the surface of a silicon positron to positronium converter with oxidized tunable nanochannels~\cite{Mariazzi2010}. Right: Pictorial illustration of positron thermalization in mesoporous material. Ionization places on a thermalization path are shown. They are composed of electrons (-), ions (+) and positron (blue line). In the upper example a positron thermalizes producing free electrons that quickly thermalize. Next, positronium is formed by recombination with a thermalized electron and Ps is localized in the void. After bouncing between the walls in the void (black arrows) it leaves the material. In the lower example a thermalized positron scatters in the material and annihilates directly into two photons (red arrows).
}
\end{figure}

In addition to this Ore mechanism with
positronium formation during the process of positron thermalization, even thermalized positrons can form positronium in condensed molecular media (dielectric liquids, polymers, molecular solids and ionic crystals).
For details, see for instance \textcite{Brandt1975, Eldrup1983,Sferlazzo1985,Wang1998,Jean2003}. 
During the positron thermalization process, a number of positive ions, free electrons, excited molecules and radicals are created. Freed electrons have a typical average kinetic energy of 10-50~eV \cite{Mogensen1974} that thermalize traveling for a few tens of nanometers in the material. (In ice this distance is roughly 30~nm \cite{Eldrup1983}.) Positronium can then be formed by the recombination of a thermalized electron and the thermalized positron. Two models of recombination have been introduced (the spur model by~\textcite{Mogensen1974} and the blob model by~\textcite{Stepanov2002}) and successfully applied to study Ps formation in solid \cite{Eldrup1983} and liquid water \cite{Stepanov2007} and, more in general, in molecular liquids and polymers \cite{Dauwe2005,Stepanov2005}. In bulk of crystalline metals or semiconductors, this bulk formation is hindered by the presence of free electrons that screen the positron-electron interaction destroying the Ps binding \cite{Schultz1988}. In such materials, positronium formation can occur only at the surface, where a thermalized positron reaching the surface picks-up an electron forming Ps \cite{Mills1978,Lynn1980}.
For unpolarized electrons and positrons forming positronium, 
each of four $e^+ e^-$ spin states 
$|\!\!\uparrow\uparrow\rangle$,
$|\!\!\downarrow\downarrow \rangle$,
$|\!\!\uparrow\downarrow \rangle$,
$|\!\!\downarrow\uparrow \rangle$
are equally populated. Formation of 
${\mbox{o-Ps}} = \left\{ |\!\!\uparrow\uparrow \rangle;\frac{1}{\sqrt{2}}(|\!\!\uparrow\downarrow \rangle +|\!\!\downarrow\uparrow \rangle);
|\!\!\downarrow\downarrow \rangle \right\}$
is three times more probable than 
p-Ps~=~$\frac{1}{\sqrt{2}}(|\!\!\uparrow\downarrow \rangle -|\!\!\downarrow\uparrow \rangle)$
so that
$f^p_{Ps} = \frac{1}{4}$ and
$f^o_{Ps} = \frac{3}{4}$. 
In materials, ${\mbox{o-Ps}}$ undergoes pickoff and conversion processes discussed below that shorten its lifetime relative to its decay in
vacuum. In the presence of voids and open volumes (see the right panel of Fig.~\ref{fig:porous-material}),
${\mbox{o-Ps}}$ formed in the material can be emitted in the porosity with a kinetic energy typically corresponding to its workfunction that is usually of the order of a few eVs~\cite{Tuomisaari1989,Nagashima1998}. Thanks to its relatively long lifetime, ${\mbox{o-Ps}}$ can lose a fraction of its energy by collisions with the walls of the pores~\cite{Chang1987,Vallery2003iz}. If the pores are interconnected and open to the surface (as illustrated in Fig.~\ref{fig:porous-material}(right)), ${\mbox{o-Ps}}$ can diffuse along the pore network and eventually be emitted into the vacuum with a significantly lower energy \cite{Vallery2003iz,Ito2005,Tanaka2006,He2007,Cassidy2010,Mariazzi2010}. 
{The Ps emission energy into the vacuum depends on the rate and duration of the energy transfer to the surrounding material that are determined by the pores shape, dimension and characteristics of their surface \cite{Vallery2003iz,Nagashima1995,Ito2005,Tanaka2006,He2007,Mariazzi2008,Cassidy2010,Mariazzi2010,Crivelli2010PRA,Mariazzi2021}.} 
In silica Ps formation occurs with an overall positron to positronium conversion efficiency up to 84\% \cite{VanPetegem2004}. Thanks to this characteristic, in recent years efficient sources of Ps have been synthesized by exploiting either silica-based disordered porous systems \cite{Cassidy2010,Liszkay2012,Consolati2013} or oxidized nanochanneled silicon targets \cite{Mariazzi2010,Mariazzi2010b}. The left panel of Fig.~\ref{fig:porous-material} shows an image of the oxidized tunable nanochannel in silicon. 

For fundamental physics experiments
such mesoporous based-silica converters are used for the production of positronium and its emission into vacuum 
%for basic physics experiments 
\cite{Cassidy2007,Cassidy2012,Wall2015,Cooper2016,Aghion2018xtd,Amsler2019erv,Gurung2020hms,Amsler2021}; see also Section IV.
Special meso-structured based-silica thin films enable emission of positronium in transmission (forward) and in reflection (backward) relative to the direction of the positron beam \cite{Andersen2015,Mariazzi2022}.

\subsection{Direct positron annihilation in matter}

A positron passing through matter may annihilate with electrons directly in flight \cite{Weber1999,Hunt2001,Cizek2012}. However, due to the fact that the cross-section of annihilation is inversely proportional to the positron velocity,
these annihilations represent only about 1\% \cite{Dryzek2007,Harpen2003zz} of the total annihilation rate. At the end of the positron thermalization path, when its energy is small (in the order of tens of eV compared to its initial energy of MeV) the annihilation rate becomes large. As explained in Section III.B and illustrated in
the right panel of Fig.~\ref{fig:porous-material}, a positron may either form Ps, or it may diffuse in the material until it is directly annihilated with an electron. The average time elapsing to direct annihilation of a thermalized positron can be quite long. For instance, in water it is much longer than the mean lifetime of p-Ps, 125 ps, and amounts to about 400-450 ps~\cite{Eldrup1972,Kotera2005}. 
{
The fraction of implanted positrons directly annihilating (${\rm f_d}$) in water has been estimated to range between ${\rm f_d}\sim0.2$~\cite{Blanco2016} and 
${\rm f_d}\sim0.6$~\cite{Harpen2003zz,Kotera2005}. In silica, where as seen in the previous section a large amount of implanted positrons form positronium, ${\rm f_d}$ can be smaller than 0.2 \cite{VanPetegem2004,Goworek2014}.  
}

\subsection{Positronium annihilation in matter}
\begin{figure}[t!]
\vspace{-0.4cm}
\centering
\includegraphics[width=0.4\textwidth]{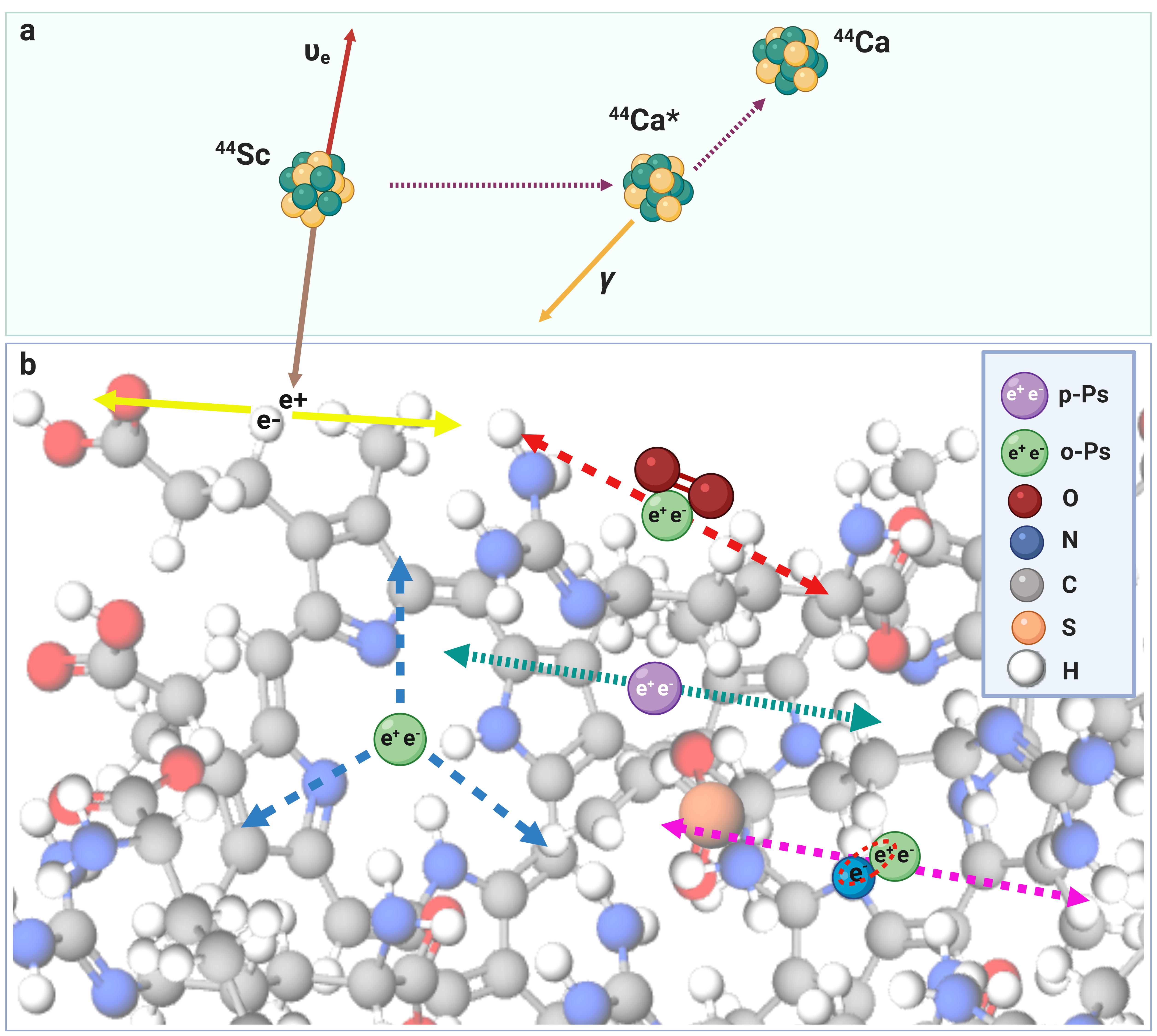}
\caption{
\label{fig:molecule_ewelina}
(a) Above, the decay of a $^{44}$Sc isotope, see Eq.~(\ref{eq:sodium-scandium}). 
(b) Below, illustration of the basic processes leading to the annihilation of a positron in the intra-molecular voids of a hemoglobin molecule. The most probable ways of annihilation are: direct annihilation into two photons (yellow solid arrows) with positron originating from the $^{44}$Sc decay, 
self-annihilation of p-Ps (green dotted arrows), self-annihilation of o-Ps (blue dashed arrows), o-Ps pick-off process (violet dotted arrows) and o-Ps conversion on O$_2$ molecule (red dashed arrows). The distances and size of atoms is shown to scale with the diameter of positronium twice as large as hydrogen.
}
\end{figure}

Positronium created in matter may annihilate via the processes shown in Fig.~\ref{fig:molecule_ewelina}: 
(i) self-annihilation in vacuum described by the decay constant $\lambda^o_0=7.04{\rm \mu s}^{-1}$ for o-Ps 
and $\lambda^p_0=7990.9{\rm \mu s}^{-1}$ for p-Ps, 
see Eqs.~(\ref{eq:G-oPs-1},\ref{eq:G-oPs-2},\ref{eq:G-pPs}); 
(ii) annihilation via pick-off process where a positron from positronium annihilates with the electron from the surrounding atoms ($\lambda_{\rm pick-off}$) -- see for example \textcite{Brandt1960,Wada2012}; 
(iii) o-Ps$\leftrightarrow$p-Ps conversion via spin exchange reactions with para-magnetic molecules such as, e.g., O$_2$ ($\lambda_{\rm conv}$) -- see for example \textcite{Ferrell1958,Kakimoto1987,Shinohara2001,Cassidy2007b} -- and 
(iv) other reactions such as oxidation ($\lambda_{\rm other}$) \cite{Stepanov2009}. 

The total decay rate is then expressed as $\lambda^{p(o)}(t,C) = 1 / \tau^{p(o)}_0 = \lambda^{p(o)}_0 + \lambda_{\rm pick-off}(t) + \lambda_{\rm conv}(C) + \lambda_{\rm other}(C)$, with dependence on time $t$ and the concentration of dissolved molecules, $C$. The pick-off rate is decreasing in time and conversion and other chemical reactions depend on the concentration of dissolved molecules.
{ 
In the biological samples, by analogy to water, 
the self-annihilation rate of p-Ps ($\lambda^p_0 = 7990.9 {\rm \mu s}^{-1}$) 
is much larger than the pick-off rate 
($\lambda_{\rm pick-off} = 512.8 {\rm \mu s}^{-1}$ in water), 
that in turn is much larger than conversion and other reactions rate 
($\lambda_{\rm conv} + \lambda_{\rm other}\approx 27 {\rm \mu s}^{-1}$ in O$_2$ saturated water),
which is larger than the o-Ps self-annihilation rate ($\lambda^0_0 = 7.0401 {\rm \mu s}^{-1}$).
In porous materials the relation changes and $\lambda_{\rm conv}\approx 25 {\rm \mu s}^{-1}$ (O$_2$ at 1 atm.) 
$> \lambda^0_0=7.0401 {\rm \mu s}^{-1} > \lambda_{\rm pick-off}\approx 1{\rm \mu s}^{-1}$.
More details are given in the caption of Fig.~\ref{fig:Ps-decay-scheme} and in the following sections.
}

\subsubsection{Annihilation via pick-off process}
{
The annihilation rate via the pick-off process ($\lambda_{\rm pick-off}$) may be treated as independent of the positronium spin,
with the same value for o-Ps and p-Ps ($\lambda^o_{\rm pick-off}$~=~$\lambda^p_{\rm pick-off}$)
\cite{Dupasquier1991}, and expressed as
$\lambda_{\rm pick-off}$~=~$\xi$($\frac{1}{4}\lambda^p_0+ \frac{3}{4}\lambda^o_0$)~$\approx$~
 $\xi$~($\frac{1}{4}\lambda_{2\gamma}+ \frac{3}{4}\lambda_{3\gamma}$). 
Here $\xi$
denotes the positron-electron contact density normalized with respect the vacuum value. $\lambda^p_0$ and $\lambda^o_0$ denote self-annihilation rate of p-Ps and o-Ps, respectively. The fractions  $\frac{1}{4}$ and $\frac{3}{4}$ originate from the spin projection combinations as explained earlier. $\lambda_{2\gamma}$ and $\lambda_{3\gamma}$ denote the decay rate into 2 and 3 photons, respectively.
Assuming, to good approximation, that 
p-Ps self-annihilates to $2\gamma$ and o-Ps to $3\gamma$
the relative rate of $3\gamma$  and $2\gamma$ pick-off annihilations is  
$3\lambda_{3\gamma}/\lambda_{2\gamma}
\approx
3 \tau_{{\rm p-Ps}} / \tau_{{\rm o-Ps}} \approx \frac{1}{378}$, where $\tau_{{\rm p-Ps}}$ and $\tau_{{\rm o-Ps}}$ denote the mean lifetime of p-Ps and o-Ps, respectively.
} %
The rate constant for pick-off annihilations ranges from a fraction of ${\rm \mu s}^{-1}$ in some mesoporous materials \cite{Saito1999,Jasinska2016} to about 
$\lambda^{\rm water}_{\rm pick-off} = 513{\rm \mu s}^{-1}-550{\rm \mu s}^{-1}$ in water~\cite{Stepanov2020,Shibuya2020}. 
The value of $513\mu s^{-1}$ is small compared to the self-annihilation rate constant of p-Ps, $\lambda^p_0 = 7990.9{\rm \mu s}^{-1}$.
Therefore, the p-Ps mean lifetime $\tau^p$
is shortened due to pick-off by only about a few picoseconds ($\tau^p_0 - \tau^p = \frac{1}{7990 {\rm \mu s}^{-1}} - \frac{1}{7990{\rm \mu s}^{-1} +513 {\rm \mu s}^{-1}}\approx 7$ ps).
In contrast, the self-annihilation rate of o-Ps with $\lambda^o_0 = 7.0401{\rm \mu s}^{-1}$ is much smaller than the pick-off rate in liquids and the o-Ps mean lifetime is significantly shortened, e.g., down to 1.8 ns in water compared to 142 ns in vacuum. In mesoporous materials the pick-off rate constant depending on the structure is of the order of 1 ${\rm \mu s}^{-1}$. It decreases the o-Ps mean lifetime by tens of nanoseconds. For example, for the IS3100 aerogel with $\tau_{\rm o-Ps} = 132$~ns~\cite{Jasinska2016}, one finds $\lambda^o_{\rm pick-off} = \frac{1}{\tau_{\rm o-Ps}}-\lambda^o_0 \approx 0.6{\rm \mu s}^{-1}$.
It is important to emphasize that the pick-off annihilation rate is not constant in time. As mentioned in section III.B and illustrated in Fig.~\ref{fig:porous-material}, positronium after formation is bouncing between the void's walls losing its energy and hence slowing down. Therefore, the average time intervals between subsequent positronium interactions with electrons from surrounding molecules are growing and the pick-off rate decreases in time. Experimentally, this may be controlled by determining the time dependence of the ratio of the 3$\gamma$ (self-annihilation) to 2$\gamma$ (pick-off annihilation).

\subsubsection{Positronium conversion and oxidation}
\label{subsec:conversion}
Positronium in matter takes part in chemical reactions with radiolytic products (e.g., {solved in water as aqueous electrons},
H$_3$O$^+$, OH-radicals) created by the positron during thermalization and in reactions with dissolved substances~\cite{Stepanov2007,Stepanov2020}. Interaction of positronium with dissolved para-magnetic molecules possessing magnetic moment, e.g., molecular oxygen 
O$_2$, may lead to the spin exchange and conversion of p-Ps into o-Ps and o-Ps into p-Ps, e.g., via the ${\rm \mbox{o-Ps} + O_2 \to \mbox{p-Ps} + O_2}$ reaction \cite{Ferrell1958,Stepanov2020,Kakimoto1987,Shinohara2001}. (Non-paramagnetic molecules as N$_2$ are not causing conversion reactions). 
In addition the O$_2$ molecule may also oxidize positronium via the ${\rm Ps + O_2} \to e^+ + {\rm O_2^{-}}$ process.
Both processes, o-Ps conversion and oxidation, contribute to the quenching of o-Ps. Conversion in some organic liquids (cyclohexane, isooctane, isopropanol) is 5 to 10 times more frequent than oxidation \cite{Stepanov2020}. 
The conversion and oxidation rate constants depend on the dissolved oxygen concentration $C_{{\rm O}_2}$: $\lambda_{conv} +\lambda_{other} = k_{O_2} \cdot C_{O_2}$, with 
value
$k_{O_2}$
for water 
measured to be
$0.0204\pm0.0008 \
{\rm \mu mol^{-1} \mu s^{-1} L}$ ~\cite{Shibuya2020} %
and  
$0.0186\pm0.0010 \
{\rm \mu mol^{-1} \mu s^{-1} L}$~\cite{Stepanov2020}.
This gives 
$\lambda_{conv} +\lambda_{other} \approx 27 {\rm \mu s}^{-1}$ with saturated O$_2$ in water $\sim 1400 \
{\rm \mu mol L^{-1}}$.
Thus the conversion rate depends linearly on the dissolved O$_2$ concentration. 
In mesoporous materials at relatively low concentrations, $< 0.05$~atm., it exceeds the pick-off rate. 
On the other hand,
at $0.25$~atm. it exceeds the self-annihilation rate of o-Ps~\cite{Zhou2015}. 
In water the conversion rate is much lower than that of pick-off, but in some organic liquids with high oxygen solubility it may become the dominant effect \cite{Stepanov2020}.
The main annihilation processes of the thermalized positron in a molecule of hemoglobin, which is the most important component of red blood cells (erythrocytes), an example relevant for medical applications discussed in Section~\ref{sec:medical-applications}, is shown in Fig.~\ref{fig:molecule_ewelina}.

\section{Fundamental physics experiments with positronium}
\label{sec:fundamental-physics}
In the last few decades, the development of techniques for trapping many positrons and forming bunches containing up to several $10^7$ positrons \cite{Surko1989,Murphy1992,Cassidy2006,Danielson2015} and the development of efficient positron to positronium converters (see Section III) is allowing a significant advancement in the field of experimental positronium physics \cite{Cassidy2018tgq}. This includes experiments with positronium spectroscopy, tests of gravity on antimatter and production of a positronium Bose-Einstein condensate. 

\subsection{Positronium Spectroscopy} 

Positronium spectroscopy 
presently focuses on precision measurements of hyperfine transitions between singlet and triplet states \cite{Cassidy2018tgq}.
Today hyperfine splittings in positronium are measured to the MHz level whereas the theoretical NRQED calculations are typically at the kHz level.
There are several interesting  ``discrepancies'' at the few standard deviations level between measurements of hyperfine splittings and NRQED predictions
\cite{Adkins:2022omi}.

Early measurements of the frequency of the 1$S$ hyperfine interaction \cite{Ritter1984mqy,Mills1983zzd} are about 3 standard deviations below the theoretical NRQED predictions \cite{Adkins1997xrp,Hoang1997ki,Czarnecki1998zv},
or about one part in 10$^5$.
More recent measurements are reported in \textcite{Ishida2011ds,Ishida2013waa} and \textcite{Miyazaki2014bla}
with the first closer to the theoretical prediction, to within 1.2 standard deviations. 
These results have raised discussion about possible systematics in the experiments \cite{Heiss2018jbl} and the theory and contributing Feynman diagrams \cite{Karshenboim2003vs}.

Motivated by the situation with the 1$S$ splitting, the ETH Z\"urich group plan an in vacuum precision measurement
to look at the $2 ^3S_1 \to 2 ^1S_0$ transition and to compare with NRQED predictions. This new experiment
will be free of systematic uncertainties inherent in previous 1$S$ HFS transition measurements \cite{Heiss2018jbl}.

Most recently \textcite{Gurung2020hms,Gurung2021xss} have measured the $2 ^3S_1 \to 2 ^3P_0$ transition on Ps emitted into vacuum from mesoporous silica targets
and determined the transition frequency $\nu_0=18501.02 \pm 0.61$ MHz.
This value differs from the NRQED prediction $\nu_0=18498.25 \pm 0.08$ MHz, where the quoted theoretical error includes an estimate of unknown higher order NRQED corrections.
The difference is about one part in $10^4$ -- a $4.5 \sigma$ effect.

The intriguing status of these measurements and their relation to theory calls for new experiments.
If there are no underestimated systematics in the experiments,
{ 
given the constraints on possible new interactions coupling to the electron discussed in Section II,
then attention will turn to QED bound state theory.
What from QED might 
be missing in 
present NRQED calculations?}
Assuming no large extra diagrams waiting to enter at next order, one might consider the input assumptions to the NRQED bound state calculations.
One effect might be a slightly underestimated harder momentum distribution of electron velocities in the positronium wavefunctions, e.g., on external lines. Alternatively, one might consider enhanced Ps resonance contributions as admixtures in the self-energy diagrams for excited states.

\subsection{Positronium in gravity tests and Bose-Einstein condensates}

{
Going beyond 
positronium spectroscopic and decay measurements
positronium also plays an important role in other fundamental physics experiments: 
tests of the  equivalence principle 
through the effect of gravity on antimatter 
and possible 
Bose-Einstein condensates 
involving antimatter. 
}

To measure the effect of gravity on
antimatter,  
{on the one hand o-Ps in excited states is being used by two experiments at CERN’s Antiproton Decelerator, AEgIS ~\cite{Doser2018zfc} and GBAR ~\cite{Perez2012, Dufour2015}, as an intermediate tool to produce antihydrogen via a charge-exchange reaction with antiprotons ~\cite{Amsler2021}}. The goal of these experiments is to measure the acceleration experienced by antihydrogen in the gravitational field of the earth. 
{The cross section of the charge-exchange reaction, for high value of the principal quantum number of o-Ps, is expected to scale with the fourth power of the principal quantum number itself \cite{Krasnicky2016}. Consequently, production of antihydrogen via charge-exchange reaction will take benefit from the efficient laser excitation to Rydberg states demonstrated in the last decade on o-Ps emitted into vacuum from silica-based converters \cite{Cassidy2012,Wall2015,Aghion2016}.} On the other hand, long-lived positronium states have been proposed as probes for direct measurements of gravity on a matter-antimatter system \cite{Mills2002,Oberthaler2002, Mariazzi2020bgc}. Both o-Ps exited to Rydberg states \cite{Cassidy2012} and to the metastable $2 ^3S$ level \cite{Amsler2019erv} have been proposed for such measurements.

Ortho-positronium has the potential to form a Bose-Einstein condensate, BEC, at high densities \cite{PhysRevB.49.454}
which, if observed, would be the first BEC involving antimatter with the experiments also providing information about high density o-Ps collisions and possible o-Ps molecule formation.

A recent suggestion \cite{PhysRevA.100.063615} involves taking a hollow spherical bubble containing thousands of spin-aligned o-Ps atoms in superfluid liquid $^4$He.
The bubble would be stable against breakup into smaller bubbles, and the Ps would form a BEC with a number density of  $\sim 10^{20}$~cm$^3$  and a BEC critical temperature $T_c \approx 300$~K. With present experimental methods bubbles might be formable containing about $10^5$ o-Ps atoms.

{For a BEC involving spin-flip from o-Ps to p-Ps}, the spontaneous radiation of positronium atoms in the BEC due to the two-photon collective annihilation decay might be used as an intense $\gamma$-ray source \cite{PhysRevLett.113.023904}.
Due to BEC coherence the number of emitted photons will grow at every place in the condensate.
For laser production with direction focused radiation an elongated condensate might be used.
Spontaneously emitted entangled and opposite directed photon pairs will be amplified, leading to an exponential buildup of a macroscopic population into end-fire modes in the elongated condensate.

\subsection{Photon Entanglement in Positronium Decays}
\label{sec:entanglement}
Quantum entanglement of the emitted photons in positronium decays is interesting as a fundamental physics issue~\cite{Acin2000,Hiesmayr2017xgx,Marek2017}. It also has interesting applications in PET devices for medical diagnostics~\cite{McNamara2014,Toghyani2016,Moskal2018pus,Hiesmayr2018rcm,Caradonna2020,Watts2021}. 
Thus far the entanglement of photons from positronium, though predicted by theory, has not been observed  experimentally. Implementation of such experiment is challenging since the polarisation of photons in MeV energy range cannot be studied using optical methods.

Photons are spin-1 particles
characterized by their 
momentum and polarization, 
with two transverse polarization states for real
photons.
In the linear polarization basis, the 2$\gamma$  state $|\psi \rangle$ originating from p-Ps can be written as
\begin{equation}
|\psi \rangle = \frac{1}{\sqrt{2}} 
\left( |H \rangle_1 \otimes |V \rangle_2 - |V\rangle_1 \otimes |H\rangle_2 \right)
\label{2gstate}
\end{equation}
where $|H \rangle$ and $|V \rangle$ denote the horizontal and
vertical polarized states, and the symbol $\otimes$ refers to a tensor product. 
The minus sign
between the two combinations reflects the parity -1 eigenvalue of ground state positronium. 
The entanglement of the 2$\gamma$ state described in Eq.~(\ref{2gstate}) manifests itself in the fact that there is no choice of basis ($|A \rangle$,$|B \rangle$) in which the state could be described by the single tensor product of $|A \rangle \otimes |B \rangle$
--
this we call entanglement. 
Moreover, Bose-symmetry and parity conservation in the decay of p-Ps imply that the state of the resulting two photons is maximally entangled 
and that the photons polarizations 
are orthogonal to each other,
$\vec{\epsilon}_1 \perp \vec{\epsilon}_2$
~\cite{Hiesmayr2018rcm}.
\begin{figure}[t!]
\vspace{-0.5cm}
  \includegraphics[width=0.48\textwidth]{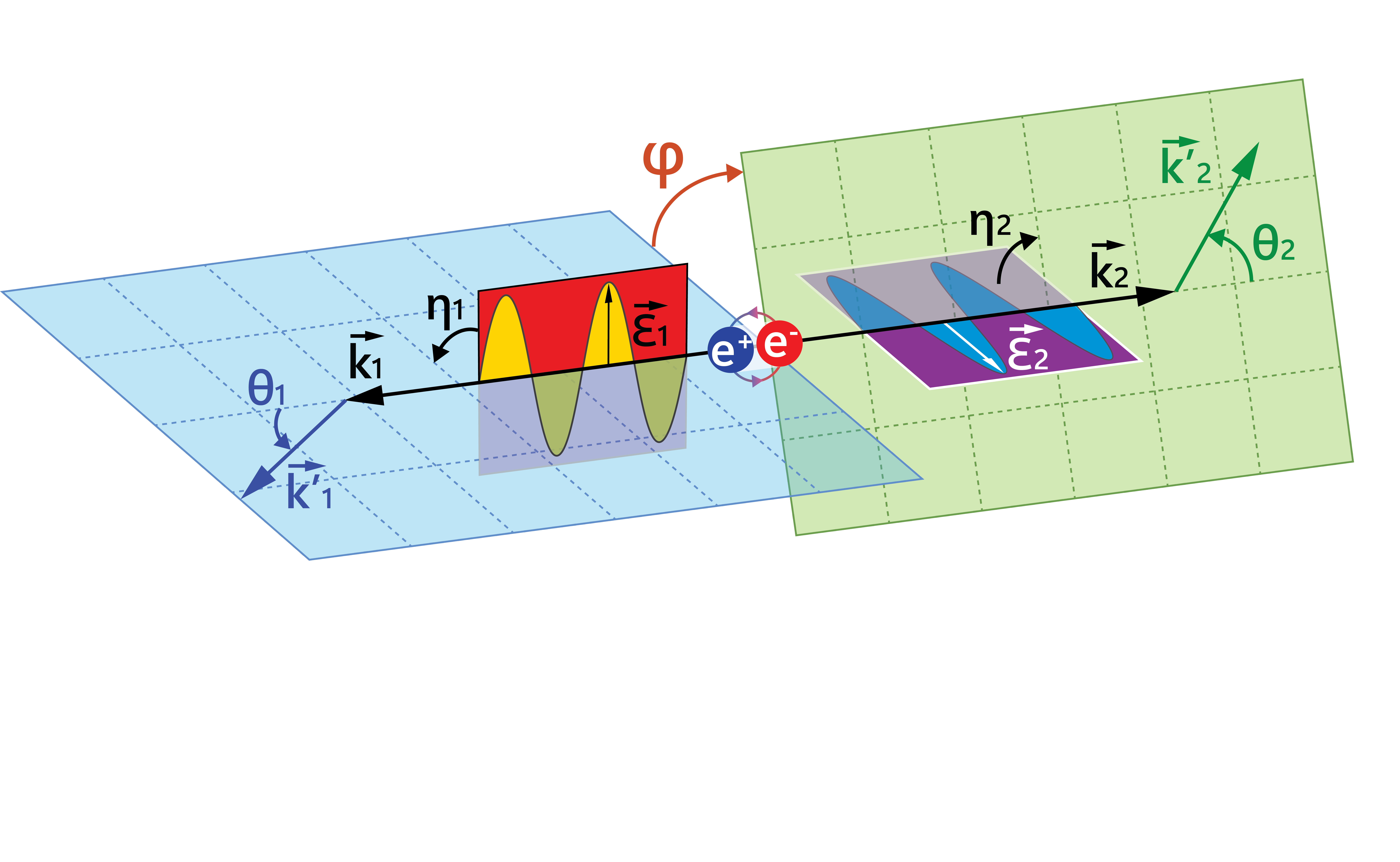}
\vspace{-2cm}
\caption{Schematic illustration of Compton scattering of two photons originating from p-Ps annihilation. Due to the momentum conservation ($\vec{k_1} = -\vec{k_2}$), the annihilation photons propagate back-to-back along the same axis in the p-Ps rest-frame. $\theta_1$ and $\theta_2$ denote the scattering angles, $\eta_1$ and $\eta_2$ denote the angles between the scattering planes and the polarisation directions $\vec{\epsilon}_1$ and $\vec{\epsilon}_2$, respectively, and $\varphi$ indicates the relative angle between the scattering planes.
\label{fig:BackToBack}  
}
\end{figure}
Photons in the energy range of MeV interact in matter predominantly with electrons through the photoelectric and Compton effects. 
Compton scattering
(Fig.~\ref{fig:BackToBack}) may be used for the estimation of the linear polarization of the incoming photon, since the scattering is not isotropic and it is most probable in the plane perpendicular to the polarization of the incoming photon~\cite{Klein1929}. 
For the 
p-Ps $\to 2\gamma$ process (Fig.~\ref{fig:BackToBack}), when each $\gamma$ interacts via Compton scattering with an electron one can estimate the angle between the polarization directions of the photons $|\eta_1 - \eta_2|$ 
by measurement of the relative angle $\varphi$ between the scattering planes~\cite{Moskal2018pus}. The distribution of $\varphi$ may serve for studies of quantum entanglement~\cite{Hiesmayr2018rcm}. The experimentally determined distributions~\cite{MoskalIEEE2018,Watts2021,Abdurashitov2022} indeed peak for $\varphi$~=~90$^\circ$ and are consistent with predictions obtained under the assumption that photons are entangled.  
However, in order to prove the entanglement of photons from p-Ps, measurements in at least two different bases are required~\cite{Hiesmayr2018rcm},
e.g., $(|H \rangle,|V \rangle)$ and $(|+45^\circ \rangle,|-45^\circ\rangle)$.
Yet, the researcher has no influence on the Compton scattering and an active basis choice cannot be realized. Therefore, so far the experimental challenge of proving the entanglement of photons from positronium decays remains open. 
However, the strong correlation between the photon's polarisation may be applied in medical diagnostics 
as discussed in Section~\ref{subsec:QET}.  

\section{Development of positronium research in biology and medicine 
}
\label{sec:biology}
As discussed in the previous sections, the predictions of positronium properties based on NRQED theory are currently many orders of magnitude more precise than the experimental results. Experimental precision is to large extent limited because positronium is produced in medium, and its properties in materials are altered with respect to the vacuum. Yet, the variation of positronium properties as a function of the nano-structure of the material  and concentration in it of paramagnetic molecules constitute the basis for its application in studies of materials as well as in studies of biological processes in living organisms and potentially also in medicine. 

Although positron 
emitting radionuclides have been used in diagnostic medicine since \textcite{Kuhl1963}
developed the foundations of medical positron emission tomography (PET) in the late 1950s, the properties of the positronium atom have never until recently been used in medicine. Only recent advances in developing multi-photon tomography~\cite{Moskal2021nature} and imaging of positronium properties~\cite{Moskal2021science} opened realistic perspectives of making use of positronium as a diagnostic indicator of the tissue pathology in clinics~\cite{MoskalStepien2022}.

In this Section we explain how positronium can help in understanding the structure of biological objects, how positronium is used in life-science and why its properties should be translated to medicine.

The method used for the studies is called positron annihilation lifetime spectroscopy (PALS) and it is based on the measurement of positron lifetime spectrum in the investigated sample.  The positron may be implemented to the sample by using positron beam or by the application of $\beta^+$ radionuclides such as, e.g., $^{22}$Na or $^{44}$Sc. Typically in the PALS method two scintillation detectors are used.  One detector for determining time of the annihilation photon originating from the electron-positron annihilation, and the other for determining the time when positron enters the sample which is established by the measurement of prompt gamma photon emitted by the excited daughter nucleus as described in Eq.~(\ref{eq:sodium-scandium}). The measured lifetime spectra enables one to extract the intensities and mean lifetime distributions of positrons undergoing annihilations due to various processes depicted in Fig.~\ref{fig:Ps-decay-scheme}.

The first measurements of intensity ($I_{\rm o-Ps}$) and the mean lifetime  ($\tau_{\rm o-Ps}$) of o-Ps were performed on samples containing organic compounds, benzene derivatives. This experiment  showed significant differences in the signal intensity ($I$) in the presence of halogen atoms (F, Cl, Br, I), which was explained by the increased electron density of such molecules~\cite{Hatcher1958}. 
An attempt in those studies was made to find a relation between the mean lifetime and the dissociation energy of the molecular bonds in simple organic compounds and no suggestion was then given that the functional groups, similar to those organic compounds, might affect the average lifetime of o-Ps ($\tau_{\rm o-Ps}$) in biological samples~\cite{Kerr1962,Brown1974}. Immediately after finding positronium properties in organic fluids, \textcite{Gustafson1970}
measured $\tau_{\rm o-Ps}$ in a biological sample (muscle).
However, he focused his attention on the order and arrangement of tissue water and not on muscle structure. Further studies on complex biological systems like biological membranes or proteins moved positronium research towards structural biology and showed that this approach was efficient in the study of biological reactions (such as electron transfer~\cite{Jean1977}), phase transition of lipids~\cite{Stinson1980,Chow1981,Jean1982} macromolecule structure~\cite{Handel1976}, hydratation~\cite{Handel1980,Gregory1992,Akiyama2007} and porosity~\cite{Pamula2006,Chamerski2017} of biological samples. Nowadays, after many years of focused research, 
PALS 
appears to be a promising technique in the investigation of the structure of macromolecules~\cite{Chen2012} and clinical samples~\cite{Moskal2021,Moskal2021science,Fulvio2022,Zgardzinska2020}.

\subsection{Positronium in biological materials and systems}
{The first applications of positronium properties in the life sciences were made by Handel in late 70s~\cite{Handel1976,Handel1980}. Having in mind that positron annihilation lifetime is sensitive to changes in free volume caused by pressure (which is rarely studied in biological systems) or by thermal expansion in the same phase, he showed significant changes in positronium lifetime during phase transitions of biological systems.}
{Covalent bonds between carbon atoms in organic molecules hold structure (architecture) of biological macromolecules contributing to the creation of free volumes (so-called molecular voids).}
Such a molecular structure (here we call it nanostructure~\cite{Pethrick1997}) is changing dynamically with temperature~\cite{Handel1976,Stinson1980,Sane2009}, and is stabilized by hydrogen {bonds}~\cite{Handel1980,Gregory1992,Kilburn2006}.
The specific feature of the shortening of $\tau_{\rm o-Ps}$ in the pick-off process has been proposed 
for probing sub-nanometer-sized local free volumes in solid materials and organic polymers to assess the size and nature of chemical environment~\cite{Pethrick1997,Dlubek2000}. Positronium mesurements in biological samples have been performed with a liquid $^{22}$Na source (such as NaCl solution) prepared in a thin-walled glass bead~\cite{Handel1976} sealed between thin mylar films~\cite{Chow1981}, Al foil~\cite{Jean1977,Jean1982} sealed between a Kapton foil~\cite{Stinson1980,Gregory1992,Pamula2006,Bura2020,Moskal2021} or directly dissolved as an open source~\cite{Sane2009}. 

In life sciences two types of bio-materials,
hydrogels and biomembranes, were studied intensively to characterize molecular structure (porosity, permeability and hydrophobicity) in the context of their biological activity~\cite{Pamula2006,Sane2009}. 
Various aspects of
PALS applied
to life sciences 
are discussed in the review by~\textcite{Chen2012},
which summarizes
recent knowledge about possible application of o-Ps in biology.

Fluidity and regularity of biological membranes are changing depending on the phospholipids (e.g., POPC - palmitoyl-oleoyl-glycero-phosphocholine) and cholesterol content, which can be observed as changes in $\tau_{\rm o-Ps}$. For example, if cholesterol content is at the high value of 40\% and DPPC (dipalmitoylphosphatidylcholine) is 60\%, $\tau_{\rm o-Ps}$  reaches the lowest value of $\sim$1.86 ns~\cite{Sane2009}. Admixture of ceramides and cholesterol ceramide interactions in DPPC membranes also influence $\tau_{\rm o-Ps}$ by changing the free volume void pattern ~\cite{Axpe2015,Garcia-Arribas2016}. 
The slope of $\tau_{\rm o-Ps}$ rises rapidly where the membrane undergoes a gel-fluid transformation, at the transition temperature of the lipid systems~\cite{Jean1982, Sane2009} and biological membranes (red cell ghosts)~\cite{Chow1981}.
In contrast to lipid systems, where  membrane permeability is regulated by lipid fluidity~\cite{Sane2009}, in hydrogels, interaction between polymers and water is a crucial process regulating their biological activity~\cite{Sane2011}. This process can be successfully studied by means of PALS~\cite{Pamula2008}. 
The dehydration process in crystallized and amorphous state of maromoleculess (trehalose) is marked significantly by sharp changes in the mean $\tau_{\rm o-Ps}$ and intensity, which are related to changes in the total free volume fraction~\cite{Kilburn2006}.

\subsection{Positronium in ex vivo research}
The first experiment on a biological sample was performed in 1970 and dedicated to studying  semicrystaline structure of water in muscle cells~\cite{Gustafson1970,Chen2012}. However, the pioneering experiments on $ex \, vivo$ samples showing that small temperature variations cause detectable changes in free voids were done on bovine liver and rabbit muscle~\cite{Elias2001}. To develop technical details of o-Ps measurements, a number of experiments were performed on human and mice skin to study differences in the mean {$\tau_{o-Ps}$} of normal cells and cancer (basal cell carcinoma and squamous cell carcinoma)~\cite{Jean2006,Jean2007}. This approach appeared to be promising and indicated a reduction of free volume at the molecular level for the skin with cancer 
while the
number of patients was the limitation to conclude about usefulness of positronium imaging in cancer diagnostics~\cite{Liu2007}. Extending these studies to skin melanoma showed that positrons annihilate at a smaller rate with increase of melanoma cells, which confirmed o-Ps utility in biomedical research~\cite{Liu2008}. In addition to human and animal tissues, also unicellular organisms and multicellular tissue-like structures (spheroids) were investigated giving promising results in positronium research ~\cite{Kubicz2015,Axpe2014,Karimi2020}.
Positronium annihilation in tissues strongly depends on water content. 
In highly hydrated organs (lens) or tissues (myoma), the mean o-Ps lifetime is below or around $\sim$2 ns~\cite{Sane2010,Zgardzinska2020}.
In adipose tissue this time is significantly increased~\cite{Moskal2021,Moskal2021science,Fulvio2022}
confirming the observation from biological systems 
that structural characteristic and molecular composition determine positronium annihilation.

\section{Medical applications of positrons and positronium}
\label{sec:medical-applications}
Non invasive imaging of the interior of the body constitutes a 
powerful diagnosis tool enabling personalized and targeted therapy.
Here we briefly report on tomographic methods based on positron and positronium annihilations inside living organisms. 
We begin with the description of Positron Emission Tomography (PET) that is a well established diagnostic method delivering information about the metabolism rate of administered pharmaceuticals, and about receptor expression on cell membranes~\cite{Humm2003,Conti2009,Vanderberghe2020,Alavi2021bioalg}. 
PET is based on the administration to the living organism of pharmaceuticals labeled with positron-emitting isotopes, e.g., Fluoro-Deoxy-Glucose (FDG) labeled with $^{18}$F or Prostate Specific Membrane Antigen (PSMA) labeled with $^{68}$Ga for metabolic and receptor imaging, respectively~\cite{Moskal2020petclin}. 
The image of the annihilation density distribution is reconstructed based on the registration of $2\gamma$ events 
originating mostly from direct annihilations, p-Ps self-annihilations and o-Ps pick-off processes~(Fig.~\ref{fig:Ps-decay-scheme}). The reconstructed annihilation density distribution corresponds to the image of the metabolic rate (glucose taken by a cell) or to the image of cancer cell expression (density of cancerous receptor on a cell)  depending on the administered radio-pharamaceuticals. 
In 2019 the first  total-body PET systems were introduced to clinics which enable dynamical imaging (filming) of all organs and tissues in the body simultaneously~\cite{Vanderberghe2020,Badawi2019,Karp2020,Surti-review2020}. So far PET detectors reconstruct only annihilation position distribution. Only recently the method of {\it positronium imaging} was developed which enables one to image in the living organisms properties of positronium such as a mean lifetime, production intensity, and the 3$\gamma$ to 2$\gamma$ decay rate ratio~\cite{MoskalIEEE2019,Moskal2019pmb,Moskal2020ejnmmi}. 
{\it Positronium imaging} requires multi-photon tomography systems enabling  registration of electron-positron annihilations not only into two photons (as in standard PET) but also decays to three photons, as well as simultaneous registration of annihilation photons and prompt photon from the radionuclide deexcitation. Such tomography systems as well as three-photon and mean lifetime image reconstruction methods were recently demonstrated by the J-PET collaboration~\cite{Moskal2021nature,Moskal2021science}.  
The first ex-vivo positronium images of healthy and cancer tissues were published by~\textcite{Moskal2021science}.
The o-Ps mean lifetime (Section~\ref{sec:positronium-in-materials}) 
tells one
about the size of intra-molecular voids (free volumes between atoms), whereas its formation probability (production intensity) informs one about the voids concentration. Both lifetime and production intensity depend on the bio-active molecule concentration. Notably, positronium may serve as a biomarker for the assessment of tissue pathology~\cite{NRP2019,Moskal2021,Moskal2021science} and may be of particular diagnostics relevance as a biomarker of the concentration of oxygen~\cite{MoskalStepien2021hypoxia} (section~\ref{sec:hypoxia}). It is important to stress that positronium mean lifetime imaging delivers information complementary to the metabolic and receptor PET images and is also complementary to anatomic (electron density distribution) and morphological (hydrogen atom density distribution) images  achievable with Computed Tomography (CT) and Magnetic Resonance (MR) tomography, respectively.
\subsection{Positron Emission Tomography}
The healthy and cancerous tissue differ, e.g., by the expression profile of receptors at the cell membranes, by the angiogenesis resulting in different concentrations of oxygen molecules, or by glucose metabolism rate.
We next explain how a cancer cell, which needs more glucose for its metabolism and unlimited divisions or has the vastness of cancerous receptors on its surface, can be distinguished 
using PET scans
from a healthy cell which is rather modest in its needs and behaviour. 

The number of PET scans has doubled within the last 10 years reaching around 2,000,000 PET scans per year in USA (2017) and  45,000 in Poland (2016)~\cite{Cegla2021}.  Typically and most commonly, $^{18}$F-Fluoro-Deoxy-Glucose (FDG) is used as a positron-emitting compound (radiopharmaceutical) in PET for testing cancer and brain metabolism. This radiotracer was developed and first tested in humans for imaging cerebral glucose metabolism in 1976 at the University of Pennsylvania~\cite{Reivich1979,Alavi2002,Alavi2021bioalg}, and is still used in around 90\% of PET scan examinations.
\begin{figure}[t!]
\centering
\includegraphics[width=0.42\textwidth]{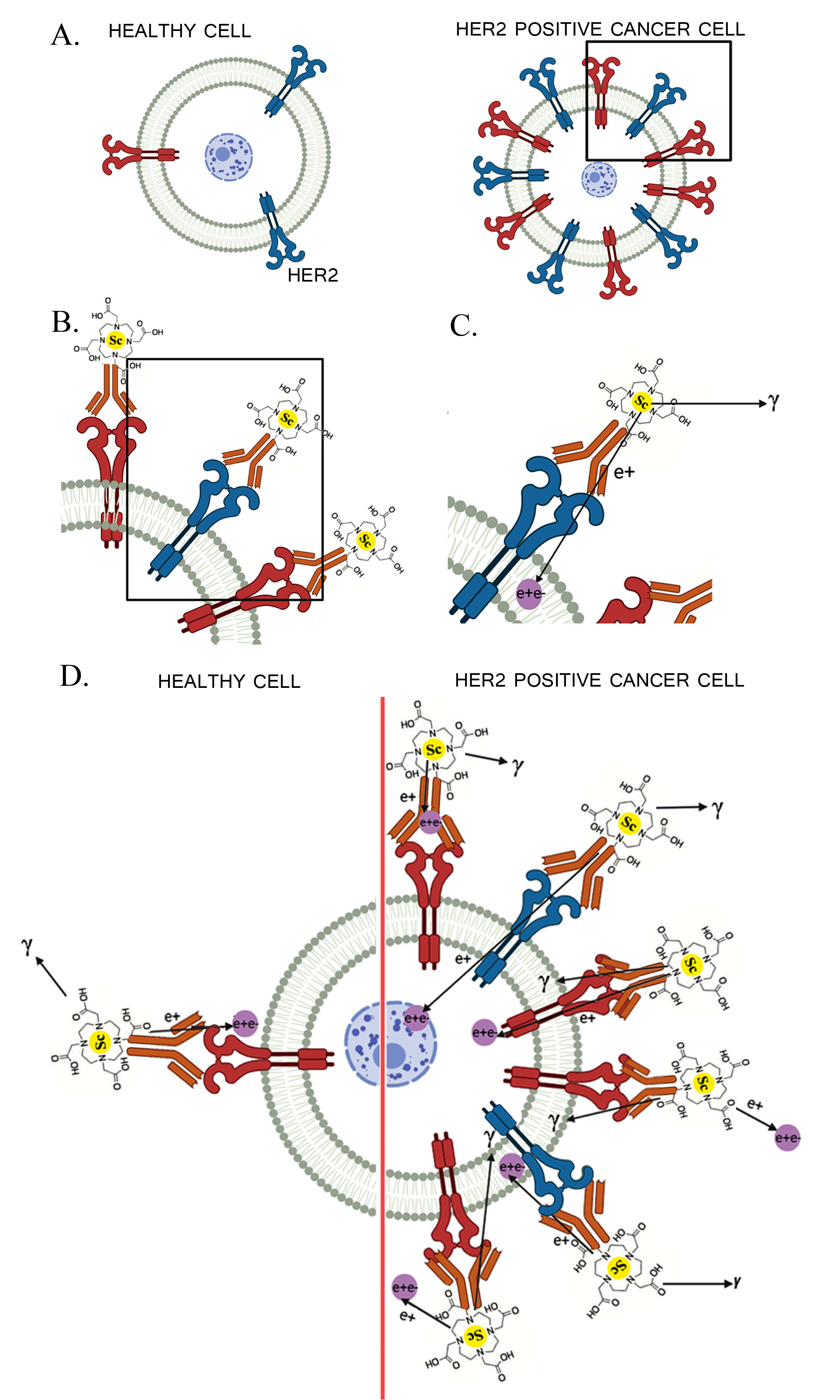}
\vspace{-0.3cm}
\caption{
Positronium imaging of
HER2-positive cancer cells.~A.~The HER2 receptor (epidermal growth factor receptor - EGFR) is scarcely expressed on the surface of healthy cells and significantly (100-times or more) overexpressed on cancer cells (e.g. breast cancer~\textcite{Venter1987,Rubin2001}). Different colours represent combinations of different units forming dimmers of the HER2 molecule. B.~Trastuzumab (herceptin), a humizied monoclonal antibody that binds to HER2, is labelled with $^{44}$Sc isotope.~C.~$^{44}$Sc isotope emits a~prompt $\gamma$ and a positron ($e^{+}$), to form a positronium atom.~D.~Positronium atoms annihilate in cells (with highest abundance in cancer), on their surface and within cell organelles (cell membranes, cytosol $nuclei$). Environmental factors like temperature, water content, and other specific tissue features like chemical and molecular composition, determine the $\tau_{\rm o-Ps}$ in the diagnosed tissue.
\label{fig:cells_receptors}
}
\end{figure}
FDG is a glucose analogue, where at the second carbon atom in a glucose ring ($C2$), the normal hydroxyl group (-OH) is substituted with the $^{18}$F isotope. The half-lifetime of the $^{18}$F isotope is 110 min which makes $^{18}$F-FDG a useful radiopharmaceutical in the diagnosis of disease processes characterized by increased glucose consumption, primarily in neoplastic diseases, for the assessment of brain metabolism or myocardial viability, in drug-resistant epilepsy, diagnosis of Alzheimer's disease spectrum (ADS), inflammatory processes, and systemic diseases~\cite{Alavi2019,Alavi2021bioalg}. 
After administration via intravenous injection, FDG is distributed through the blood stream within minutes and is actively transported into the cells by specific glucose transporters - membrane proteins which contribute in glucose uptake (mostly GLUT1 and GLUT3~\textcite{Marom2001,Avril2004}). Normally, once phosphorylated, a glucose molecule continues along the glycolytic pathway (glycolysis) for energy production. However, FDG cannot undergo glycolysis because the -OH group is substituted with the $^{18}$F atom. Only after $^{18}$F decays radioactively, fluorine at the $C2$ position is converted to $^{18}$O. After picking up a proton (H$^{+}$) from a hydronium ion {(H$_3$O$^{+}$)} in its aqueous environment, the FDG molecule becomes glucose-6-phosphate labeled with harmless nonradioactive ``heavy oxygen'' in the -OH group at the $C2$ position, ready to be metabolized in the cell~\cite{Krolicki2021}.
Another approach used for PET imaging applies a radiotracer for direct labelling of a target cell. In breast cancers approximately 20-30\% of cases overexpress the HER2 receptor (human epidermal growth factor receptor family), which results from HER2-gene amplification~\cite{Rubin2001,Sawyers2009}. In around 90\% of HER2-positive cancer cells, up to several hundred HER2-gene copies are generated to produce over 100 times more protein receptor in a cancer cell relative to
a healthy cell~\cite{Venter1987,Zabaglo2013}. This makes the HER2 protein a suitable and ideal biomarker for treatments and diagnosis of HER2-positive cancer not only in breast, but also in gastric, bladder, pancreatic and ovarian cancers~\cite{Sawyers2009,Yan2015}. Several groups of molecules targeting HER2 have been developed for molecular imaging with radiotracers used in PET.
Among them, 
the designed humanized monoclonal antibody against HER2 protein (trastuzumab) has been used in multiple clinical trials~\cite{Henry2018}. 
Currently, clinical trials with HER2 targeting radiotracers use radionuclides emitting additional prompt $\gamma$ during $\beta^+$-decay, which would enable  determination of positronium mean lifetime, as proposed recently in~\textcite{Moskal2020petclin}.
Using $\beta^+\gamma$ emitters for targeting HER2 opens up new possibilities for positronium imaging in breast cancer diagnostics and treatment~(Fig.~\ref{fig:cells_receptors}).

\subsection{Positronium Imaging}
 \label{sec:positronium_imaging}
 \begin{figure*}[t]
\centering
\includegraphics[width=0.54\textwidth]{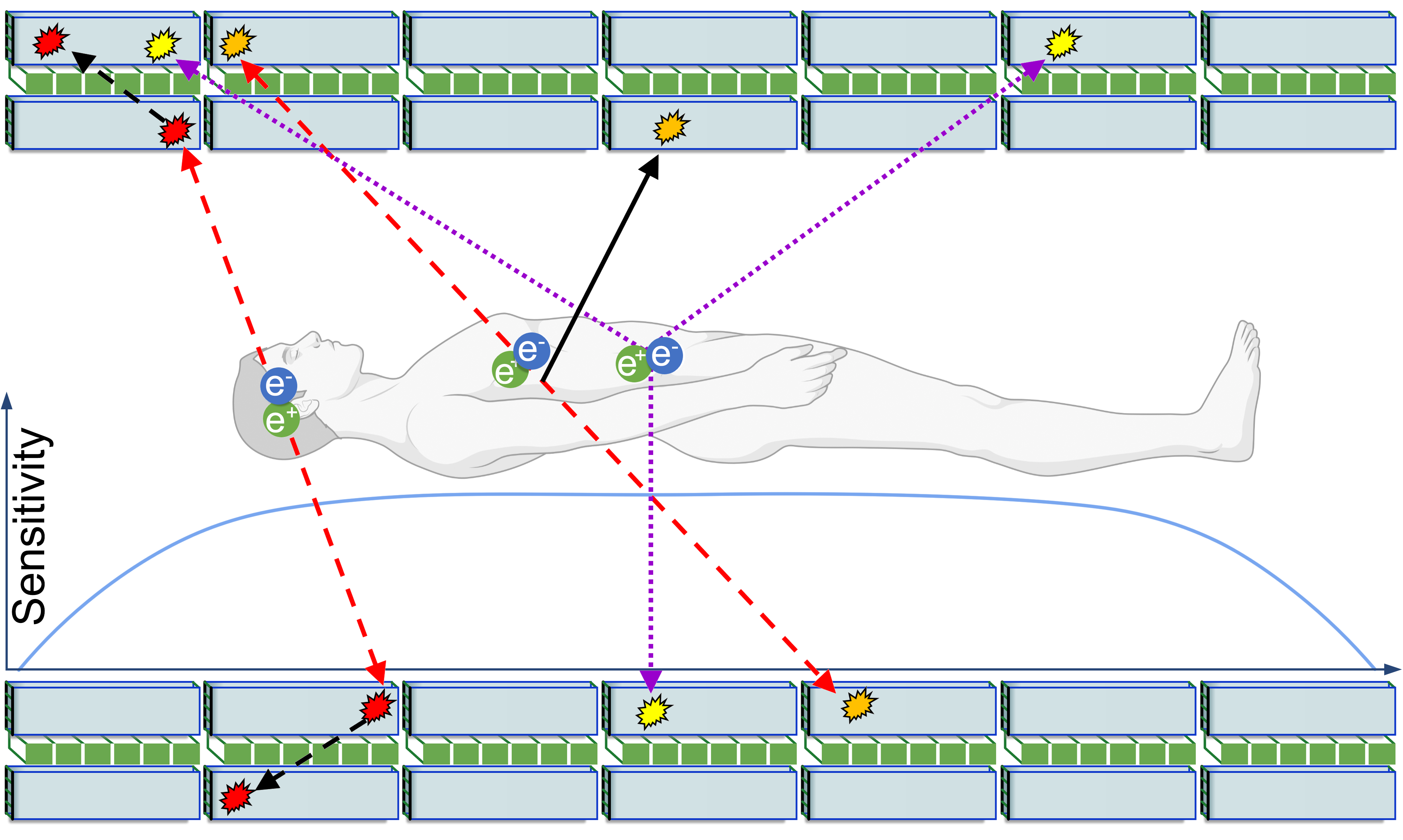}
\includegraphics[width=0.34\textwidth]{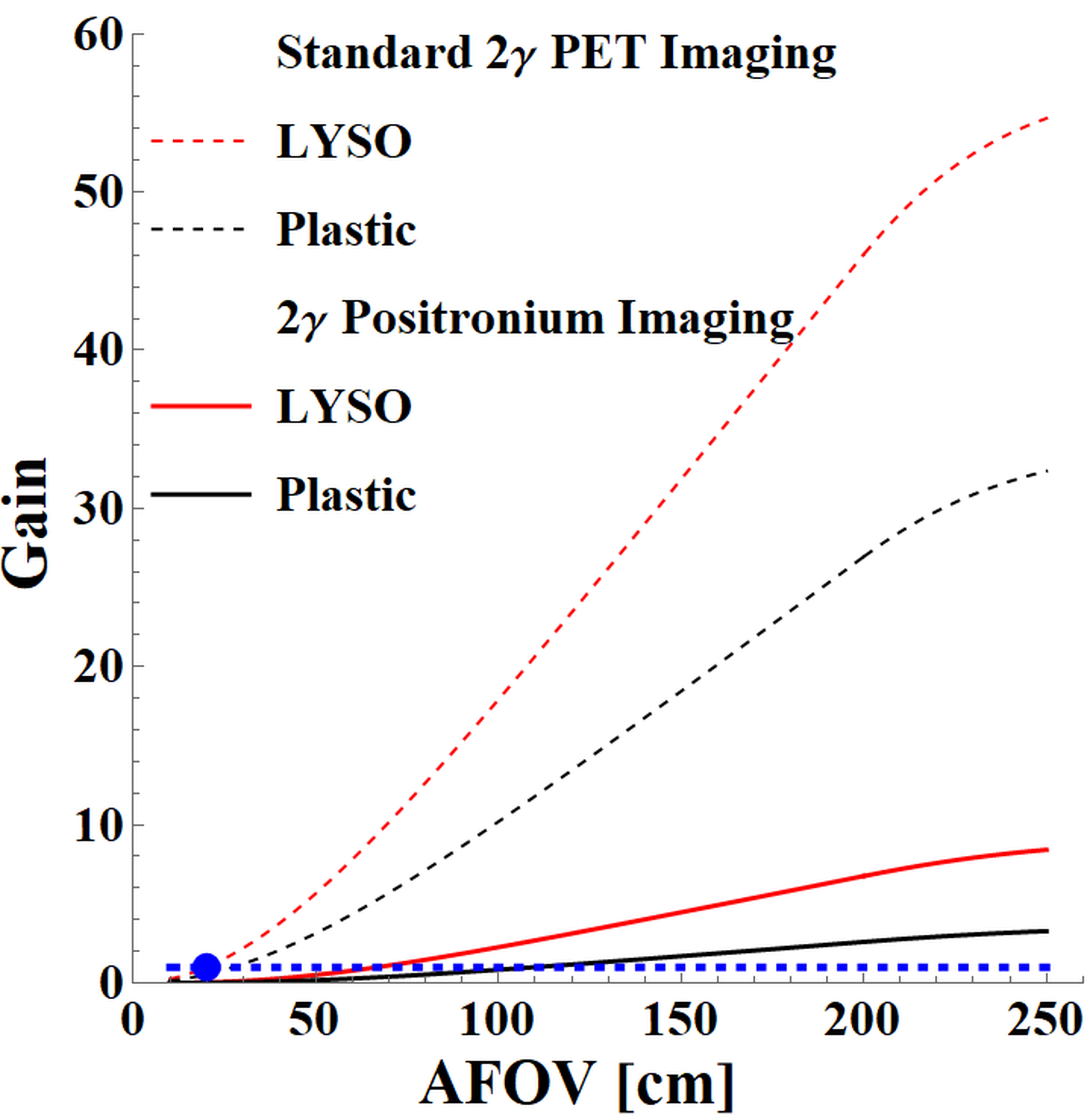}
\caption{
(Left) Scheme of the example of total-body PET for the positronium and quantum entanglement imaging
showing an axial cross section of the tomograph design 
composed of two detection layers~\cite{Moskal2020petclin}.  The single
detection module consists of a scintillator and WLS strips read out by SiPM matrices.
Here elements are presented not to scale. Dashed 
red arrows indicate
example lines of response 
originating from $e^+e^-$ annihilation. $2\gamma$ originating from the brain scatter twice in the plastic scintillator and are shown as an example of event usable for  quantum entanglement tomography discussed in section~\ref{subsec:QET}.
Violet dotted arrows indicate $3\gamma$ decay and dashed red arrows together with solid black arrow indicate annihilation and the prompt photon useful for positronium imaging discussed in section~\ref{sec:positronium_imaging}. Superimposed charts indicate the sensitivity (in arbitrary
units) along the 
axial field of view,
AFOV.
(Right) Sensitivity for the $2\gamma$ positronium imaging (2  annihilation $\gamma$ plus prompt photon) compared to the sensitivity for the standard $2\gamma$ PET imaging. Results for crystal (LYSO) PET and plastic PET are shown as a function of the scanner's AFOV~\cite{Moskal2020petclin}. The sensitivity gain is shown with respect to the 20 cm AFOV LYSO PET (indicated with horizontal blue dotted line).
\label{fig:total-body}
}
\end{figure*}
During PET diagnosis of a
living organism, annihilation of positrons proceeds via formation of positronium in about as much as 40\% of cases~\cite{Harpen2003zz,Jasinska2017,Kotera2005}.
This makes newly invented positronium imaging ~\cite{Moskal2021science} a promising method for the $in \, vivo$ assessment of tissue pathology. 
Positronium imaging may be defined as a spatially resolved reconstruction of positronium properties in living organisms~\cite{MoskalIEEE2019}. Information about positronium mean lifetime may be directly inferred by the application of $\beta^+\gamma$ emitters such as $^{44}$Sc,
which enable one to determine the positronium lifetime in the organism by measurement of the time of the emission of the prompt photon and the time of annihilation~(Fig.~\ref{fig:total-body}). Coincident detection of both prompt and annihilation photons and registrations of their positions and times of interaction in the tomograph allows one
to reconstruct the position of annihilation and lifetime of positronium in each image element (voxel) separately (on a voxel by voxel basis). 
For the reconstruction of the annihilation position and time, both $3\gamma$ self-annihilation of o-Ps~\cite{Moskal2021nature,Moskal2019pmb,Gajos2016} as well as $2\gamma$ pick-off and conversion processes of \mbox{o-Ps}~\cite{Moskal2021science,Moskal2020ejnmmi} may be applied. A first multi-photon PET detector enabling positronium imaging was constructed based on plastic scinillators~\cite{Moskal2014nim,Niedzwiecki2017,Dulski2021}. 
It recently provided $ex \, vivo$ $2\gamma$ positronium images of phantoms (objects designed to test the imaging performance)
built from cardiac myxoma cancer tissues and adipose tissues~\cite{Moskal2021science}, as well as $3\gamma$ images of the extended cylindrical phantoms~\cite{Moskal2021nature}. 
 Positronium mean lifetime imaging based on two photons
 is more than 300 times more effective than that based on three photons because: (i) in the tissue, due to the pick-off and conversion processes, o-Ps decays about 70 times, viz. $(\frac{\tau^o_0}{\tau_{\rm tissue}} - 1$), more frequently to $2\gamma$ than to $3\gamma$,
 (ii) the attenuation of $3\gamma$ events in the body is much larger (more than 4 times) with respect to $2\gamma$, both due to higher number of photons and their lower energies, and (iii) the efficiency for the detection and selection of $3\gamma$ is lower than for $2\gamma$. The right panel of Fig.~\ref{fig:total-body} shows a comparison of sensitivity for standard PET imaging and 
 $2\gamma$ positronium imaging~\cite{Moskal2020petclin}. The sensitivity for positronium imaging is lower since it requires registration of prompt photon in addition to two annihilation photons.  However, the sensitivity is increasing with the growth of the axial field of view, and the figure indicates that total-body PET systems (with 200~cm long scanner) will provide even higher sensitivity for positronium imaging then current (20~cm long) scanners provide for standard PET diagnostics.  
 Using the standard whole-body PET protocol, total-body PET sensitivity enables one to achieve determination of the positronium lifetime with the precision of about 20~ps for the 2~cm~$\times$~2~cm~$\times$~2~cm voxels~\cite{Moskal2020ejnmmi,Moskal2021science}, and 2~ps when averaging over the whole organs~\cite{MoskalStepien2021hypoxia} (see section~\ref{sec:hypoxia}).
 The time resolution for determining mean o-Ps lifetime in the tissue depends predominantly on the value of the mean o-Ps lifetime
 and may be estimated as $\tau_{tissue}$/$\sqrt{N}$, where $N$ denotes the number of events in a given voxel of the positronium image~\cite{Moskal2020ejnmmi}. 
 Assuming that $\tau_{tissue}$~=~2~ns, it can be estimated that with 10$^4$ registered events per $cm^3$ 
 (as expected for the total-body PET systems~\cite{Moskal2020ejnmmi}), a resolution of $\sigma \approx 20$~{ps} is achievable.
 Interpretation of positronium properties as a diagnostic parameter will still require systematic research. The resolution of 20~ps obtained in first in-vitro images~\cite{Moskal2021science} and expected for total-body PET systems~\cite{Moskal2020ejnmmi} is sufficient to distinguish between the healthy and cancer tissues for which differences larger than 50~ps (in the range of 50ps~-~200ps~\cite{Jasinska2017,Zgardzinska2020}) or even 700~ps~\cite{Moskal2021science,Moskal2021} are observed.
 Recently, both the new methods for precise analysis and decomposition of positron annihilation lifetime 
 spectra~\cite{Dulski-Avalanche-2020,Shibua-positronium-2022}, as well as new positronium image reconstruction methods were developed using maximum likelihood image estimation, the latter resulting in spatial resolution of the image better than 4~mm~\cite{Qi-positronium-2022,Zhu-positronium-2022}.
 
 These results 
indicate that positronium imaging may be introduced in clinics for the assessment of tissue alterations at the molecular level before they lead to the functional and morphological changes~\cite{MoskalStepien2022}.  The practical diagnostic benefits of positronium imaging will be a subject of long-standing research and are yet to be determined. Here we hypothesized that when applied to brain diagnostics, positronium imaging with its potential for the in-vivo assessment of the changes of the nano-structure of tissues may become an early diagnostics indicator for neuro-degenerative diseases such as Dementia, Alzheimer or Parkinson. 
\subsection{Positronium as a biomarker of hypoxia} 
\label{sec:hypoxia}
The decay rate of ortho-positronium in the tissue due to the conversion processes on paramagnetic molecules is linearly proportional to the  concentration on these molecules (see discussion in section~\ref{subsec:conversion}). Therefore, positronium may be used for oxygen concentration assessment in the tissue~\cite{MoskalStepien2021hypoxia,Shibuya2020,Stepanov2020,Zare-Biganeh-2022}. 
In this section the possibility of in-vivo sensing of oxygen concentration by means of positronium mean lifetime determination is considered. 

\begin{figure}[t!]
\centering
\includegraphics[width=0.45\textwidth]{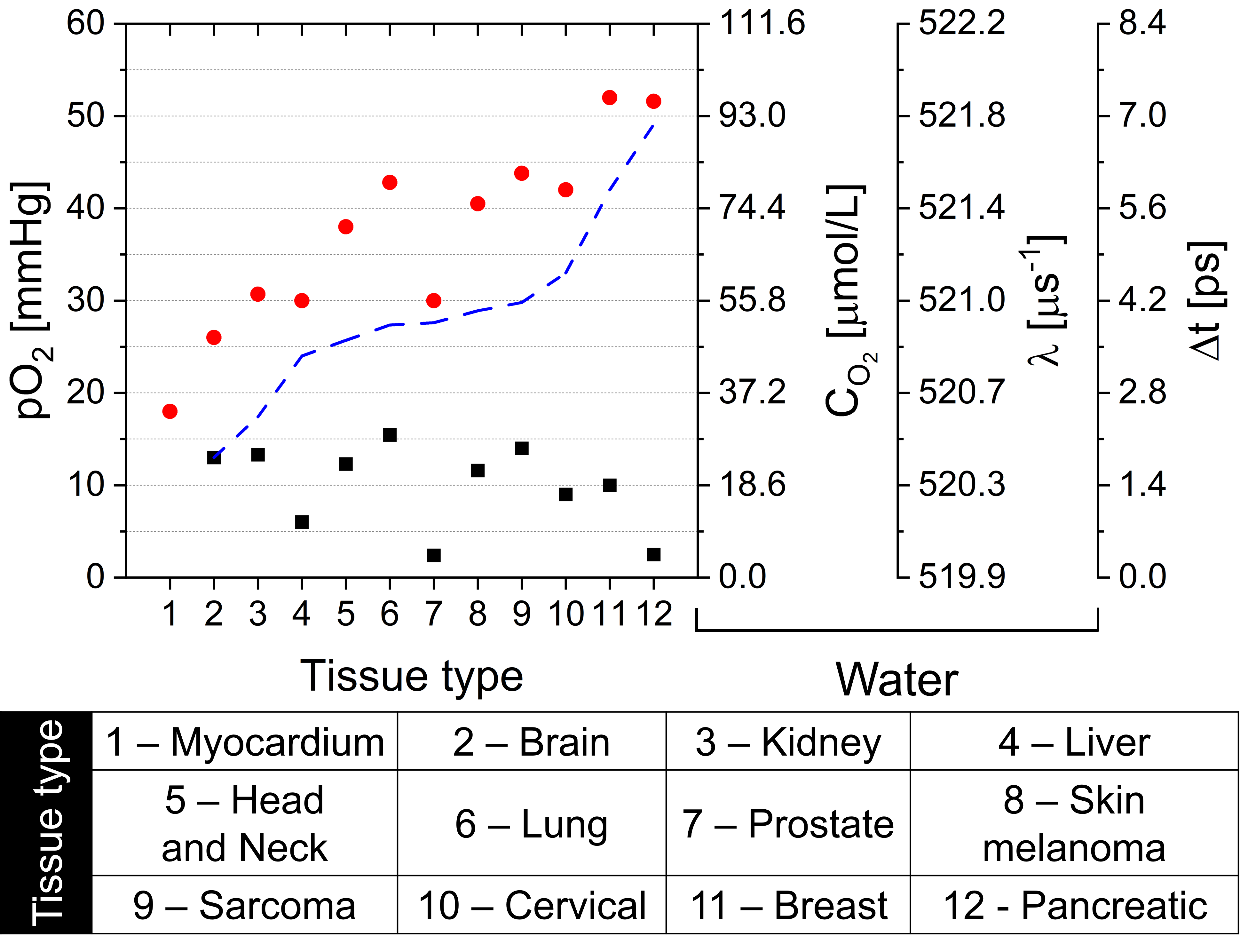}
\caption{Comparison of partial pressure of oxygen molecules  (pO$_{2}$) in healthy and cancer tissues. The horizontal axis indicates a tissue type (1-12) explained 
below the graph. 
The right axes refers to water and indicate: the concentration of oxygen ($C_{{\rm O}_2}$), the o-Ps annihilation rate $\lambda^{o}$ in water, and the change of the o-Ps mean lifetime $\Delta t$ due the concentration of oxygen molecules dissolved in water. Normal pO$_{2}$ for healthy tissues (red circles) and hypoxia in cancer tissues (black squares) are shown based on 
experimental data summarized and reviewed in  medians which do not show real patients variability~\cite{Vaupel2007,Vaupel2021,McKeown2014,Swartz2020}. For head and neck tumours, sarcoma and normal subcutaneous tissue data were compiled from the studies  ~\cite{Nordsmark1994,Becker1998}. 
For a kidney and melanoma there were separate studies~\cite{Lartigau1997,Lawrentschuk2005}.
The tissue type is numbered according to the increasing partial pressure difference between healthy and cancer tissue,
indicated by
the blue dashed line.
\label{fig:hypoxia}
}
\end{figure}

A normal level of oxygen concentration in the cells is referred to as normoxia while 
hypoxia is defined as a state or condition, in which oxygen supply is not sufficient to support physiological processes in tissue and organs. Local hypoxia is usually a result of vessels occlusion (arteries or arterioles) to cause stroke, myocardial infarction or other organ injury leading to cell death namely necrosis~\cite{McKeown2014}. In solid tumors hypoxia is often observed and leads to the development of an aggressive phenotype, acquired treatment resistance, and is associated with a poor prognosis~\cite{Vaupel2021,Brahimi2007,Krolicki2021,McKeown2014}.
Therefore, $in \, vivo$ assessment of the degree of hypoxia would be advantageous for the personalised cancer therapy~\cite{Cramer-Vaupel-hypoxia-2022}. Recently, it was argued that the possibilities of application of positronium imaging with total-body PET scanners opens perspectives for the application of positronium as a biomarker for in-vivo assessment of the degree of hypoxia~\cite{MoskalStepien2021hypoxia}.  
Fig.~\ref{fig:hypoxia} demonstrates that the partial pressure of oxygen (pO$_2$) in cancer tissues is significantly lower than in corresponding healthy tissue. 
The differences 
vary between 10 mmHg (for the brain) and 50 mmHg (for the pancreas). 
    
The experimentally established relationship (for water) between the partial oxygen pressure pO$_2$ and the o-Ps decay rate constant $\lambda^o$ reads~\cite{Shibuya2020}: ${\rm pO}_2 {\rm [mmHg]} = 26.3(11) \cdot (\lambda^o - 519.9(16) {\rm \mu s}^{-1})$
where 519.9${\rm \mu s}^{-1}$ accounts for o-Ps self-annihilation and pick-off rate in water. 
This relation indicates (as shown in Fig.~\ref{fig:hypoxia}) 
that the differences of pO$_2$ in the range from 10~mmHg to 50~mmHg result in the change of ortho-positronium mean lifetime in water by about 
2~ps to 7~ps. 
Estimation for water is the most pessimistic since for organic liquids (e.g. cyclohexane, isooctane, isopropanol) these mean oPs lifetime changes are larger~\cite{Shibuya2020,Stepanov2020,Stepanov2021}.
 These estimations indicate that in order to apply positronium as a biomarker for hypoxia an extremaly high (few ps) mean lifetime resolution determination is required.
  Resolution of 20~ps was already obtained in the first experimental positronium images for $ex \, vivo$ studies of cardiac myxoma tumors~\cite{Moskal2021science,Moskal2021} with about $10^4$ registered o-Ps annihilations. 
{
With 100 times more registered o-Ps annihilations, $10^6$, 2 ps resolution would be achievable. Such number of annihilations can be collected by means of the  total-body PET system for organs with the volume larger than 100~cm$^3$ (e.g. pancreas or liver). Therefore, identification of hypoxia (organ averaged) using positronium as a biomarker may become feasible with total-body PET systems
}
\subsection{Quantum Entanglement Tomography}
\label{subsec:QET}
\begin{figure}[b!]
\vspace{-0.75cm}
\centering
\includegraphics[width=0.38\textwidth]{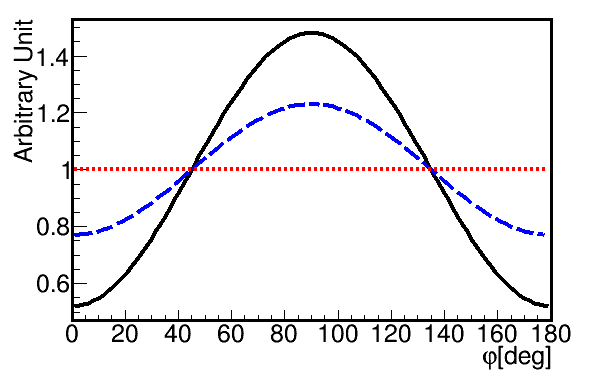}
\caption{\label{fig:phi-distribution} 
Distribution of the angle 
$\varphi$
between scattering planes of the photons
emitted in p-Ps$\, \to 2 \gamma$
when the photons are each scattered at $\theta \sim 82^\circ$; for definition of these angles see 
Fig.~\ref{fig:BackToBack}.
The black solid curve corresponds to the photon pair being entangled, 
the dashed blue curve to independent 
Compton interaction of the two photons,
and the red dotted line indicates 
when the photon polarizations are  uncorrelated, e.g., when the photons originate from two different p-Ps decays.
}
\end{figure}
Photons originating from the decay of positronium are expected to be quantum entangled in polarization and exhibit non-local correlations as discussed in subsection~\ref{sec:entanglement}. 
These correlations may be used for the improvement of the quality of PET image reconstruction~\cite{McNamara2014,Toghyani2016}, and for the elaboration of new quantum biomarkers by using entanglement witnesses as a diagnostic indicators~\cite{Hiesmayr2017xgx,Hiesmayr2018rcm}. 
The latter may work provided that the type and degree of quantum entanglement of photons from the decay of positronium 
 is  
 affected by the molecular environment in cells. 
 This is a topic of current investigation
 \cite{Hiesmayr2017xgx,Caradonna2020,Sharma2022puzzle}
 requiring new experimental input.

Fig.~\ref{fig:phi-distribution} compares the distribution of
the angle $\varphi$  between scattering planes calculated under the assumption that photons from the p-Ps\ $\to2\gamma$ process are entangled (black solid curve), for the case when scattering of one photon is completely independent of the scattering of the other photon (blue dashed curve), and for the case 
when photons originate from different annihilation processes (red dotted curve). 
Recently it was shown experimentally by~\textcite{MoskalIEEE2018,Watts2021,Abdurashitov2022} that indeed the $\varphi$ distribution for $2\gamma$ annihilations is enhanced at $\varphi=90^\circ$ as expected for the quantum entangled state of $2\gamma$.

The image quality of standard $2\gamma$ PET may be improved by reduction of the fraction of events for which any of the photons was scattered in the patient, or for which photons originate from different annihilation events. This may be achieved by selecting events for which the angle between polarization direction of the two photons is close to $\varphi=90^\circ$~\cite{McNamara2014,Toghyani2016,Moskal2018pus,Watts2021}. 
Application of the selection based on the relative angle between the scattering planes will decrease the fraction of unwanted scatter and random events~\cite{Toghyani2016}
relative to
the fraction of useful events. 
This will challenge the designs of PET systems, especially since
the visibility of the quantum correlation is maximal for scattering angles around $\theta_1 = \theta_2 $~$\sim$~82$^\circ$ while the scattering cross sections is the highest for forward scattered photons ($\theta_1 = \theta_2$~$\sim$~0$^\circ$)~\cite{Klein1929}.

\subsection{Roadmap for multi-photon total-body positron and positronium tomography}
\label{subsec:TBPET}
Positron emission tomography is presently experiencing a quantitative change in diagnosis 
paradigm~\cite{Alavi2021bioalg,Vanderberghe2020,Surti-review2020,Moskal2020petclin,Krolicki2021,Djekidel2022}. 
With the advent of total-body PET systems~\cite{Badawi2019,Karp2020,Quadra2022,uEXPLORER-Chiny-2022,Niedzwiecki2017}, covering the whole human body (with the detector length of about 2~m), the simultaneous imaging of the metabolism rate of all organs and tissues in the body becomes possible. This opens possibilities for studies (in physiology, biochemistry and pharmacology) of the kinetics of administered pharmaceuticals in the whole body, and in determining pharmaceuticals' uptake correlations in close and distance 
organs~\cite{Badawi2019,zhang2020}.
High sensitivity of total-body PET systems~\cite{Vanderberghe2020,Surti-review2020} (up to factor of $\sim$40 higher with respect to standard 20~cm long PET, see Fig.~\ref{fig:total-body}) enables also dynamic and kinetic model based parametric imaging~\cite{Feng2021,Wang2022}, and therefore increases the diagnosis specificity in differentiating between the healthy and cancer tissues. In parallel, recent development in PET technology resulted in the first multi-photon (multi-gamma) PET system~\cite{Moskal2021nature} capable of positronium imaging~\cite{Moskal2021science} based on registration of two~\cite{Moskal2020ejnmmi} and three photons~\cite{Moskal2019pmb} from positronium annihilations and the prompt photon from deexcitation of isotopes attached to the pharamaceuticals. Moreover, there is a continuous development of new materials~\cite{Lecoq2022,Lecoq2022IEEE} 
and new systems and techniques~\cite{Kwon2021,Ota2019,Gundacker2019,Jensen-Turtos2022,Tao-Levin2021,Cates-Levin2019} for improving the time and spatial resolution to the point where imaging by direct determination of the density distribution of annihilation points 
will become possible. The direct PET image  of the 2D brain phantom was experimentally demonstrated with the spatial resolution of 4.8~mm~\cite{Kwon2021}.  The new generation of total-body PET systems will combine high sensitivity with multi-photon imaging, and next also with high timing resolution. The technology for the multi-photon total-body imaging is known and it is at the stage of translation into clinics~\cite{MoskalStepien2022}.
The annihilation photons' detection technology for PET is also developing towards a more cost-effective solution focusing on plastic scintillators ~\cite{Moskal2020petclin} and sparse detector configurations~\cite{Karakatsanis2022}.
It is important to stress that the total-body multi-photon PET systems will enable also high precision studies of fundamental positronium decays~\cite{Moskal2016moj} by increasing efficiency for the studies of 3-photon positronium decays by more than order of magnitude
with respect to present detectors~\cite{Moskal2021nature}.   

\section{Conclusions and Perspectives}
\label{sec:conclusion}

Positronium, 
the bound state of $e^- e^+$, is interesting both in 
fundamental physics and in applications
from 
antimatter research 
through to biology and medicine.
The properties of positronium formation and decay in medium 
depend on the chemical environment and
open new windows of opportunity in biological and medical science.
Effective mean decay rates are sensitive to the health of biological tissue with exciting prospects 
to revolutionize 
next generation 
total-body PET scanning
through positronium imaging as a new tool in medical diagnosis.

Traditional PET is based on the parameters of the concentration of the radiopharmaceutical and does not take into account 
changes in the positronium annihilation mechanism due to the chemical environment.
The average lifetime of positronium, due to its sensitivity to changes in the nanostructure, will allow to take into account an additional parameter in the reconstruction of the histopathological image.

In this Colloquium we have surveyed prime topics in positronium
physics and new prospects for medical applications. 
We conclude with a summary of key issues and open questions 
where next generation
experiments should yield vital information:
\begin{itemize}
    \item     
Pushing limits of QED bound state theory, anomalies 
between data in spectroscopic measurements and theory call for new precision measurements 
of positronium.
Are there missed systematics or might the data be pointing
to new
(bound state) physics waiting to be understood?
\item
Studies of gravity on antimatter will provide new tests of the equivalence
principle. 
Does gravity couple equally to matter 
as to antimatter?
\item
Might next generation experiments searching for invisible decays of positronium help in understanding dark matter?
\item
About 40\% of the positrons in conventional PET scans are formed via positronium intermediate states.
Can this be 
developed through positronium imaging as a practical tool for medical diagnostics?
Might 
imaging diagnostics
be further enhanced by sensing quantum entanglement of emitted photons?

\item
Traditional histopathological imaging requires punctual tissue sampling, which is always somewhat invasive for the patient. 
More traditional
liquid biopsy, which is based on taking a sample of blood or other body fluid, 
does not give the possibility of locating the lesion. 
Might 
virtual biopsy with positronium  imaging 
be able to 
tell whether the 
tissue is cancerous or not
without need for invasive incision?

\item
Possible differences expected in mean o-Ps 
lifetime between healthy and cancer tissues are predominantly due to the structural changes caused by the increased over-expression of receptors, cell-cycle controlling molecules and other changes in metabolic pathways (due to inherited or acquired mutations) and to some extent by the changes of the concentration of the molecular oxygen dissolved in cells. The influence of the oxygen concentration on the mean o-Ps 
lifetime may enable organ-averaged
 identification of hypoxia with positronium as a biomarker.
How might this be translated to clinics?
\end{itemize}

\section*{Acknowledgments}
We thank
E. Czerwi{\'n}ski,
W. Krzemie{\'n},
C. Malbrunot and
E. Perez del Rio
for helpful discussions, and
M. Durak-Kozica,
E. Kubicz, 
D. Kumar,
S. Parzych
and
Shivani
for help with preparation of the figures.
We acknowledge support by the Foundation for Polish Science through the TEAM POIR.04.04.00-00-4204/17 programme, the National Science Centre of Poland through grants no. 2019/35/B/ST2/03562, 2019/33/B/NZ3/01004, 2021/42/A/ST2/00423 and 2021/43/B/ST2/02150,
the Jagiellonian University via project CRP/0641.221.2020, as well as the SciMat and qLife Priority Research Area budget under the program Excellence Initiative - Research University at Jagiellonian University.

\bibliography{apssamp}

%merlin.mbs apsrmp4-1.bst 2010-07-25 4.21a (PWD, AO, DPC) hacked
%Control: author (11) reversed first initials
\providecommand{\noopsort}[1]{}\providecommand{\singleletter}[1]{#1}%
\begin{thebibliography}{271}%
\makeatletter
\providecommand \@ifxundefined [1]{%
 \@ifx{#1\undefined}
}%
\providecommand \@ifnum [1]{%
 \ifnum #1\expandafter \@firstoftwo
 \else \expandafter \@secondoftwo
 \fi
}%
\providecommand \@ifx [1]{%
 \ifx #1\expandafter \@firstoftwo
 \else \expandafter \@secondoftwo
 \fi
}%
\providecommand \natexlab [1]{#1}%
\providecommand \enquote  [1]{``#1''}%
\providecommand \bibnamefont  [1]{#1}%
\providecommand \bibfnamefont [1]{#1}%
\providecommand \citenamefont [1]{#1}%
\providecommand \href@noop [0]{\@secondoftwo}%
\providecommand \href [0]{\begingroup \@sanitize@url \@href}%
\providecommand \@href[1]{\@@startlink{#1}\@@href}%
\providecommand \@@href[1]{\endgroup#1\@@endlink}%
\providecommand \@sanitize@url [0]{\catcode `\\12\catcode `\$12\catcode
  `\&12\catcode `\#12\catcode `\^12\catcode `\_12\catcode `\%12\relax}%
\providecommand \@@startlink[1]{}%
\providecommand \@@endlink[0]{}%
\providecommand \url  [0]{\begingroup\@sanitize@url \@url }%
\providecommand \@url [1]{\endgroup\@href {#1}{\urlprefix }}%
\providecommand \urlprefix  [0]{URL }%
\providecommand \Eprint [0]{\href }%
\providecommand \doibase [0]{http://dx.doi.org/}%
\providecommand \selectlanguage [0]{\@gobble}%
\providecommand \bibinfo  [0]{\@secondoftwo}%
\providecommand \bibfield  [0]{\@secondoftwo}%
\providecommand \translation [1]{[#1]}%
\providecommand \BibitemOpen [0]{}%
\providecommand \bibitemStop [0]{}%
\providecommand \bibitemNoStop [0]{.\EOS\space}%
\providecommand \EOS [0]{\spacefactor3000\relax}%
\providecommand \BibitemShut  [1]{\csname bibitem#1\endcsname}%
\let\auto@bib@innerbib\@empty
%</preamble>
\bibitem [{\citenamefont {Abdurashitov}\ \emph {et~al.}(2022)\citenamefont
  {Abdurashitov} \emph {et~al.}}]{Abdurashitov2022}%
  \BibitemOpen
  \bibfield  {author} {\bibinfo {author} {\bibnamefont {Abdurashitov},
  \bibfnamefont {D.}},  \emph {et~al.}} (\bibinfo {year} {2022}),\ \href
  {\doibase https://doi.org/10.1088/1748-0221/17/03/P03010} {\bibfield
  {journal} {\bibinfo  {journal} {JINST}\ }\textbf {\bibinfo {volume} {17}},\
  \bibinfo {pages} {P03010}}\BibitemShut {NoStop}%
\bibitem [{\citenamefont {Acin}\ \emph {et~al.}(2001)\citenamefont {Acin},
  \citenamefont {Latorre},\ and\ \citenamefont {Pascual}}]{Acin2000}%
  \BibitemOpen
  \bibfield  {author} {\bibinfo {author} {\bibnamefont {Acin}, \bibfnamefont
  {A.}}, \bibinfo {author} {\bibfnamefont {J.~I.}\ \bibnamefont {Latorre}}, \
  and\ \bibinfo {author} {\bibfnamefont {P.}~\bibnamefont {Pascual}}} (\bibinfo
  {year} {2001}),\ \href {\doibase 10.1103/PhysRevA.63.042107} {\bibfield
  {journal} {\bibinfo  {journal} {Phys. Rev. A}\ }\textbf {\bibinfo {volume}
  {63}},\ \bibinfo {pages} {042107}}\BibitemShut {NoStop}%
\bibitem [{\citenamefont {Adkins}\ \emph {et~al.}(2022)\citenamefont {Adkins},
  \citenamefont {Cassidy},\ and\ \citenamefont
  {P\'erez-R\'\i{}os}}]{Adkins:2022omi}%
  \BibitemOpen
  \bibfield  {author} {\bibinfo {author} {\bibnamefont {Adkins}, \bibfnamefont
  {G.~S.}}, \bibinfo {author} {\bibfnamefont {D.~B.}\ \bibnamefont {Cassidy}},
  \ and\ \bibinfo {author} {\bibfnamefont {J.}~\bibnamefont
  {P\'erez-R\'\i{}os}}} (\bibinfo {year} {2022}),\ \href {\doibase
  10.1016/j.physrep.2022.05.002} {\bibfield  {journal} {\bibinfo  {journal}
  {Phys. Rept.}\ }\textbf {\bibinfo {volume} {975}},\ \bibinfo {pages}
  {1}}\BibitemShut {NoStop}%
\bibitem [{\citenamefont {Adkins}\ \emph {et~al.}(1997)\citenamefont {Adkins},
  \citenamefont {Fell},\ and\ \citenamefont {Mitrikov}}]{Adkins1997xrp}%
  \BibitemOpen
  \bibfield  {author} {\bibinfo {author} {\bibnamefont {Adkins}, \bibfnamefont
  {G.~S.}}, \bibinfo {author} {\bibfnamefont {R.~N.}\ \bibnamefont {Fell}}, \
  and\ \bibinfo {author} {\bibfnamefont {P.~M.}\ \bibnamefont {Mitrikov}}}
  (\bibinfo {year} {1997}),\ \href {\doibase 10.1103/PhysRevLett.79.3383}
  {\bibfield  {journal} {\bibinfo  {journal} {Phys. Rev. Lett.}\ }\textbf
  {\bibinfo {volume} {79}},\ \bibinfo {pages} {3383}}\BibitemShut {NoStop}%
%%CITATION = PRLTA,79,3383;%%
\bibitem [{\citenamefont {Adkins}\ \emph {et~al.}(2002)\citenamefont {Adkins},
  \citenamefont {Fell},\ and\ \citenamefont {Sapirstein}}]{Adkins2002fg}%
  \BibitemOpen
  \bibfield  {author} {\bibinfo {author} {\bibnamefont {Adkins}, \bibfnamefont
  {G.~S.}}, \bibinfo {author} {\bibfnamefont {R.~N.}\ \bibnamefont {Fell}}, \
  and\ \bibinfo {author} {\bibfnamefont {J.}~\bibnamefont {Sapirstein}}}
  (\bibinfo {year} {2002}),\ \href@noop {} {\bibfield  {journal} {\bibinfo
  {journal} {Annals Phys.}\ }\textbf {\bibinfo {volume} {295}},\ \bibinfo
  {pages} {136}}\BibitemShut {NoStop}%
\bibitem [{\citenamefont {Adkins}\ \emph {et~al.}(2003)\citenamefont {Adkins},
  \citenamefont {McGovern}, \citenamefont {Fell},\ and\ \citenamefont
  {Sapirstein}}]{Adkins2003eh}%
  \BibitemOpen
  \bibfield  {author} {\bibinfo {author} {\bibnamefont {Adkins}, \bibfnamefont
  {G.~S.}}, \bibinfo {author} {\bibfnamefont {N.~M.}\ \bibnamefont {McGovern}},
  \bibinfo {author} {\bibfnamefont {R.~N.}\ \bibnamefont {Fell}}, \ and\
  \bibinfo {author} {\bibfnamefont {J.}~\bibnamefont {Sapirstein}}} (\bibinfo
  {year} {2003}),\ \href {\doibase 10.1103/PhysRevA.68.032512} {\bibfield
  {journal} {\bibinfo  {journal} {Phys. Rev. A}\ }\textbf {\bibinfo {volume}
  {68}},\ \bibinfo {pages} {032512}}\BibitemShut {NoStop}%
\bibitem [{\citenamefont {Aghion}\ \emph {et~al.}(2016)\citenamefont {Aghion}
  \emph {et~al.}}]{Aghion2016}%
  \BibitemOpen
  \bibfield  {author} {\bibinfo {author} {\bibnamefont {Aghion}, \bibfnamefont
  {S.}},  \emph {et~al.} (\bibinfo {collaboration} {AEgIS})} (\bibinfo {year}
  {2016}),\ \href {\doibase 10.1103/PhysRevA.94.012507} {\bibfield  {journal}
  {\bibinfo  {journal} {Phys. Rev. A}\ }\textbf {\bibinfo {volume} {94}},\
  \bibinfo {pages} {012507}}\BibitemShut {NoStop}%
\bibitem [{\citenamefont {Aghion}\ \emph {et~al.}(2018)\citenamefont {Aghion}
  \emph {et~al.}}]{Aghion2018xtd}%
  \BibitemOpen
  \bibfield  {author} {\bibinfo {author} {\bibnamefont {Aghion}, \bibfnamefont
  {S.}},  \emph {et~al.} (\bibinfo {collaboration} {AEgIS})} (\bibinfo {year}
  {2018}),\ \href {\doibase 10.1103/PhysRevA.98.013402} {\bibfield  {journal}
  {\bibinfo  {journal} {Phys. Rev. A}\ }\textbf {\bibinfo {volume} {98}},\
  \bibinfo {pages} {013402}}\BibitemShut {NoStop}%
%%CITATION = ARXIV:1802.07012;%%
\bibitem [{\citenamefont {Akiyama}\ \emph {et~al.}(2007)\citenamefont
  {Akiyama}, \citenamefont {Shibahara}, \citenamefont {Takeda}, \citenamefont
  {Izumi}, \citenamefont {Honda}, \citenamefont {Tagawa},\ and\ \citenamefont
  {Nishijima}}]{Akiyama2007}%
  \BibitemOpen
  \bibfield  {author} {\bibinfo {author} {\bibnamefont {Akiyama}, \bibfnamefont
  {Y.}}, \bibinfo {author} {\bibfnamefont {Y.}~\bibnamefont {Shibahara}},
  \bibinfo {author} {\bibfnamefont {S.}~\bibnamefont {Takeda}}, \bibinfo
  {author} {\bibfnamefont {Y.}~\bibnamefont {Izumi}}, \bibinfo {author}
  {\bibfnamefont {Y.}~\bibnamefont {Honda}}, \bibinfo {author} {\bibfnamefont
  {S.}~\bibnamefont {Tagawa}}, \ and\ \bibinfo {author} {\bibfnamefont
  {S.}~\bibnamefont {Nishijima}}} (\bibinfo {year} {2007}),\ \href {\doibase
  http://dx.doi.org/10.1002/polb.21188} {\bibfield  {journal} {\bibinfo
  {journal} {J.Polym. Sci. Part B: Poly. Phys.}\ }\textbf {\bibinfo {volume}
  {45}},\ \bibinfo {pages} {2031.}}\BibitemShut {Stop}%
\bibitem [{\citenamefont {Al-Ramadhan}\ and\ \citenamefont
  {Gidley}(1994)}]{AlRamadhan1994zz}%
  \BibitemOpen
  \bibfield  {author} {\bibinfo {author} {\bibnamefont {Al-Ramadhan},
  \bibfnamefont {A.~H.}}, \ and\ \bibinfo {author} {\bibfnamefont
  {D.}~\bibnamefont {Gidley}}} (\bibinfo {year} {1994}),\ \href {\doibase
  10.1103/PhysRevLett.72.1632} {\bibfield  {journal} {\bibinfo  {journal}
  {Phys. Rev. Lett.}\ }\textbf {\bibinfo {volume} {72}},\ \bibinfo {pages}
  {1632}}\BibitemShut {NoStop}%
\bibitem [{\citenamefont {Alavi}\ and\ \citenamefont
  {Reivich}(2002)}]{Alavi2002}%
  \BibitemOpen
  \bibfield  {author} {\bibinfo {author} {\bibnamefont {Alavi}, \bibfnamefont
  {A.}}, \ and\ \bibinfo {author} {\bibfnamefont {M.}~\bibnamefont {Reivich}}}
  (\bibinfo {year} {2002}),\ \href {\doibase
  https://doi.org/10.1053/snuc.2002.29269} {\bibfield  {journal} {\bibinfo
  {journal} {Semin. Nucl. Med.}\ }\textbf {\bibinfo {volume} {32}},\ \bibinfo
  {pages} {2.}}\BibitemShut {Stop}%
\bibitem [{\citenamefont {Alavi}\ \emph {et~al.}(2021)\citenamefont {Alavi},
  \citenamefont {Werner}, \citenamefont {Stepien},\ and\ \citenamefont
  {Moskal}}]{Alavi2021bioalg}%
  \BibitemOpen
  \bibfield  {author} {\bibinfo {author} {\bibnamefont {Alavi}, \bibfnamefont
  {A.}}, \bibinfo {author} {\bibfnamefont {T.}~\bibnamefont {Werner}}, \bibinfo
  {author} {\bibfnamefont {E.}~\bibnamefont {Stepien}}, \ and\ \bibinfo
  {author} {\bibfnamefont {P.}~\bibnamefont {Moskal}}} (\bibinfo {year}
  {2021}),\ \href {\doibase https://doi.org/10.1515/bams-2021-0186} {\bibfield
  {journal} {\bibinfo  {journal} {Bio-Algorithms and Med-Systems}\ }\textbf
  {\bibinfo {volume} {17}},\ \bibinfo {pages} {203}}\BibitemShut {NoStop}%
\bibitem [{\citenamefont {Amsler}\ \emph {et~al.}(2019)\citenamefont {Amsler}
  \emph {et~al.}}]{Amsler2019erv}%
  \BibitemOpen
  \bibfield  {author} {\bibinfo {author} {\bibnamefont {Amsler}, \bibfnamefont
  {C.}},  \emph {et~al.} (\bibinfo {collaboration} {AEgIS})} (\bibinfo {year}
  {2019}),\ \href {\doibase 10.1103/PhysRevA.99.033405} {\bibfield  {journal}
  {\bibinfo  {journal} {Phys. Rev.}\ }\textbf {\bibinfo {volume} {A99}},\
  \bibinfo {pages} {033405}}\BibitemShut {NoStop}%
%%CITATION = ARXIV:1808.01808;%%
\bibitem [{\citenamefont {Amsler}\ \emph {et~al.}(2021)\citenamefont {Amsler}
  \emph {et~al.}}]{Amsler2021}%
  \BibitemOpen
  \bibfield  {author} {\bibinfo {author} {\bibnamefont {Amsler}, \bibfnamefont
  {C.}},  \emph {et~al.} (\bibinfo {collaboration} {AEgIS})} (\bibinfo {year}
  {2021}),\ \href {\doibase 10.1038/s42005-020-00494-z} {\bibfield  {journal}
  {\bibinfo  {journal} {Comm. Phys.}\ }\textbf {\bibinfo {volume} {4}},\
  \bibinfo {pages} {19}}\BibitemShut {NoStop}%
\bibitem [{\citenamefont {Andersen}\ \emph {et~al.}(2015)\citenamefont
  {Andersen}, \citenamefont {Cassidy}, \citenamefont {Chevallieri},
  \citenamefont {Cooper}, \citenamefont {Deller}, \citenamefont {Wall},\ and\
  \citenamefont {Uggerhoj}}]{Andersen2015}%
  \BibitemOpen
  \bibfield  {author} {\bibinfo {author} {\bibnamefont {Andersen},
  \bibfnamefont {S.~L.}}, \bibinfo {author} {\bibfnamefont {D.}~\bibnamefont
  {Cassidy}}, \bibinfo {author} {\bibfnamefont {J.}~\bibnamefont
  {Chevallieri}}, \bibinfo {author} {\bibfnamefont {B.}~\bibnamefont {Cooper}},
  \bibinfo {author} {\bibfnamefont {A.}~\bibnamefont {Deller}}, \bibinfo
  {author} {\bibfnamefont {T.}~\bibnamefont {Wall}}, \ and\ \bibinfo {author}
  {\bibfnamefont {U.}~\bibnamefont {Uggerhoj}}} (\bibinfo {year} {2015}),\
  \href@noop {} {\bibfield  {journal} {\bibinfo  {journal} {J. Phys. B: At.
  Mol. Opt. Phys.}\ }\textbf {\bibinfo {volume} {48}},\ \bibinfo {pages}
  {204003}}\BibitemShut {NoStop}%
\bibitem [{\citenamefont {Anderson}(1933)}]{Anderson1933mb}%
  \BibitemOpen
  \bibfield  {author} {\bibinfo {author} {\bibnamefont {Anderson},
  \bibfnamefont {C.~D.}}} (\bibinfo {year} {1933}),\ \href {\doibase
  10.1103/PhysRev.43.491} {\bibfield  {journal} {\bibinfo  {journal} {Phys.
  Rev.}\ }\textbf {\bibinfo {volume} {43}},\ \bibinfo {pages}
  {491}}\BibitemShut {NoStop}%
%%CITATION = PHRVA,43,491;%%
\bibitem [{\citenamefont {Andreev}\ \emph {et~al.}(2018)\citenamefont {Andreev}
  \emph {et~al.}}]{Andreev2018ayy}%
  \BibitemOpen
  \bibfield  {author} {\bibinfo {author} {\bibnamefont {Andreev}, \bibfnamefont
  {V.}},  \emph {et~al.} (\bibinfo {collaboration} {ACME})} (\bibinfo {year}
  {2018}),\ \href@noop {} {\bibfield  {journal} {\bibinfo  {journal} {Nature}\
  }\textbf {\bibinfo {volume} {562}}~(\bibinfo {number} {7727}),\ \bibinfo
  {pages} {355}}\BibitemShut {NoStop}%
\bibitem [{\citenamefont {Aoyama}\ \emph {et~al.}(2018)\citenamefont {Aoyama},
  \citenamefont {Kinoshita},\ and\ \citenamefont {Nio}}]{Aoyama2017uqe}%
  \BibitemOpen
  \bibfield  {author} {\bibinfo {author} {\bibnamefont {Aoyama}, \bibfnamefont
  {T.}}, \bibinfo {author} {\bibfnamefont {T.}~\bibnamefont {Kinoshita}}, \
  and\ \bibinfo {author} {\bibfnamefont {M.}~\bibnamefont {Nio}}} (\bibinfo
  {year} {2018}),\ \href {\doibase 10.1103/PhysRevD.97.036001} {\bibfield
  {journal} {\bibinfo  {journal} {Phys. Rev. D}\ }\textbf {\bibinfo {volume}
  {97}},\ \bibinfo {pages} {036001}}\BibitemShut {NoStop}%
\bibitem [{\citenamefont {Avachat}\ \emph {et~al.}(2022)\citenamefont
  {Avachat}, \citenamefont {Leja}, \citenamefont {Mahmoud}, \citenamefont
  {Anastasio}, \citenamefont {Sivaguru},\ and\ \citenamefont
  {A.}}]{Fulvio2022}%
  \BibitemOpen
  \bibfield  {author} {\bibinfo {author} {\bibnamefont {Avachat}, \bibfnamefont
  {A.}}, \bibinfo {author} {\bibfnamefont {A.}~\bibnamefont {Leja}}, \bibinfo
  {author} {\bibfnamefont {K.}~\bibnamefont {Mahmoud}}, \bibinfo {author}
  {\bibfnamefont {M.}~\bibnamefont {Anastasio}}, \bibinfo {author}
  {\bibfnamefont {M.}~\bibnamefont {Sivaguru}}, \ and\ \bibinfo {author}
  {\bibfnamefont {D.~F.}\ \bibnamefont {A.}}} (\bibinfo {year} {2022}),\ \href
  {\doibase https://doi.org/10.21203/rs.3.rs-1657111/v1} {\
  https://doi.org/10.21203/rs.3.rs-1657111/v1}\BibitemShut {NoStop}%
\bibitem [{\citenamefont {Avetissian}\ \emph {et~al.}(2014)\citenamefont
  {Avetissian}, \citenamefont {Avetissian},\ and\ \citenamefont
  {Mkrtchian}}]{PhysRevLett.113.023904}%
  \BibitemOpen
  \bibfield  {author} {\bibinfo {author} {\bibnamefont {Avetissian},
  \bibfnamefont {H.~K.}}, \bibinfo {author} {\bibfnamefont {A.~K.}\
  \bibnamefont {Avetissian}}, \ and\ \bibinfo {author} {\bibfnamefont {G.~F.}\
  \bibnamefont {Mkrtchian}}} (\bibinfo {year} {2014}),\ \href {\doibase
  10.1103/PhysRevLett.113.023904} {\bibfield  {journal} {\bibinfo  {journal}
  {Phys. Rev. Lett.}\ }\textbf {\bibinfo {volume} {113}},\ \bibinfo {pages}
  {023904}}\BibitemShut {NoStop}%
\bibitem [{\citenamefont {Avril}(2004)}]{Avril2004}%
  \BibitemOpen
  \bibfield  {author} {\bibinfo {author} {\bibnamefont {Avril}, \bibfnamefont
  {N.}}} (\bibinfo {year} {2004}),\ \href@noop {} {\bibfield  {journal}
  {\bibinfo  {journal} {J. Nucl. Med.}\ }\textbf {\bibinfo {volume} {45}},\
  \bibinfo {pages} {930.}}\BibitemShut {Stop}%
\bibitem [{\citenamefont {Axpe}\ \emph {et~al.}(2014)\citenamefont {Axpe},
  \citenamefont {Lopez-Euba}, \citenamefont {Castellanos-Rubio}, \citenamefont
  {Merida}, \citenamefont {Garcia}, \citenamefont {Plaza-Izurieta},
  \citenamefont {Fernandez-Jimenez}, \citenamefont {Plazaola},\ and\
  \citenamefont {Bilbao}}]{Axpe2014}%
  \BibitemOpen
  \bibfield  {author} {\bibinfo {author} {\bibnamefont {Axpe}, \bibfnamefont
  {E.}}, \bibinfo {author} {\bibfnamefont {T.}~\bibnamefont {Lopez-Euba}},
  \bibinfo {author} {\bibfnamefont {A.}~\bibnamefont {Castellanos-Rubio}},
  \bibinfo {author} {\bibfnamefont {D.}~\bibnamefont {Merida}}, \bibinfo
  {author} {\bibfnamefont {J.}~\bibnamefont {Garcia}}, \bibinfo {author}
  {\bibfnamefont {L.}~\bibnamefont {Plaza-Izurieta}}, \bibinfo {author}
  {\bibfnamefont {N.}~\bibnamefont {Fernandez-Jimenez}}, \bibinfo {author}
  {\bibfnamefont {F.}~\bibnamefont {Plazaola}}, \ and\ \bibinfo {author}
  {\bibfnamefont {J.}~\bibnamefont {Bilbao}}} (\bibinfo {year} {2014}),\ \href
  {\doibase doi.org.10.1371/journal.pone.0083838} {\bibfield  {journal}
  {\bibinfo  {journal} {PLoS One}\ }\textbf {\bibinfo {volume} {19}},\ \bibinfo
  {pages} {e83838}}\BibitemShut {NoStop}%
\bibitem [{\citenamefont {Axpe}\ \emph {et~al.}(2015)\citenamefont {Axpe} \emph
  {et~al.}}]{Axpe2015}%
  \BibitemOpen
  \bibfield  {author} {\bibinfo {author} {\bibnamefont {Axpe}, \bibfnamefont
  {E.}},  \emph {et~al.}} (\bibinfo {year} {2015}),\ \href {\doibase
  https://10.1039/C5RA05142H} {\bibfield  {journal} {\bibinfo  {journal} {RSC
  Adv}\ }\textbf {\bibinfo {volume} {5}},\ \bibinfo {pages}
  {44282.}}\BibitemShut {Stop}%
\bibitem [{\citenamefont {Badawi}\ \emph {et~al.}(2019)\citenamefont {Badawi}
  \emph {et~al.}}]{Badawi2019}%
  \BibitemOpen
  \bibfield  {author} {\bibinfo {author} {\bibnamefont {Badawi}, \bibfnamefont
  {R.~D.}},  \emph {et~al.}} (\bibinfo {year} {2019}),\ \href@noop {}
  {\bibfield  {journal} {\bibinfo  {journal} {J. Nucl. Med.}\ }\textbf
  {\bibinfo {volume} {60}},\ \bibinfo {pages} {299}}\BibitemShut {NoStop}%
\bibitem [{\citenamefont {Bass}(2019)}]{Bass2019ibo}%
  \BibitemOpen
  \bibfield  {author} {\bibinfo {author} {\bibnamefont {Bass}, \bibfnamefont
  {S.~D.}}} (\bibinfo {year} {2019}),\ \href {\doibase
  10.5506/APhysPolB.50.1319} {\bibfield  {journal} {\bibinfo  {journal} {Acta
  Phys. Polon. B}\ }\textbf {\bibinfo {volume} {50}},\ \bibinfo {pages}
  {1319}}\BibitemShut {NoStop}%
\bibitem [{\citenamefont {Becker}\ \emph {et~al.}(1998)\citenamefont {Becker},
  \citenamefont {Hansgen}, \citenamefont {Bloching}, \citenamefont {Weigel},
  \citenamefont {Lautenschlager},\ and\ \citenamefont {Dunst}}]{Becker1998}%
  \BibitemOpen
  \bibfield  {author} {\bibinfo {author} {\bibnamefont {Becker}, \bibfnamefont
  {A.}}, \bibinfo {author} {\bibfnamefont {G.}~\bibnamefont {Hansgen}},
  \bibinfo {author} {\bibfnamefont {M.}~\bibnamefont {Bloching}}, \bibinfo
  {author} {\bibfnamefont {C.}~\bibnamefont {Weigel}}, \bibinfo {author}
  {\bibfnamefont {C.}~\bibnamefont {Lautenschlager}}, \ and\ \bibinfo {author}
  {\bibfnamefont {J.}~\bibnamefont {Dunst}}} (\bibinfo {year} {1998}),\ \href
  {\doibase 10.1016/s0360-3016(98)00182-5} {\bibfield  {journal} {\bibinfo
  {journal} {Int. J. Radiat. Oncol. Biol. Phys.}\ }\textbf {\bibinfo {volume}
  {42}},\ \bibinfo {pages} {35}}\BibitemShut {NoStop}%
\bibitem [{\citenamefont {Berko}\ and\ \citenamefont
  {Pendleton}(1980)}]{Berko1980}%
  \BibitemOpen
  \bibfield  {author} {\bibinfo {author} {\bibnamefont {Berko}, \bibfnamefont
  {S.}}, \ and\ \bibinfo {author} {\bibfnamefont {H.}~\bibnamefont
  {Pendleton}}} (\bibinfo {year} {1980}),\ \href {\doibase
  10.1146/annurev.ns.30.120180.002551} {\bibfield  {journal} {\bibinfo
  {journal} {Ann. Rev. of Nucl. and Part. Sci.}\ }\textbf {\bibinfo {volume}
  {30}},\ \bibinfo {pages} {543}}\BibitemShut {NoStop}%
\bibitem [{\citenamefont {Bernreuther}\ \emph {et~al.}(1988)\citenamefont
  {Bernreuther}, \citenamefont {Low}, \citenamefont {Ma},\ and\ \citenamefont
  {Nachtmann}}]{Bernreuther1988tt}%
  \BibitemOpen
  \bibfield  {author} {\bibinfo {author} {\bibnamefont {Bernreuther},
  \bibfnamefont {W.}}, \bibinfo {author} {\bibfnamefont {U.}~\bibnamefont
  {Low}}, \bibinfo {author} {\bibfnamefont {J.}~\bibnamefont {Ma}}, \ and\
  \bibinfo {author} {\bibfnamefont {O.}~\bibnamefont {Nachtmann}}} (\bibinfo
  {year} {1988}),\ \href@noop {} {\bibfield  {journal} {\bibinfo  {journal} {Z.
  Phys. C}\ }\textbf {\bibinfo {volume} {41}},\ \bibinfo {pages}
  {143}}\BibitemShut {NoStop}%
\bibitem [{\citenamefont {Blanco}\ \emph {et~al.}(2013)\citenamefont {Blanco},
  \citenamefont {Muñoz}, \citenamefont {Almeida}, \citenamefont {Ferreira~da
  Silva}, \citenamefont {Limão-Vieira}, \citenamefont {Fuss}, \citenamefont
  {Sanz},\ and\ \citenamefont {García}}]{Blanco2013}%
  \BibitemOpen
  \bibfield  {author} {\bibinfo {author} {\bibnamefont {Blanco}, \bibfnamefont
  {F.}}, \bibinfo {author} {\bibfnamefont {A.}~\bibnamefont {Muñoz}}, \bibinfo
  {author} {\bibfnamefont {D.}~\bibnamefont {Almeida}}, \bibinfo {author}
  {\bibfnamefont {F.}~\bibnamefont {Ferreira~da Silva}}, \bibinfo {author}
  {\bibfnamefont {P.}~\bibnamefont {Limão-Vieira}}, \bibinfo {author}
  {\bibfnamefont {M.}~\bibnamefont {Fuss}}, \bibinfo {author} {\bibfnamefont
  {A.}~\bibnamefont {Sanz}}, \ and\ \bibinfo {author} {\bibfnamefont
  {G.}~\bibnamefont {García}}} (\bibinfo {year} {2013}),\ \href {\doibase
  10.1140/epjd/e2013-40276-1} {\bibfield  {journal} {\bibinfo  {journal} {Eur.
  Phys. J. D}\ }\textbf {\bibinfo {volume} {67}},\ \bibinfo {pages}
  {199}}\BibitemShut {NoStop}%
\bibitem [{\citenamefont {Blanco}\ \emph {et~al.}(2016)\citenamefont {Blanco},
  \citenamefont {Roldán}, \citenamefont {Krupa}, \citenamefont {McEachran},
  \citenamefont {White}, \citenamefont {Marjanović}, \citenamefont
  {Petrović}, \citenamefont {Brunger}, \citenamefont {Machacek}, \citenamefont
  {Buckman}, \citenamefont {Sullivan}, \citenamefont {Chiari}, \citenamefont
  {Limão-Vieira},\ and\ \citenamefont {García}}]{Blanco2016}%
  \BibitemOpen
  \bibfield  {author} {\bibinfo {author} {\bibnamefont {Blanco}, \bibfnamefont
  {F.}}, \bibinfo {author} {\bibfnamefont {A.}~\bibnamefont {Roldán}},
  \bibinfo {author} {\bibfnamefont {K.}~\bibnamefont {Krupa}}, \bibinfo
  {author} {\bibfnamefont {R.}~\bibnamefont {McEachran}}, \bibinfo {author}
  {\bibfnamefont {R.}~\bibnamefont {White}}, \bibinfo {author} {\bibfnamefont
  {S.}~\bibnamefont {Marjanović}}, \bibinfo {author} {\bibfnamefont
  {Z.}~\bibnamefont {Petrović}}, \bibinfo {author} {\bibfnamefont
  {M.}~\bibnamefont {Brunger}}, \bibinfo {author} {\bibfnamefont
  {J.}~\bibnamefont {Machacek}}, \bibinfo {author} {\bibfnamefont
  {S.}~\bibnamefont {Buckman}}, \bibinfo {author} {\bibfnamefont
  {J.}~\bibnamefont {Sullivan}}, \bibinfo {author} {\bibfnamefont
  {L.}~\bibnamefont {Chiari}}, \bibinfo {author} {\bibfnamefont
  {P.}~\bibnamefont {Limão-Vieira}}, \ and\ \bibinfo {author} {\bibfnamefont
  {G.}~\bibnamefont {García}}} (\bibinfo {year} {2016}),\ \href {\doibase
  10.1140/epjd/e2013-40276-1} {\bibfield  {journal} {\bibinfo  {journal} {J. of
  Phys. B}\ }\textbf {\bibinfo {volume} {49}},\ \bibinfo {pages}
  {145001}}\BibitemShut {NoStop}%
\bibitem [{\citenamefont {Brahimi-Horn}\ \emph {et~al.}(2007)\citenamefont
  {Brahimi-Horn}, \citenamefont {Chiche},\ and\ \citenamefont
  {Pouyssegur}}]{Brahimi2007}%
  \BibitemOpen
  \bibfield  {author} {\bibinfo {author} {\bibnamefont {Brahimi-Horn},
  \bibfnamefont {C.}}, \bibinfo {author} {\bibfnamefont {J.}~\bibnamefont
  {Chiche}}, \ and\ \bibinfo {author} {\bibfnamefont {J.}~\bibnamefont
  {Pouyssegur}}} (\bibinfo {year} {2007}),\ \href@noop {} {\bibfield  {journal}
  {\bibinfo  {journal} {J. Mol. Med.}\ }\textbf {\bibinfo {volume} {85}},\
  \bibinfo {pages} {1301}}\BibitemShut {NoStop}%
\bibitem [{\citenamefont {Brandt}\ \emph {et~al.}(1960)\citenamefont {Brandt},
  \citenamefont {Berko},\ and\ \citenamefont {Walker}}]{Brandt1960}%
  \BibitemOpen
  \bibfield  {author} {\bibinfo {author} {\bibnamefont {Brandt}, \bibfnamefont
  {W.}}, \bibinfo {author} {\bibfnamefont {S.}~\bibnamefont {Berko}}, \ and\
  \bibinfo {author} {\bibfnamefont {W.}~\bibnamefont {Walker}}} (\bibinfo
  {year} {1960}),\ \href {\doibase 10.1103/PhysRev.120.1289} {\bibfield
  {journal} {\bibinfo  {journal} {Phys. Rev.}\ }\textbf {\bibinfo {volume}
  {120}},\ \bibinfo {pages} {1289}}\BibitemShut {NoStop}%
\bibitem [{\citenamefont {Brandt}\ and\ \citenamefont
  {Wilkenfeld}(1975)}]{Brandt1975}%
  \BibitemOpen
  \bibfield  {author} {\bibinfo {author} {\bibnamefont {Brandt}, \bibfnamefont
  {W.}}, \ and\ \bibinfo {author} {\bibfnamefont {J.}~\bibnamefont
  {Wilkenfeld}}} (\bibinfo {year} {1975}),\ \href {\doibase
  10.1103/PhysRevB.12.2579} {\bibfield  {journal} {\bibinfo  {journal} {Phys.
  Rev. B}\ }\textbf {\bibinfo {volume} {12}},\ \bibinfo {pages}
  {2579}}\BibitemShut {NoStop}%
\bibitem [{\citenamefont {Brown}(1974)}]{Brown1974}%
  \BibitemOpen
  \bibfield  {author} {\bibinfo {author} {\bibnamefont {Brown}, \bibfnamefont
  {B.~J.}}} (\bibinfo {year} {1974}),\ \href@noop {} {\bibfield  {journal}
  {\bibinfo  {journal} {Aust. J. Chem}\ }\textbf {\bibinfo {volume} {27}},\
  \bibinfo {pages} {1125.}}\BibitemShut {Stop}%
\bibitem [{\citenamefont {Bura}\ \emph {et~al.}(2000)\citenamefont {Bura},
  \citenamefont {Dulski}, \citenamefont {Kubicz}, \citenamefont {Malczak},
  \citenamefont {Pedziwiatr}, \citenamefont {Szczepanek}, \citenamefont
  {Stepien},\ and\ \citenamefont {Moskal}}]{Bura2020}%
  \BibitemOpen
  \bibfield  {author} {\bibinfo {author} {\bibnamefont {Bura}, \bibfnamefont
  {Z.}}, \bibinfo {author} {\bibfnamefont {K.}~\bibnamefont {Dulski}}, \bibinfo
  {author} {\bibfnamefont {E.}~\bibnamefont {Kubicz}}, \bibinfo {author}
  {\bibfnamefont {P.}~\bibnamefont {Malczak}}, \bibinfo {author} {\bibfnamefont
  {M.}~\bibnamefont {Pedziwiatr}}, \bibinfo {author} {\bibfnamefont
  {M.}~\bibnamefont {Szczepanek}}, \bibinfo {author} {\bibfnamefont {E.~L.}\
  \bibnamefont {Stepien}}, \ and\ \bibinfo {author} {\bibfnamefont
  {P.}~\bibnamefont {Moskal}}} (\bibinfo {year} {2000}),\ \href {\doibase
  https://doi.org/10.5506/APhysPolB.51.377} {\bibfield  {journal} {\bibinfo
  {journal} {Act. Phys. Pol. B}\ }\textbf {\bibinfo {volume} {51}},\ \bibinfo
  {pages} {377}}\BibitemShut {NoStop}%
\bibitem [{\citenamefont {Caradonna}\ \emph {et~al.}(2019)\citenamefont
  {Caradonna}, \citenamefont {Reutens}, \citenamefont {Takahashi},
  \citenamefont {Takeda},\ and\ \citenamefont {Vegh}}]{Caradonna2020}%
  \BibitemOpen
  \bibfield  {author} {\bibinfo {author} {\bibnamefont {Caradonna},
  \bibfnamefont {P.}}, \bibinfo {author} {\bibfnamefont {D.}~\bibnamefont
  {Reutens}}, \bibinfo {author} {\bibfnamefont {T.}~\bibnamefont {Takahashi}},
  \bibinfo {author} {\bibfnamefont {S.}~\bibnamefont {Takeda}}, \ and\ \bibinfo
  {author} {\bibfnamefont {V.}~\bibnamefont {Vegh}}} (\bibinfo {year} {2019}),\
  \href@noop {} {\bibfield  {journal} {\bibinfo  {journal} {J. Phys. Commun.}\
  }\textbf {\bibinfo {volume} {3}},\ \bibinfo {pages} {105005}}\BibitemShut
  {NoStop}%
\bibitem [{\citenamefont {Cassidy}\ \emph {et~al.}(2010)\citenamefont
  {Cassidy}, \citenamefont {Crivelli}, \citenamefont {Hisakado}, \citenamefont
  {Liszkay}, \citenamefont {Meligne}, \citenamefont {Perez}, \citenamefont
  {Tom},\ and\ \citenamefont {Mills}}]{Cassidy2010}%
  \BibitemOpen
  \bibfield  {author} {\bibinfo {author} {\bibnamefont {Cassidy}, \bibfnamefont
  {D.}}, \bibinfo {author} {\bibfnamefont {P.}~\bibnamefont {Crivelli}},
  \bibinfo {author} {\bibfnamefont {T.}~\bibnamefont {Hisakado}}, \bibinfo
  {author} {\bibfnamefont {L.}~\bibnamefont {Liszkay}}, \bibinfo {author}
  {\bibfnamefont {V.}~\bibnamefont {Meligne}}, \bibinfo {author} {\bibfnamefont
  {P.}~\bibnamefont {Perez}}, \bibinfo {author} {\bibfnamefont
  {H.}~\bibnamefont {Tom}}, \ and\ \bibinfo {author} {\bibfnamefont
  {A.}~\bibnamefont {Mills}}} (\bibinfo {year} {2010}),\ \href {\doibase
  10.1103/PhysRevA.81.012715} {\bibfield  {journal} {\bibinfo  {journal} {Phys.
  Rev. A}\ }\textbf {\bibinfo {volume} {81}},\ \bibinfo {pages}
  {012715}}\BibitemShut {NoStop}%
\bibitem [{\citenamefont {Cassidy}\ \emph {et~al.}(2006)\citenamefont
  {Cassidy}, \citenamefont {Deng},\ and\ \citenamefont
  {Greaves}}]{Cassidy2006}%
  \BibitemOpen
  \bibfield  {author} {\bibinfo {author} {\bibnamefont {Cassidy}, \bibfnamefont
  {D.}}, \bibinfo {author} {\bibfnamefont {S.}~\bibnamefont {Deng}}, \ and\
  \bibinfo {author} {\bibfnamefont {R.}~\bibnamefont {Greaves}}} (\bibinfo
  {year} {2006}),\ \href {\doibase 10.1063/1.2221509} {\bibfield  {journal}
  {\bibinfo  {journal} {Rev of Sci Inst.}\ }\textbf {\bibinfo {volume} {77}},\
  \bibinfo {pages} {073106}}\BibitemShut {NoStop}%
\bibitem [{\citenamefont {Cassidy}\ \emph {et~al.}(2012)\citenamefont
  {Cassidy}, \citenamefont {Hisakado}, \citenamefont {Tom},\ and\ \citenamefont
  {Mills}}]{Cassidy2012}%
  \BibitemOpen
  \bibfield  {author} {\bibinfo {author} {\bibnamefont {Cassidy}, \bibfnamefont
  {D.}}, \bibinfo {author} {\bibfnamefont {T.}~\bibnamefont {Hisakado}},
  \bibinfo {author} {\bibfnamefont {H.}~\bibnamefont {Tom}}, \ and\ \bibinfo
  {author} {\bibfnamefont {A.}~\bibnamefont {Mills}}} (\bibinfo {year}
  {2012}),\ \href {\doibase 10.1103/PhysRevLett.108.043401} {\bibfield
  {journal} {\bibinfo  {journal} {Phys. Rev. Lett.}\ }\textbf {\bibinfo
  {volume} {108}},\ \bibinfo {pages} {043401}}\BibitemShut {NoStop}%
\bibitem [{\citenamefont {Cassidy}\ and\ \citenamefont
  {Mills}(2007)}]{Cassidy2007}%
  \BibitemOpen
  \bibfield  {author} {\bibinfo {author} {\bibnamefont {Cassidy}, \bibfnamefont
  {D.}}, \ and\ \bibinfo {author} {\bibfnamefont {A.}~\bibnamefont {Mills}}}
  (\bibinfo {year} {2007}),\ \href {\doibase 10.1038/nature06094} {\bibfield
  {journal} {\bibinfo  {journal} {Nature}\ }\textbf {\bibinfo {volume} {449}},\
  \bibinfo {pages} {195}}\BibitemShut {NoStop}%
\bibitem [{\citenamefont {Cassidy}\ \emph {et~al.}(2007)\citenamefont
  {Cassidy}, \citenamefont {Yokoyama}, \citenamefont {Deng}, \citenamefont
  {Griscom}, \citenamefont {Miyadera}, \citenamefont {Tom}, \citenamefont
  {Varma},\ and\ \citenamefont {Mills}}]{Cassidy2007b}%
  \BibitemOpen
  \bibfield  {author} {\bibinfo {author} {\bibnamefont {Cassidy}, \bibfnamefont
  {D.}}, \bibinfo {author} {\bibfnamefont {K.}~\bibnamefont {Yokoyama}},
  \bibinfo {author} {\bibfnamefont {S.}~\bibnamefont {Deng}}, \bibinfo {author}
  {\bibfnamefont {D.}~\bibnamefont {Griscom}}, \bibinfo {author} {\bibfnamefont
  {H.}~\bibnamefont {Miyadera}}, \bibinfo {author} {\bibfnamefont
  {H.}~\bibnamefont {Tom}}, \bibinfo {author} {\bibfnamefont {C.}~\bibnamefont
  {Varma}}, \ and\ \bibinfo {author} {\bibfnamefont {A.}~\bibnamefont {Mills}}}
  (\bibinfo {year} {2007}),\ \href {\doibase 10.1103/PhysRevB.75.085415}
  {\bibfield  {journal} {\bibinfo  {journal} {Phys. Rev. B}\ }\textbf {\bibinfo
  {volume} {75}},\ \bibinfo {pages} {085415}}\BibitemShut {NoStop}%
\bibitem [{\citenamefont {Cassidy}(2018)}]{Cassidy2018tgq}%
  \BibitemOpen
  \bibfield  {author} {\bibinfo {author} {\bibnamefont {Cassidy}, \bibfnamefont
  {D.~B.}}} (\bibinfo {year} {2018}),\ \href {\doibase
  10.1140/epjd/e2018-80721-y} {\bibfield  {journal} {\bibinfo  {journal} {Eur.
  Phys. J. D}\ }\textbf {\bibinfo {volume} {72}}~(\bibinfo {number} {3}),\
  \bibinfo {pages} {53}}\BibitemShut {NoStop}%
\bibitem [{\citenamefont {Caswell}\ and\ \citenamefont
  {Lepage}(1986)}]{Caswell1985ui}%
  \BibitemOpen
  \bibfield  {author} {\bibinfo {author} {\bibnamefont {Caswell}, \bibfnamefont
  {W.~E.}}, \ and\ \bibinfo {author} {\bibfnamefont {G.~P.}\ \bibnamefont
  {Lepage}}} (\bibinfo {year} {1986}),\ \href {\doibase
  10.1016/0370-2693(86)91297-9} {\bibfield  {journal} {\bibinfo  {journal}
  {Phys. Lett.}\ }\textbf {\bibinfo {volume} {167B}},\ \bibinfo {pages}
  {437}}\BibitemShut {NoStop}%
%%CITATION = PHLTA,167B,437;%%
\bibitem [{\citenamefont {Cates}\ and\ \citenamefont
  {Levin}(2019)}]{Cates-Levin2019}%
  \BibitemOpen
  \bibfield  {author} {\bibinfo {author} {\bibnamefont {Cates}, \bibfnamefont
  {J.}}, \ and\ \bibinfo {author} {\bibfnamefont {C.}~\bibnamefont {Levin}}}
  (\bibinfo {year} {2019}),\ \href@noop {} {\bibfield  {journal} {\bibinfo
  {journal} {Phys. Med. Biol.}\ }\textbf {\bibinfo {volume} {64}},\ \bibinfo
  {pages} {175016}}\BibitemShut {NoStop}%
\bibitem [{\citenamefont {Ceg{\l}a}\ and\ \citenamefont
  {Piotrowski}(2021)}]{Cegla2021}%
  \BibitemOpen
  \bibfield  {author} {\bibinfo {author} {\bibnamefont {Ceg{\l}a},
  \bibfnamefont {P.}}, \ and\ \bibinfo {author} {\bibfnamefont
  {T.}~\bibnamefont {Piotrowski}}} (\bibinfo {year} {2021}),\ \href {\doibase
  https://doi.org/10.1515/bams-2021-0145} {\bibfield  {journal} {\bibinfo
  {journal} {Bio-Algorithms and Med-Systems}\ }\textbf {\bibinfo {volume}
  {17}},\ \bibinfo {pages} {259}}\BibitemShut {NoStop}%
\bibitem [{\citenamefont {Chamerski}\ \emph {et~al.}(2017)\citenamefont
  {Chamerski}, \citenamefont {Korzekwa}, \citenamefont {Filipecki},
  \citenamefont {Shpotyuk}, \citenamefont {Stopa}, \citenamefont {Jelen},\ and\
  \citenamefont {Sitarz}}]{Chamerski2017}%
  \BibitemOpen
  \bibfield  {author} {\bibinfo {author} {\bibnamefont {Chamerski},
  \bibfnamefont {K.}}, \bibinfo {author} {\bibfnamefont {W.}~\bibnamefont
  {Korzekwa}}, \bibinfo {author} {\bibfnamefont {J.}~\bibnamefont {Filipecki}},
  \bibinfo {author} {\bibfnamefont {O.}~\bibnamefont {Shpotyuk}}, \bibinfo
  {author} {\bibfnamefont {M.}~\bibnamefont {Stopa}}, \bibinfo {author}
  {\bibfnamefont {P.}~\bibnamefont {Jelen}}, \ and\ \bibinfo {author}
  {\bibfnamefont {M.}~\bibnamefont {Sitarz}}} (\bibinfo {year} {2017}),\ \href
  {\doibase doi.org.10.1016/j.saa.2017.10.051} {\bibfield  {journal} {\bibinfo
  {journal} {Nanoscale Res. Lett.}\ }\textbf {\bibinfo {volume} {12}},\
  \bibinfo {pages} {303}}\BibitemShut {NoStop}%
\bibitem [{\citenamefont {Champion}(2005)}]{Champion2005}%
  \BibitemOpen
  \bibfield  {author} {\bibinfo {author} {\bibnamefont {Champion},
  \bibfnamefont {C.}}} (\bibinfo {year} {2005}),\ \href@noop {} {\bibfield
  {journal} {\bibinfo  {journal} {Braz. Arch. Biol. Technol.}\ }\textbf
  {\bibinfo {volume} {48}},\ \bibinfo {pages} {191}}\BibitemShut {NoStop}%
\bibitem [{\citenamefont {Chang}\ \emph {et~al.}(1987)\citenamefont {Chang},
  \citenamefont {Xu},\ and\ \citenamefont {Zeng}}]{Chang1987}%
  \BibitemOpen
  \bibfield  {author} {\bibinfo {author} {\bibnamefont {Chang}, \bibfnamefont
  {T.}}, \bibinfo {author} {\bibfnamefont {M.}~\bibnamefont {Xu}}, \ and\
  \bibinfo {author} {\bibfnamefont {X.}~\bibnamefont {Zeng}}} (\bibinfo {year}
  {1987}),\ \href {\doibase 10.1016/0375-9601(87)90458-0} {\bibfield  {journal}
  {\bibinfo  {journal} {Phys. Lett. A}\ }\textbf {\bibinfo {volume} {126}},\
  \bibinfo {pages} {189}}\BibitemShut {NoStop}%
\bibitem [{\citenamefont {Charlton}\ \emph {et~al.}(2021)\citenamefont
  {Charlton} \emph {et~al.}}]{Charlton2021}%
  \BibitemOpen
  \bibfield  {author} {\bibinfo {author} {\bibnamefont {Charlton},
  \bibfnamefont {M.}},  \emph {et~al.} (\bibinfo {collaboration} {GBAR})}
  (\bibinfo {year} {2021}),\ \href {\doibase 10.1016/j.nima.2020.164657}
  {\bibfield  {journal} {\bibinfo  {journal} {Nucl. Inst. and Meth. A}\
  }\textbf {\bibinfo {volume} {985}},\ \bibinfo {pages} {164657}}\BibitemShut
  {NoStop}%
\bibitem [{\citenamefont {Chen}\ \emph {et~al.}(2012)\citenamefont {Chen},
  \citenamefont {VanHorn},\ and\ \citenamefont {Jean}}]{Chen2012}%
  \BibitemOpen
  \bibfield  {author} {\bibinfo {author} {\bibnamefont {Chen}, \bibfnamefont
  {H.}}, \bibinfo {author} {\bibfnamefont {J.}~\bibnamefont {VanHorn}}, \ and\
  \bibinfo {author} {\bibfnamefont {Y.}~\bibnamefont {Jean}}} (\bibinfo {year}
  {2012}),\ \href {\doibase
  https://doi.org/10.4028/www.scientific.net/DDF.331.275} {\bibfield  {journal}
  {\bibinfo  {journal} {Defect. Diffusion. Forum.}\ }\textbf {\bibinfo {volume}
  {331}},\ \bibinfo {pages} {275}}\BibitemShut {NoStop}%
\bibitem [{\citenamefont {Choi{\'n}ski}\ and\ \citenamefont
  {{\L}yczko}(2021)}]{Choinski2021}%
  \BibitemOpen
  \bibfield  {author} {\bibinfo {author} {\bibnamefont {Choi{\'n}ski},
  \bibfnamefont {J.}}, \ and\ \bibinfo {author} {\bibfnamefont
  {M.}~\bibnamefont {{\L}yczko}}} (\bibinfo {year} {2021}),\ \href {\doibase
  https://doi.org/10.1515/bams-2021-0136} {\bibfield  {journal} {\bibinfo
  {journal} {Bio-Algorithms and Med-Systems}\ }\textbf {\bibinfo {volume}
  {17}},\ \bibinfo {pages} {241}}\BibitemShut {NoStop}%
\bibitem [{\citenamefont {Chow}\ \emph {et~al.}(1981)\citenamefont {Chow},
  \citenamefont {Chuang},\ and\ \citenamefont {Tseng}}]{Chow1981}%
  \BibitemOpen
  \bibfield  {author} {\bibinfo {author} {\bibnamefont {Chow}, \bibfnamefont
  {E.~I.~H.}}, \bibinfo {author} {\bibfnamefont {S.}~\bibnamefont {Chuang}}, \
  and\ \bibinfo {author} {\bibfnamefont {P.}~\bibnamefont {Tseng}}} (\bibinfo
  {year} {1981}),\ \href {\doibase
  http://dx.doi.org/10.1016/0005-2736(81)90344-8} {\bibfield  {journal}
  {\bibinfo  {journal} {Biochim. Biophys. Acta Rep.}\ }\textbf {\bibinfo
  {volume} {646}},\ \bibinfo {pages} {356.}}\BibitemShut {Stop}%
\bibitem [{\citenamefont {Coleman}(2003)}]{Coleman2003}%
  \BibitemOpen
  \bibfield  {author} {\bibinfo {author} {\bibnamefont {Coleman}, \bibfnamefont
  {P.}}} (\bibinfo {year} {2003}),\ \href@noop {} {\emph {\bibinfo {title}
  {{Principles And Applications Of Positron And Positronium Chemistry}}}}\
  (\bibinfo  {publisher} {Jean,Y., Mallon, P.E.,Schrader, D.M., World
  Scientific, New Jersey, London, Singapore, Hong Kong})\BibitemShut {NoStop}%
\bibitem [{\citenamefont {Consolati}\ \emph {et~al.}(2013)\citenamefont
  {Consolati}, \citenamefont {Ferragut}, \citenamefont {Galarneau},
  \citenamefont {Di~Renzo},\ and\ \citenamefont {Quasso}}]{Consolati2013}%
  \BibitemOpen
  \bibfield  {author} {\bibinfo {author} {\bibnamefont {Consolati},
  \bibfnamefont {G.}}, \bibinfo {author} {\bibfnamefont {R.}~\bibnamefont
  {Ferragut}}, \bibinfo {author} {\bibfnamefont {A.}~\bibnamefont {Galarneau}},
  \bibinfo {author} {\bibfnamefont {F.}~\bibnamefont {Di~Renzo}}, \ and\
  \bibinfo {author} {\bibfnamefont {F.}~\bibnamefont {Quasso}}} (\bibinfo
  {year} {2013}),\ \href@noop {} {\bibfield  {journal} {\bibinfo  {journal}
  {Chem. Soc. Rev.}\ }\textbf {\bibinfo {volume} {42}},\ \bibinfo {pages}
  {3821}}\BibitemShut {NoStop}%
\bibitem [{\citenamefont {Conti}(2009)}]{Conti2009}%
  \BibitemOpen
  \bibfield  {author} {\bibinfo {author} {\bibnamefont {Conti}, \bibfnamefont
  {M.}}} (\bibinfo {year} {2009}),\ \href@noop {} {\bibfield  {journal}
  {\bibinfo  {journal} {Phys. Med.}\ }\textbf {\bibinfo {volume} {25}},\
  \bibinfo {pages} {1}}\BibitemShut {NoStop}%
\bibitem [{\citenamefont {Cooper}\ \emph {et~al.}(2016)\citenamefont {Cooper},
  \citenamefont {Alonso}, \citenamefont {Deller}, \citenamefont {Liszkay},\
  and\ \citenamefont {Cassidy}}]{Cooper2016}%
  \BibitemOpen
  \bibfield  {author} {\bibinfo {author} {\bibnamefont {Cooper}, \bibfnamefont
  {B.}}, \bibinfo {author} {\bibfnamefont {A.}~\bibnamefont {Alonso}}, \bibinfo
  {author} {\bibfnamefont {A.}~\bibnamefont {Deller}}, \bibinfo {author}
  {\bibfnamefont {L.}~\bibnamefont {Liszkay}}, \ and\ \bibinfo {author}
  {\bibfnamefont {D.}~\bibnamefont {Cassidy}}} (\bibinfo {year} {2016}),\ \href
  {\doibase 10.1103/PhysRevB.93.125305} {\bibfield  {journal} {\bibinfo
  {journal} {Phys. Rev. B}\ }\textbf {\bibinfo {volume} {93}},\ \bibinfo
  {pages} {125305}}\BibitemShut {NoStop}%
\bibitem [{\citenamefont {Cramer}\ and\ \citenamefont
  {Vaupel}(2022)}]{Cramer-Vaupel-hypoxia-2022}%
  \BibitemOpen
  \bibfield  {author} {\bibinfo {author} {\bibnamefont {Cramer}, \bibfnamefont
  {T.}}, \ and\ \bibinfo {author} {\bibfnamefont {P.}~\bibnamefont {Vaupel}}}
  (\bibinfo {year} {2022}),\ \href {\doibase 10.1016/j.jhep.2021.12.028}
  {\bibfield  {journal} {\bibinfo  {journal} {Journal of Hepatology}\ }\textbf
  {\bibinfo {volume} {76}},\ \bibinfo {pages} {975}}\BibitemShut {NoStop}%
\bibitem [{\citenamefont {Crivelli}\ \emph {et~al.}(2010)\citenamefont
  {Crivelli}, \citenamefont {Gendotti}, \citenamefont {Rubbia}, \citenamefont
  {Liszkay}, \citenamefont {Perez},\ and\ \citenamefont
  {Corbel}}]{Crivelli2010PRA}%
  \BibitemOpen
  \bibfield  {author} {\bibinfo {author} {\bibnamefont {Crivelli},
  \bibfnamefont {P.}}, \bibinfo {author} {\bibfnamefont {U.}~\bibnamefont
  {Gendotti}}, \bibinfo {author} {\bibfnamefont {A.}~\bibnamefont {Rubbia}},
  \bibinfo {author} {\bibfnamefont {L.}~\bibnamefont {Liszkay}}, \bibinfo
  {author} {\bibfnamefont {P.}~\bibnamefont {Perez}}, \ and\ \bibinfo {author}
  {\bibfnamefont {C.}~\bibnamefont {Corbel}}} (\bibinfo {year} {2010}),\ \href
  {\doibase 10.1103/PhysRevA.81.052703} {\bibfield  {journal} {\bibinfo
  {journal} {Phys. Rev. A}\ }\textbf {\bibinfo {volume} {81}},\ \bibinfo
  {pages} {052703}}\BibitemShut {NoStop}%
\bibitem [{\citenamefont {Czarnecki}\ \emph {et~al.}(1999)\citenamefont
  {Czarnecki}, \citenamefont {Melnikov},\ and\ \citenamefont
  {Yelkhovsky}}]{Czarnecki1998zv}%
  \BibitemOpen
  \bibfield  {author} {\bibinfo {author} {\bibnamefont {Czarnecki},
  \bibfnamefont {A.}}, \bibinfo {author} {\bibfnamefont {K.}~\bibnamefont
  {Melnikov}}, \ and\ \bibinfo {author} {\bibfnamefont {A.}~\bibnamefont
  {Yelkhovsky}}} (\bibinfo {year} {1999}),\ \href {\doibase
  10.1103/PhysRevLett.82.311} {\bibfield  {journal} {\bibinfo  {journal} {Phys.
  Rev. Lett.}\ }\textbf {\bibinfo {volume} {82}},\ \bibinfo {pages}
  {311}}\BibitemShut {NoStop}%
%%CITATION = HEP-PH/9809341;%%
\bibitem [{\citenamefont {Danielson}\ \emph {et~al.}(2015)\citenamefont
  {Danielson}, \citenamefont {Dubin}, \citenamefont {Greaves},\ and\
  \citenamefont {Surko}}]{Danielson2015}%
  \BibitemOpen
  \bibfield  {author} {\bibinfo {author} {\bibnamefont {Danielson},
  \bibfnamefont {J.}}, \bibinfo {author} {\bibfnamefont {D.}~\bibnamefont
  {Dubin}}, \bibinfo {author} {\bibfnamefont {R.}~\bibnamefont {Greaves}}, \
  and\ \bibinfo {author} {\bibfnamefont {C.}~\bibnamefont {Surko}}} (\bibinfo
  {year} {2015}),\ \href {\doibase 10.1103/RevModPhys.87.247} {\bibfield
  {journal} {\bibinfo  {journal} {Rev. Mod. Phys.}\ }\textbf {\bibinfo {volume}
  {87}},\ \bibinfo {pages} {247}}\BibitemShut {NoStop}%
\bibitem [{\citenamefont {Dauwe}\ \emph {et~al.}(2005)\citenamefont {Dauwe},
  \citenamefont {van Waeyenberge},\ and\ \citenamefont
  {de~Baerdemaeker}}]{Dauwe2005}%
  \BibitemOpen
  \bibfield  {author} {\bibinfo {author} {\bibnamefont {Dauwe}, \bibfnamefont
  {C.}}, \bibinfo {author} {\bibfnamefont {B.}~\bibnamefont {van Waeyenberge}},
  \ and\ \bibinfo {author} {\bibfnamefont {J.}~\bibnamefont {de~Baerdemaeker}}}
  (\bibinfo {year} {2005}),\ \href {\doibase 10.12693/APhysPolA.107.623}
  {\bibfield  {journal} {\bibinfo  {journal} {Acta Phys. Pol. A}\ }\textbf
  {\bibinfo {volume} {107}},\ \bibinfo {pages} {623}}\BibitemShut {NoStop}%
\bibitem [{\citenamefont {Deutsch}(1951)}]{Deutsch1951zza}%
  \BibitemOpen
  \bibfield  {author} {\bibinfo {author} {\bibnamefont {Deutsch}, \bibfnamefont
  {M.}}} (\bibinfo {year} {1951}),\ \href {\doibase 10.1103/PhysRev.82.455}
  {\bibfield  {journal} {\bibinfo  {journal} {Phys. Rev.}\ }\textbf {\bibinfo
  {volume} {82}},\ \bibinfo {pages} {455}}\BibitemShut {NoStop}%
%%CITATION = PHRVA,82,455;%%
\bibitem [{\citenamefont {Dirac}(1931)}]{Dirac1931kp}%
  \BibitemOpen
  \bibfield  {author} {\bibinfo {author} {\bibnamefont {Dirac}, \bibfnamefont
  {P.~A.~M.}}} (\bibinfo {year} {1931}),\ \href {\doibase
  10.1098/rspa.1931.0130} {\bibfield  {journal} {\bibinfo  {journal} {Proc.
  Roy. Soc. Lond.}\ }\textbf {\bibinfo {volume} {A133}}~(\bibinfo {number}
  {821}),\ \bibinfo {pages} {60}}\BibitemShut {NoStop}%
%%CITATION = PRSLA,A133,60;%%
\bibitem [{\citenamefont {Djekidel}\ \emph {et~al.}(2022)\citenamefont
  {Djekidel}, \citenamefont {AlSadi}, \citenamefont {Abi~Akl}, \citenamefont
  {Vandenberghe},\ and\ \citenamefont {O.}}]{Djekidel2022}%
  \BibitemOpen
  \bibfield  {author} {\bibinfo {author} {\bibnamefont {Djekidel},
  \bibfnamefont {M.}}, \bibinfo {author} {\bibfnamefont {R.}~\bibnamefont
  {AlSadi}}, \bibinfo {author} {\bibfnamefont {M.}~\bibnamefont {Abi~Akl}},
  \bibinfo {author} {\bibfnamefont {S.}~\bibnamefont {Vandenberghe}}, \ and\
  \bibinfo {author} {\bibfnamefont {B.}~\bibnamefont {O.}}} (\bibinfo {year}
  {2022}),\ \href {\doibase 10.1007/s00259-022-05873-y} {\bibfield  {journal}
  {\bibinfo  {journal} {Eur. J. Nucl. Med. Mol. Imaging}\ }\textbf {\bibinfo
  {volume} {49}},\ \bibinfo {pages} {3624}}\BibitemShut {NoStop}%
\bibitem [{\citenamefont {Dlubek}\ \emph {et~al.}(2000)\citenamefont {Dlubek},
  \citenamefont {Fretwell},\ and\ \citenamefont {Alam}}]{Dlubek2000}%
  \BibitemOpen
  \bibfield  {author} {\bibinfo {author} {\bibnamefont {Dlubek}, \bibfnamefont
  {G.}}, \bibinfo {author} {\bibfnamefont {H.}~\bibnamefont {Fretwell}}, \ and\
  \bibinfo {author} {\bibfnamefont {M.}~\bibnamefont {Alam}}} (\bibinfo {year}
  {2000}),\ \href {\doibase https://doi.org/10.1021/ma9904215} {\bibfield
  {journal} {\bibinfo  {journal} {Macromolecules}\ }\textbf {\bibinfo {volume}
  {33}},\ \bibinfo {pages} {187.}}\BibitemShut {Stop}%
\bibitem [{\citenamefont {Doser}\ \emph {et~al.}(2018)\citenamefont {Doser}
  \emph {et~al.}}]{Doser2018zfc}%
  \BibitemOpen
  \bibfield  {author} {\bibinfo {author} {\bibnamefont {Doser}, \bibfnamefont
  {M.}},  \emph {et~al.}} (\bibinfo {year} {2018}),\ \href {\doibase
  10.1098/rsta.2017.0274} {\bibfield  {journal} {\bibinfo  {journal} {Phil.
  Trans. Roy. Soc. Lond.}\ }\textbf {\bibinfo {volume} {A376}}~(\bibinfo
  {number} {2116}),\ \bibinfo {pages} {20170274}}\BibitemShut {NoStop}%
%%CITATION = PTRSA,A376,20170274;%%
\bibitem [{\citenamefont {Dryzek}\ \emph {et~al.}(2007)\citenamefont {Dryzek},
  \citenamefont {Suzuki},\ and\ \citenamefont {Yu}}]{Dryzek2007}%
  \BibitemOpen
  \bibfield  {author} {\bibinfo {author} {\bibnamefont {Dryzek}, \bibfnamefont
  {J.}}, \bibinfo {author} {\bibfnamefont {T.}~\bibnamefont {Suzuki}}, \ and\
  \bibinfo {author} {\bibfnamefont {R.}~\bibnamefont {Yu}}} (\bibinfo {year}
  {2007}),\ \href {\doibase 10.1016/j.radphyschem.2006.03.054} {\bibfield
  {journal} {\bibinfo  {journal} {Radiat. Phys. Chem.}\ }\textbf {\bibinfo
  {volume} {76}},\ \bibinfo {pages} {297}}\BibitemShut {NoStop}%
\bibitem [{\citenamefont {Dufour}\ \emph {et~al.}(2015)\citenamefont {Dufour},
  \citenamefont {Cassidy}, \citenamefont {Crivelli}, \citenamefont {Debu},
  \citenamefont {Lambrecht}, \citenamefont {A.~Nesvizhevsky}, \citenamefont
  {Reynaud}, \citenamefont {Voronin},\ and\ \citenamefont {Wall}}]{Dufour2015}%
  \BibitemOpen
  \bibfield  {author} {\bibinfo {author} {\bibnamefont {Dufour}, \bibfnamefont
  {G.}}, \bibinfo {author} {\bibfnamefont {D.}~\bibnamefont {Cassidy}},
  \bibinfo {author} {\bibfnamefont {P.}~\bibnamefont {Crivelli}}, \bibinfo
  {author} {\bibfnamefont {P.}~\bibnamefont {Debu}}, \bibinfo {author}
  {\bibnamefont {Lambrecht}}, \bibinfo {author} {\bibfnamefont
  {V.}~\bibnamefont {A.~Nesvizhevsky}}, \bibinfo {author} {\bibfnamefont
  {S.}~\bibnamefont {Reynaud}}, \bibinfo {author} {\bibfnamefont
  {A.}~\bibnamefont {Voronin}}, \ and\ \bibinfo {author} {\bibfnamefont
  {T.}~\bibnamefont {Wall}}} (\bibinfo {year} {2015}),\ \href {\doibase
  10.1155/2015/379642} {\bibfield  {journal} {\bibinfo  {journal} {Adv. in High
  Ener. Phys.}\ }\textbf {\bibinfo {volume} {2015}},\ \bibinfo {pages}
  {379642}}\BibitemShut {NoStop}%
\bibitem [{\citenamefont {Dulski}(2020)}]{Dulski-Avalanche-2020}%
  \BibitemOpen
  \bibfield  {author} {\bibinfo {author} {\bibnamefont {Dulski}, \bibfnamefont
  {K.}}} (\bibinfo {year} {2020}),\ \href@noop {} {\bibfield  {journal}
  {\bibinfo  {journal} {Acta Phys. Pol. A}\ }\textbf {\bibinfo {volume}
  {137}},\ \bibinfo {pages} {167}}\BibitemShut {NoStop}%
\bibitem [{\citenamefont {Dulski}\ \emph {et~al.}(2021)\citenamefont {Dulski}
  \emph {et~al.}}]{Dulski2021}%
  \BibitemOpen
  \bibfield  {author} {\bibinfo {author} {\bibnamefont {Dulski}, \bibfnamefont
  {K.}},  \emph {et~al.}} (\bibinfo {year} {2021}),\ \href@noop {} {\bibfield
  {journal} {\bibinfo  {journal} {Nucl. Instrum. Meth. A}\ }\textbf {\bibinfo
  {volume} {1008}},\ \bibinfo {pages} {165452}}\BibitemShut {NoStop}%
\bibitem [{\citenamefont {Dupasquier}\ \emph {et~al.}(1991)\citenamefont
  {Dupasquier}, \citenamefont {De~Natale},\ and\ \citenamefont
  {Rolando}}]{Dupasquier1991}%
  \BibitemOpen
  \bibfield  {author} {\bibinfo {author} {\bibnamefont {Dupasquier},
  \bibfnamefont {A.}}, \bibinfo {author} {\bibfnamefont {P.}~\bibnamefont
  {De~Natale}}, \ and\ \bibinfo {author} {\bibfnamefont {A.}~\bibnamefont
  {Rolando}}} (\bibinfo {year} {1991}),\ \href@noop {} {\bibfield  {journal}
  {\bibinfo  {journal} {Phys. Rev. B}\ }\textbf {\bibinfo {volume} {43}},\
  \bibinfo {pages} {10036}}\BibitemShut {NoStop}%
\bibitem [{\citenamefont {Dupasquier}\ and\ \citenamefont
  {Zecca}(1985)}]{Dupasquier1985}%
  \BibitemOpen
  \bibfield  {author} {\bibinfo {author} {\bibnamefont {Dupasquier},
  \bibfnamefont {A.}}, \ and\ \bibinfo {author} {\bibfnamefont
  {A.}~\bibnamefont {Zecca}}} (\bibinfo {year} {1985}),\ \href {\doibase
  10.1007/BF02724348} {\bibfield  {journal} {\bibinfo  {journal} {Riv. Nuovo
  Cim.}\ }\textbf {\bibinfo {volume} {8}},\ \bibinfo {pages} {1}}\BibitemShut
  {NoStop}%
\bibitem [{\citenamefont {Eldrup}\ and\ \citenamefont
  {Mogensen}(1972)}]{Eldrup1972}%
  \BibitemOpen
  \bibfield  {author} {\bibinfo {author} {\bibnamefont {Eldrup}, \bibfnamefont
  {M.}}, \ and\ \bibinfo {author} {\bibfnamefont {O.}~\bibnamefont {Mogensen}}}
  (\bibinfo {year} {1972}),\ \href {\doibase 10.1063/1.1677990} {\bibfield
  {journal} {\bibinfo  {journal} {J. Chem. Phys.}\ }\textbf {\bibinfo {volume}
  {57}},\ \bibinfo {pages} {495}}\BibitemShut {NoStop}%
\bibitem [{\citenamefont {Eldrup}\ \emph {et~al.}(1983)\citenamefont {Eldrup},
  \citenamefont {Vehanen}, \citenamefont {Schultz},\ and\ \citenamefont
  {Lynn}}]{Eldrup1983}%
  \BibitemOpen
  \bibfield  {author} {\bibinfo {author} {\bibnamefont {Eldrup}, \bibfnamefont
  {M.}}, \bibinfo {author} {\bibfnamefont {A.}~\bibnamefont {Vehanen}},
  \bibinfo {author} {\bibfnamefont {P.~J.}\ \bibnamefont {Schultz}}, \ and\
  \bibinfo {author} {\bibfnamefont {K.}~\bibnamefont {Lynn}}} (\bibinfo {year}
  {1983}),\ \href {\doibase 10.1103/PhysRevLett.51.2007} {\bibfield  {journal}
  {\bibinfo  {journal} {Phys. Rev. Lett.}\ }\textbf {\bibinfo {volume} {51}},\
  \bibinfo {pages} {2007}}\BibitemShut {NoStop}%
\bibitem [{\citenamefont {Eldrup}\ \emph {et~al.}(1985)\citenamefont {Eldrup},
  \citenamefont {Vehanen}, \citenamefont {Schultz},\ and\ \citenamefont
  {Lynn}}]{Eldrup1985}%
  \BibitemOpen
  \bibfield  {author} {\bibinfo {author} {\bibnamefont {Eldrup}, \bibfnamefont
  {M.}}, \bibinfo {author} {\bibfnamefont {A.}~\bibnamefont {Vehanen}},
  \bibinfo {author} {\bibfnamefont {P.~J.}\ \bibnamefont {Schultz}}, \ and\
  \bibinfo {author} {\bibfnamefont {K.}~\bibnamefont {Lynn}}} (\bibinfo {year}
  {1985}),\ \href {\doibase 10.1103/PhysRevB.32.7048} {\bibfield  {journal}
  {\bibinfo  {journal} {Phys. Rev. B}\ }\textbf {\bibinfo {volume} {32}},\
  \bibinfo {pages} {7048}}\BibitemShut {NoStop}%
\bibitem [{\citenamefont {Elias}\ \emph {et~al.}(2001)\citenamefont {Elias},
  \citenamefont {Al-Mashhadani},\ and\ \citenamefont
  {Al-Shiebani}}]{Elias2001}%
  \BibitemOpen
  \bibfield  {author} {\bibinfo {author} {\bibnamefont {Elias}, \bibfnamefont
  {M.~M.}}, \bibinfo {author} {\bibfnamefont {A.}~\bibnamefont
  {Al-Mashhadani}}, \ and\ \bibinfo {author} {\bibfnamefont {Z.}~\bibnamefont
  {Al-Shiebani}}} (\bibinfo {year} {2001}),\ \href@noop {} {\bibfield
  {journal} {\bibinfo  {journal} {Dirasat. Pure. Sci.}\ }\textbf {\bibinfo
  {volume} {28}},\ \bibinfo {pages} {240.}}\BibitemShut {Stop}%
\bibitem [{\citenamefont {Feng}\ \emph {et~al.}(1981)\citenamefont {Feng},
  \citenamefont {Pratt},\ and\ \citenamefont {Tseng}}]{Feng1981}%
  \BibitemOpen
  \bibfield  {author} {\bibinfo {author} {\bibnamefont {Feng}, \bibfnamefont
  {I.~J.}}, \bibinfo {author} {\bibfnamefont {R.~H.}\ \bibnamefont {Pratt}}, \
  and\ \bibinfo {author} {\bibfnamefont {H.~K.}\ \bibnamefont {Tseng}}}
  (\bibinfo {year} {1981}),\ \href {\doibase 10.1103/PhysRevA.24.1358}
  {\bibfield  {journal} {\bibinfo  {journal} {Phys. Rev. A}\ }\textbf {\bibinfo
  {volume} {24}},\ \bibinfo {pages} {1358}}\BibitemShut {NoStop}%
\bibitem [{\citenamefont {Feng}\ \emph {et~al.}(2021)\citenamefont {Feng} \emph
  {et~al.}}]{Feng2021}%
  \BibitemOpen
  \bibfield  {author} {\bibinfo {author} {\bibnamefont {Feng}, \bibfnamefont
  {T.}},  \emph {et~al.}} (\bibinfo {year} {2021}),\ \href {\doibase
  10.2967/jnumed.119.238113} {\bibfield  {journal} {\bibinfo  {journal} {J.
  Nucl. Med}\ }\textbf {\bibinfo {volume} {62}},\ \bibinfo {pages}
  {738}}\BibitemShut {NoStop}%
\bibitem [{\citenamefont {Ferrell}(1958)}]{Ferrell1958}%
  \BibitemOpen
  \bibfield  {author} {\bibinfo {author} {\bibnamefont {Ferrell}, \bibfnamefont
  {R.}}} (\bibinfo {year} {1958}),\ \href {\doibase 10.1103/PhysRev.110.1355}
  {\bibfield  {journal} {\bibinfo  {journal} {Phys. Rev.}\ }\textbf {\bibinfo
  {volume} {110}},\ \bibinfo {pages} {1355}}\BibitemShut {NoStop}%
\bibitem [{\citenamefont {Filosofov}\ \emph {et~al.}(2010)\citenamefont
  {Filosofov} \emph {et~al.}}]{Filosofov2010}%
  \BibitemOpen
  \bibfield  {author} {\bibinfo {author} {\bibnamefont {Filosofov},
  \bibfnamefont {D.}},  \emph {et~al.}} (\bibinfo {year} {2010}),\ \href@noop
  {} {\bibfield  {journal} {\bibinfo  {journal} {Radiochim. Acta}\ }\textbf
  {\bibinfo {volume} {98}},\ \bibinfo {pages} {149}}\BibitemShut {NoStop}%
\bibitem [{\citenamefont {Gajos}\ \emph {et~al.}(2016)\citenamefont {Gajos}
  \emph {et~al.}}]{Gajos2016}%
  \BibitemOpen
  \bibfield  {author} {\bibinfo {author} {\bibnamefont {Gajos}, \bibfnamefont
  {A.}},  \emph {et~al.}} (\bibinfo {year} {2016}),\ \href@noop {} {\bibfield
  {journal} {\bibinfo  {journal} {Nucl. Instrum. Meth. A}\ }\textbf {\bibinfo
  {volume} {819}},\ \bibinfo {pages} {54}}\BibitemShut {NoStop}%
\bibitem [{\citenamefont {Garcia-Arribas}\ \emph {et~al.}(2016)\citenamefont
  {Garcia-Arribas} \emph {et~al.}}]{Garcia-Arribas2016}%
  \BibitemOpen
  \bibfield  {author} {\bibinfo {author} {\bibnamefont {Garcia-Arribas},
  \bibfnamefont {A.~B.}},  \emph {et~al.}} (\bibinfo {year} {2016}),\ \href
  {\doibase doi.org/10.1021/acs.langmuir.6b00927} {\bibfield  {journal}
  {\bibinfo  {journal} {Langmuir}\ }\textbf {\bibinfo {volume} {32}},\ \bibinfo
  {pages} {5434.}}\BibitemShut {Stop}%
\bibitem [{\citenamefont {Gninenko}\ \emph {et~al.}(2002)\citenamefont
  {Gninenko}, \citenamefont {Krasnikov},\ and\ \citenamefont
  {Rubbia}}]{Gninenko2002jn}%
  \BibitemOpen
  \bibfield  {author} {\bibinfo {author} {\bibnamefont {Gninenko},
  \bibfnamefont {S.~N.}}, \bibinfo {author} {\bibfnamefont {N.}~\bibnamefont
  {Krasnikov}}, \ and\ \bibinfo {author} {\bibfnamefont {A.}~\bibnamefont
  {Rubbia}}} (\bibinfo {year} {2002}),\ \href@noop {} {\bibfield  {journal}
  {\bibinfo  {journal} {Mod. Phys. Lett. A}\ }\textbf {\bibinfo {volume}
  {17}},\ \bibinfo {pages} {1713}}\BibitemShut {NoStop}%
\bibitem [{\citenamefont {Goworek}(2014)}]{Goworek2014}%
  \BibitemOpen
  \bibfield  {author} {\bibinfo {author} {\bibnamefont {Goworek}, \bibfnamefont
  {T.}}} (\bibinfo {year} {2014}),\ \href {\doibase 10.2478/umcschem-2013-0012}
  {\bibfield  {journal} {\bibinfo  {journal} {Annales Universitatis Mariae
  Curie-Sk{\l}odowska Lublin – Polonia Section AA}\ }\textbf {\bibinfo
  {volume} {LXIX}},\ \bibinfo {pages} {1}}\BibitemShut {NoStop}%
\bibitem [{\citenamefont {Gregory}\ \emph {et~al.}(1992)\citenamefont
  {Gregory}, \citenamefont {Chai},\ and\ \citenamefont {Su}}]{Gregory1992}%
  \BibitemOpen
  \bibfield  {author} {\bibinfo {author} {\bibnamefont {Gregory}, \bibfnamefont
  {R.~B.}}, \bibinfo {author} {\bibfnamefont {K.}~\bibnamefont {Chai}}, \ and\
  \bibinfo {author} {\bibfnamefont {W.}~\bibnamefont {Su}}} (\bibinfo {year}
  {1992}),\ \href {\doibase
  http://dx.doi.org/10.4028/www.scientific.net/MSF.105-110.1577} {\bibfield
  {journal} {\bibinfo  {journal} {Mater. Sci. Forum}\ }\textbf {\bibinfo
  {volume} {105-110}},\ \bibinfo {pages} {1577.}}\BibitemShut {Stop}%
\bibitem [{\citenamefont {Gullikson}\ and\ \citenamefont
  {Mills}(1986)}]{Gullikson1986}%
  \BibitemOpen
  \bibfield  {author} {\bibinfo {author} {\bibnamefont {Gullikson},
  \bibfnamefont {E.}}, \ and\ \bibinfo {author} {\bibfnamefont
  {A.}~\bibnamefont {Mills}}} (\bibinfo {year} {1986}),\ \href {\doibase
  10.1103/PhysRevLett.57.376} {\bibfield  {journal} {\bibinfo  {journal} {Phys.
  Rev. Lett.}\ }\textbf {\bibinfo {volume} {57}},\ \bibinfo {pages}
  {376}}\BibitemShut {NoStop}%
\bibitem [{\citenamefont {Gundacker}\ \emph {et~al.}(2019)\citenamefont
  {Gundacker} \emph {et~al.}}]{Gundacker2019}%
  \BibitemOpen
  \bibfield  {author} {\bibinfo {author} {\bibnamefont {Gundacker},
  \bibfnamefont {S.}},  \emph {et~al.}} (\bibinfo {year} {2019}),\ \href@noop
  {} {\bibfield  {journal} {\bibinfo  {journal} {Phys. Med. Biol.}\ }\textbf
  {\bibinfo {volume} {64}},\ \bibinfo {pages} {055012}}\BibitemShut {NoStop}%
\bibitem [{\citenamefont {Gurung}\ \emph {et~al.}(2020)\citenamefont {Gurung},
  \citenamefont {Babij}, \citenamefont {Hogan},\ and\ \citenamefont
  {Cassidy}}]{Gurung2020hms}%
  \BibitemOpen
  \bibfield  {author} {\bibinfo {author} {\bibnamefont {Gurung}, \bibfnamefont
  {L.}}, \bibinfo {author} {\bibfnamefont {T.~J.}\ \bibnamefont {Babij}},
  \bibinfo {author} {\bibfnamefont {S.~D.}\ \bibnamefont {Hogan}}, \ and\
  \bibinfo {author} {\bibfnamefont {D.~B.}\ \bibnamefont {Cassidy}}} (\bibinfo
  {year} {2020}),\ \href {\doibase 10.1103/PhysRevLett.125.073002} {\bibfield
  {journal} {\bibinfo  {journal} {Phys. Rev. Lett.}\ }\textbf {\bibinfo
  {volume} {125}}~(\bibinfo {number} {7}),\ \bibinfo {pages}
  {073002}}\BibitemShut {NoStop}%
%%CITATION = PRLTA,125,073002;%%
\bibitem [{\citenamefont {Gurung}\ \emph {et~al.}(2021)\citenamefont {Gurung},
  \citenamefont {Babij}, \citenamefont {Pérez-Ríos}, \citenamefont {Hogan},\
  and\ \citenamefont {Cassidy}}]{Gurung2021xss}%
  \BibitemOpen
  \bibfield  {author} {\bibinfo {author} {\bibnamefont {Gurung}, \bibfnamefont
  {L.}}, \bibinfo {author} {\bibfnamefont {T.~J.}\ \bibnamefont {Babij}},
  \bibinfo {author} {\bibfnamefont {J.}~\bibnamefont {Pérez-Ríos}}, \bibinfo
  {author} {\bibfnamefont {S.~D.}\ \bibnamefont {Hogan}}, \ and\ \bibinfo
  {author} {\bibfnamefont {D.~B.}\ \bibnamefont {Cassidy}}} (\bibinfo {year}
  {2021}),\ \href {\doibase 10.1103/PhysRevA.103.042805} {\bibfield  {journal}
  {\bibinfo  {journal} {Phys. Rev.}\ }\textbf {\bibinfo {volume} {A103}},\
  \bibinfo {pages} {042805}}\BibitemShut {NoStop}%
%%CITATION = PHRVA,A103,042805;%%
\bibitem [{\citenamefont {Gustafson}(1970)}]{Gustafson1970}%
  \BibitemOpen
  \bibfield  {author} {\bibinfo {author} {\bibnamefont {Gustafson},
  \bibfnamefont {D.~R.}}} (\bibinfo {year} {1970}),\ \href {\doibase
  https://doi.org/10.1016/S0006-3495(70)86304-4} {\bibfield  {journal}
  {\bibinfo  {journal} {Biophys J.}\ }\textbf {\bibinfo {volume} {10}},\
  \bibinfo {pages} {316.}}\BibitemShut {Stop}%
\bibitem [{\citenamefont {Handel}\ \emph {et~al.}(1976)\citenamefont {Handel},
  \citenamefont {Graf},\ and\ \citenamefont {Glass}}]{Handel1976}%
  \BibitemOpen
  \bibfield  {author} {\bibinfo {author} {\bibnamefont {Handel}, \bibfnamefont
  {E.~D.}}, \bibinfo {author} {\bibfnamefont {G.}~\bibnamefont {Graf}}, \ and\
  \bibinfo {author} {\bibfnamefont {J.}~\bibnamefont {Glass}}} (\bibinfo {year}
  {1976}),\ \href {\doibase http://dx.doi.org.10.1021/ja00424a073} {\bibfield
  {journal} {\bibinfo  {journal} {J. Am. Chem. Soc.}\ }\textbf {\bibinfo
  {volume} {98}},\ \bibinfo {pages} {2360.}}\BibitemShut {Stop}%
\bibitem [{\citenamefont {Handel}\ \emph {et~al.}(1980)\citenamefont {Handel},
  \citenamefont {Graf},\ and\ \citenamefont {Glass}}]{Handel1980}%
  \BibitemOpen
  \bibfield  {author} {\bibinfo {author} {\bibnamefont {Handel}, \bibfnamefont
  {E.~D.}}, \bibinfo {author} {\bibfnamefont {G.}~\bibnamefont {Graf}}, \ and\
  \bibinfo {author} {\bibfnamefont {J.}~\bibnamefont {Glass}}} (\bibinfo {year}
  {1980}),\ \href {\doibase http://dx.doi.10.org.1016/S0006-3495(80)85010-7}
  {\bibfield  {journal} {\bibinfo  {journal} {Biophys J.}\ }\textbf {\bibinfo
  {volume} {32}},\ \bibinfo {pages} {697.}}\BibitemShut {Stop}%
\bibitem [{\citenamefont {Hanneke}\ \emph {et~al.}(2008)\citenamefont
  {Hanneke}, \citenamefont {Fogwell},\ and\ \citenamefont
  {Gabrielse}}]{Hanneke2008tm}%
  \BibitemOpen
  \bibfield  {author} {\bibinfo {author} {\bibnamefont {Hanneke}, \bibfnamefont
  {D.}}, \bibinfo {author} {\bibfnamefont {S.}~\bibnamefont {Fogwell}}, \ and\
  \bibinfo {author} {\bibfnamefont {G.}~\bibnamefont {Gabrielse}}} (\bibinfo
  {year} {2008}),\ \href {\doibase 10.1103/PhysRevLett.100.120801} {\bibfield
  {journal} {\bibinfo  {journal} {Phys. Rev. Lett.}\ }\textbf {\bibinfo
  {volume} {100}},\ \bibinfo {pages} {120801}}\BibitemShut {NoStop}%
\bibitem [{\citenamefont {Hansen}\ and\ \citenamefont
  {Ingerslev-Jensen}(1983)}]{Hansen1983}%
  \BibitemOpen
  \bibfield  {author} {\bibinfo {author} {\bibnamefont {Hansen}, \bibfnamefont
  {H.}}, \ and\ \bibinfo {author} {\bibfnamefont {U.}~\bibnamefont
  {Ingerslev-Jensen}}} (\bibinfo {year} {1983}),\ \href {\doibase
  10.1088/0022-3727/16/7/026} {\bibfield  {journal} {\bibinfo  {journal} {J. of
  Phys. D: App. Phys.}\ }\textbf {\bibinfo {volume} {16}},\ \bibinfo {pages}
  {1353}}\BibitemShut {NoStop}%
\bibitem [{\citenamefont {Harpen}(2004)}]{Harpen2003zz}%
  \BibitemOpen
  \bibfield  {author} {\bibinfo {author} {\bibnamefont {Harpen}, \bibfnamefont
  {M.~D.}}} (\bibinfo {year} {2004}),\ \href {\doibase 10.1118/1.1630494}
  {\bibfield  {journal} {\bibinfo  {journal} {Med. Phys.}\ }\textbf {\bibinfo
  {volume} {31}},\ \bibinfo {pages} {57}}\BibitemShut {NoStop}%
%%CITATION = MPHYA,31,57;%%
\bibitem [{\citenamefont {Hatcher}\ \emph {et~al.}(1958)\citenamefont
  {Hatcher}, \citenamefont {Millett},\ and\ \citenamefont
  {Brown}}]{Hatcher1958}%
  \BibitemOpen
  \bibfield  {author} {\bibinfo {author} {\bibnamefont {Hatcher}, \bibfnamefont
  {C.~R.}}, \bibinfo {author} {\bibfnamefont {W.}~\bibnamefont {Millett}}, \
  and\ \bibinfo {author} {\bibfnamefont {L.}~\bibnamefont {Brown}}} (\bibinfo
  {year} {1958}),\ \href {\doibase https://doi.org/10.1103/PhysRev.111.12}
  {\bibfield  {journal} {\bibinfo  {journal} {Phys. Rev.}\ }\textbf {\bibinfo
  {volume} {111}},\ \bibinfo {pages} {12}}\BibitemShut {NoStop}%
\bibitem [{\citenamefont {Hawari}\ \emph {et~al.}(2011)\citenamefont {Hawari},
  \citenamefont {Gidley}, \citenamefont {Moxom}, \citenamefont {Hathaway},\
  and\ \citenamefont {Mukherjee}}]{Hawari2011}%
  \BibitemOpen
  \bibfield  {author} {\bibinfo {author} {\bibnamefont {Hawari}, \bibfnamefont
  {A.}}, \bibinfo {author} {\bibfnamefont {D.}~\bibnamefont {Gidley}}, \bibinfo
  {author} {\bibfnamefont {J.}~\bibnamefont {Moxom}}, \bibinfo {author}
  {\bibfnamefont {A.}~\bibnamefont {Hathaway}}, \ and\ \bibinfo {author}
  {\bibfnamefont {S.}~\bibnamefont {Mukherjee}}} (\bibinfo {year} {2011}),\
  \href {\doibase 10.1088/1742-6596/262/1/012024} {\bibfield  {journal}
  {\bibinfo  {journal} {J. of Phys.: Conf. Ser.}\ }\textbf {\bibinfo {volume}
  {262}},\ \bibinfo {pages} {012024}}\BibitemShut {NoStop}%
\bibitem [{\citenamefont {He}\ \emph {et~al.}(2007)\citenamefont {He},
  \citenamefont {Ohdaira}, \citenamefont {Oshima}, \citenamefont {Muramatsu},
  \citenamefont {Kinomura}, \citenamefont {Suzuki}, \citenamefont {Oka},\ and\
  \citenamefont {Kobayashi}}]{He2007}%
  \BibitemOpen
  \bibfield  {author} {\bibinfo {author} {\bibnamefont {He}, \bibfnamefont
  {C.}}, \bibinfo {author} {\bibfnamefont {T.}~\bibnamefont {Ohdaira}},
  \bibinfo {author} {\bibfnamefont {N.}~\bibnamefont {Oshima}}, \bibinfo
  {author} {\bibfnamefont {M.}~\bibnamefont {Muramatsu}}, \bibinfo {author}
  {\bibfnamefont {A.}~\bibnamefont {Kinomura}}, \bibinfo {author}
  {\bibfnamefont {R.}~\bibnamefont {Suzuki}}, \bibinfo {author} {\bibfnamefont
  {T.}~\bibnamefont {Oka}}, \ and\ \bibinfo {author} {\bibfnamefont
  {Y.}~\bibnamefont {Kobayashi}}} (\bibinfo {year} {2007}),\ \href {\doibase
  10.1103/PhysRevB.75.195404} {\bibfield  {journal} {\bibinfo  {journal} {Phys.
  Rev. B}\ }\textbf {\bibinfo {volume} {75}},\ \bibinfo {pages}
  {195404}}\BibitemShut {NoStop}%
\bibitem [{\citenamefont {Heiss}\ \emph {et~al.}(2018)\citenamefont {Heiss},
  \citenamefont {Wichmann}, \citenamefont {Rubbia},\ and\ \citenamefont
  {Crivelli}}]{Heiss2018jbl}%
  \BibitemOpen
  \bibfield  {author} {\bibinfo {author} {\bibnamefont {Heiss}, \bibfnamefont
  {M.}}, \bibinfo {author} {\bibfnamefont {G.}~\bibnamefont {Wichmann}},
  \bibinfo {author} {\bibfnamefont {A.}~\bibnamefont {Rubbia}}, \ and\ \bibinfo
  {author} {\bibfnamefont {P.}~\bibnamefont {Crivelli}}} (\bibinfo {year}
  {2018}),\ \href {\doibase 10.1088/1742-6596/1138/1/012007} {\bibfield
  {journal} {\bibinfo  {journal} {J. Phys. Conf. Ser.}\ }\textbf {\bibinfo
  {volume} {1138}}~(\bibinfo {number} {1}),\ \bibinfo {pages}
  {012007}}\BibitemShut {NoStop}%
%%CITATION = ARXIV:1805.05886;%%
\bibitem [{\citenamefont {Henry}\ \emph {et~al.}(2018)\citenamefont {Henry},
  \citenamefont {Ulaner},\ and\ \citenamefont {Lewis}}]{Henry2018}%
  \BibitemOpen
  \bibfield  {author} {\bibinfo {author} {\bibnamefont {Henry}, \bibfnamefont
  {K.~E.}}, \bibinfo {author} {\bibfnamefont {G.}~\bibnamefont {Ulaner}}, \
  and\ \bibinfo {author} {\bibfnamefont {J.}~\bibnamefont {Lewis}}} (\bibinfo
  {year} {2018}),\ \href {\doibase doi.org/10.1016/j.cpet.2018.02.010}
  {\bibfield  {journal} {\bibinfo  {journal} {PET Clin.}\ }\textbf {\bibinfo
  {volume} {13}},\ \bibinfo {pages} {423}}\BibitemShut {NoStop}%
\bibitem [{\citenamefont {Hernandez}\ \emph {et~al.}(2014)\citenamefont
  {Hernandez}, \citenamefont {Valdovinos}, \citenamefont {Yang}, \citenamefont
  {Chakravarty}, \citenamefont {Hong}, \citenamefont {Barnhart},\ and\
  \citenamefont {Cai}}]{Hernandez2014}%
  \BibitemOpen
  \bibfield  {author} {\bibinfo {author} {\bibnamefont {Hernandez},
  \bibfnamefont {R.}}, \bibinfo {author} {\bibfnamefont {H.}~\bibnamefont
  {Valdovinos}}, \bibinfo {author} {\bibfnamefont {Y.}~\bibnamefont {Yang}},
  \bibinfo {author} {\bibfnamefont {R.}~\bibnamefont {Chakravarty}}, \bibinfo
  {author} {\bibfnamefont {H.}~\bibnamefont {Hong}}, \bibinfo {author}
  {\bibfnamefont {T.}~\bibnamefont {Barnhart}}, \ and\ \bibinfo {author}
  {\bibfnamefont {W.}~\bibnamefont {Cai}}} (\bibinfo {year} {2014}),\ \href
  {\doibase 10.1021/mp500343j} {\bibfield  {journal} {\bibinfo  {journal} {Mol.
  Pharmaceutics}\ }\textbf {\bibinfo {volume} {11}},\ \bibinfo {pages}
  {2954}}\BibitemShut {NoStop}%
\bibitem [{\citenamefont {Hiesmayr}\ and\ \citenamefont
  {Moskal}(2017)}]{Hiesmayr2017xgx}%
  \BibitemOpen
  \bibfield  {author} {\bibinfo {author} {\bibnamefont {Hiesmayr},
  \bibfnamefont {B.~C.}}, \ and\ \bibinfo {author} {\bibfnamefont
  {P.}~\bibnamefont {Moskal}}} (\bibinfo {year} {2017}),\ \href {\doibase
  10.1038/s41598-017-15356-y} {\bibfield  {journal} {\bibinfo  {journal} {Sci.
  Rep.}\ }\textbf {\bibinfo {volume} {7}}~(\bibinfo {number} {1}),\ \bibinfo
  {pages} {15349}}\BibitemShut {NoStop}%
%%CITATION = ARXIV:1706.06505;%%
\bibitem [{\citenamefont {Hiesmayr}\ and\ \citenamefont
  {Moskal}(2019)}]{Hiesmayr2018rcm}%
  \BibitemOpen
  \bibfield  {author} {\bibinfo {author} {\bibnamefont {Hiesmayr},
  \bibfnamefont {B.~C.}}, \ and\ \bibinfo {author} {\bibfnamefont
  {P.}~\bibnamefont {Moskal}}} (\bibinfo {year} {2019}),\ \href {\doibase
  10.1038/s41598-019-44570-z} {\bibfield  {journal} {\bibinfo  {journal} {Sci.
  Rep.}\ }\textbf {\bibinfo {volume} {9}}~(\bibinfo {number} {1}),\ \bibinfo
  {pages} {8166}}\BibitemShut {NoStop}%
\bibitem [{\citenamefont {Hoang}\ \emph {et~al.}(1997)\citenamefont {Hoang},
  \citenamefont {Labelle},\ and\ \citenamefont {Zebarjad}}]{Hoang1997ki}%
  \BibitemOpen
  \bibfield  {author} {\bibinfo {author} {\bibnamefont {Hoang}, \bibfnamefont
  {A.~H.}}, \bibinfo {author} {\bibfnamefont {P.}~\bibnamefont {Labelle}}, \
  and\ \bibinfo {author} {\bibfnamefont {S.~M.}\ \bibnamefont {Zebarjad}}}
  (\bibinfo {year} {1997}),\ \href {\doibase 10.1103/PhysRevLett.79.3387}
  {\bibfield  {journal} {\bibinfo  {journal} {Phys. Rev. Lett.}\ }\textbf
  {\bibinfo {volume} {79}},\ \bibinfo {pages} {3387}}\BibitemShut {NoStop}%
\bibitem [{\citenamefont {Howell}\ \emph {et~al.}(1982)\citenamefont {Howell},
  \citenamefont {Alvarez},\ and\ \citenamefont {M.}}]{Howell1982}%
  \BibitemOpen
  \bibfield  {author} {\bibinfo {author} {\bibnamefont {Howell}, \bibfnamefont
  {R.~H.}}, \bibinfo {author} {\bibfnamefont {R.~A.}\ \bibnamefont {Alvarez}},
  \ and\ \bibinfo {author} {\bibfnamefont {S.}~\bibnamefont {M.}}} (\bibinfo
  {year} {1982}),\ \href {\doibase 10.1063/1.93215} {\bibfield  {journal}
  {\bibinfo  {journal} {Appl. Phys. Lett.}\ }\textbf {\bibinfo {volume} {40}},\
  \bibinfo {pages} {751}}\BibitemShut {NoStop}%
\bibitem [{\citenamefont {Hu}\ \emph {et~al.}(2022)\citenamefont {Hu} \emph
  {et~al.}}]{uEXPLORER-Chiny-2022}%
  \BibitemOpen
  \bibfield  {author} {\bibinfo {author} {\bibnamefont {Hu}, \bibfnamefont
  {H.}},  \emph {et~al.}} (\bibinfo {year} {2022}),\ \href {\doibase
  10.21203/rs.3.rs-1920965/v1} {\bibfield  {journal} {\bibinfo  {journal}
  {EJNMMI Physics}\ }10.21203/rs.3.rs-1920965/v1}\BibitemShut {NoStop}%
\bibitem [{\citenamefont {Hugenschmidt}(2016)}]{Hugenschmidt2016}%
  \BibitemOpen
  \bibfield  {author} {\bibinfo {author} {\bibnamefont {Hugenschmidt},
  \bibfnamefont {C.}}} (\bibinfo {year} {2016}),\ \href {\doibase
  10.1016/j.surfrep.2016.09.002} {\bibfield  {journal} {\bibinfo  {journal}
  {Surface Science Reports}\ }\textbf {\bibinfo {volume} {71}},\ \bibinfo
  {pages} {547}}\BibitemShut {NoStop}%
\bibitem [{\citenamefont {Hugenschmidt}\ \emph {et~al.}(2008)\citenamefont
  {Hugenschmidt}, \citenamefont {Löwe}, \citenamefont {Mayer}, \citenamefont
  {Piochacz}, \citenamefont {Pikart}, \citenamefont {Repper}, \citenamefont
  {Stadlbauer},\ and\ \citenamefont {Schreckenbach}}]{Hugenschmidt2008}%
  \BibitemOpen
  \bibfield  {author} {\bibinfo {author} {\bibnamefont {Hugenschmidt},
  \bibfnamefont {C.}}, \bibinfo {author} {\bibfnamefont {B.}~\bibnamefont
  {Löwe}}, \bibinfo {author} {\bibfnamefont {J.}~\bibnamefont {Mayer}},
  \bibinfo {author} {\bibfnamefont {C.}~\bibnamefont {Piochacz}}, \bibinfo
  {author} {\bibfnamefont {P.}~\bibnamefont {Pikart}}, \bibinfo {author}
  {\bibfnamefont {R.}~\bibnamefont {Repper}}, \bibinfo {author} {\bibfnamefont
  {M.}~\bibnamefont {Stadlbauer}}, \ and\ \bibinfo {author} {\bibfnamefont
  {K.}~\bibnamefont {Schreckenbach}}} (\bibinfo {year} {2008}),\ \href
  {\doibase 10.1016/j.nima.2008.05.038} {\bibfield  {journal} {\bibinfo
  {journal} {Nucl. Inst. and Meth. A}\ }\textbf {\bibinfo {volume} {593}},\
  \bibinfo {pages} {616}}\BibitemShut {NoStop}%
\bibitem [{\citenamefont {Humm}\ \emph {et~al.}(2003)\citenamefont {Humm} \emph
  {et~al.}}]{Humm2003}%
  \BibitemOpen
  \bibfield  {author} {\bibinfo {author} {\bibnamefont {Humm}, \bibfnamefont
  {J.~L.}},  \emph {et~al.}} (\bibinfo {year} {2003}),\ \href@noop {}
  {\bibfield  {journal} {\bibinfo  {journal} {Eur. J. Nucl. Med. Mol. Imaging}\
  }\textbf {\bibinfo {volume} {30}},\ \bibinfo {pages} {1574}}\BibitemShut
  {NoStop}%
\bibitem [{\citenamefont {Hunt}\ \emph {et~al.}(2001)\citenamefont {Hunt},
  \citenamefont {Cassidy}, \citenamefont {Sterne}, \citenamefont {Cowan},
  \citenamefont {Howell}, \citenamefont {Lynn},\ and\ \citenamefont
  {Golovchenko}}]{Hunt2001}%
  \BibitemOpen
  \bibfield  {author} {\bibinfo {author} {\bibnamefont {Hunt}, \bibfnamefont
  {A.}}, \bibinfo {author} {\bibfnamefont {D.}~\bibnamefont {Cassidy}},
  \bibinfo {author} {\bibfnamefont {P.}~\bibnamefont {Sterne}}, \bibinfo
  {author} {\bibfnamefont {T.}~\bibnamefont {Cowan}}, \bibinfo {author}
  {\bibfnamefont {R.}~\bibnamefont {Howell}}, \bibinfo {author} {\bibfnamefont
  {K.}~\bibnamefont {Lynn}}, \ and\ \bibinfo {author} {\bibfnamefont
  {J.}~\bibnamefont {Golovchenko}}} (\bibinfo {year} {2001}),\ \href {\doibase
  10.1103/PhysRevLett.86.5612} {\bibfield  {journal} {\bibinfo  {journal}
  {Phys. Rev. Lett.}\ }\textbf {\bibinfo {volume} {86}},\ \bibinfo {pages}
  {5612}}\BibitemShut {NoStop}%
\bibitem [{\citenamefont {Ishida}\ \emph {et~al.}(2014)\citenamefont {Ishida},
  \citenamefont {Namba}, \citenamefont {Asai}, \citenamefont {Kobayashi},
  \citenamefont {Saito}, \citenamefont {Yoshida}, \citenamefont {Tanaka},\ and\
  \citenamefont {Yamamoto}}]{Ishida2013waa}%
  \BibitemOpen
  \bibfield  {author} {\bibinfo {author} {\bibnamefont {Ishida}, \bibfnamefont
  {A.}}, \bibinfo {author} {\bibfnamefont {T.}~\bibnamefont {Namba}}, \bibinfo
  {author} {\bibfnamefont {S.}~\bibnamefont {Asai}}, \bibinfo {author}
  {\bibfnamefont {T.}~\bibnamefont {Kobayashi}}, \bibinfo {author}
  {\bibfnamefont {H.}~\bibnamefont {Saito}}, \bibinfo {author} {\bibfnamefont
  {M.}~\bibnamefont {Yoshida}}, \bibinfo {author} {\bibfnamefont
  {K.}~\bibnamefont {Tanaka}}, \ and\ \bibinfo {author} {\bibfnamefont
  {A.}~\bibnamefont {Yamamoto}}} (\bibinfo {year} {2014}),\ \href {\doibase
  10.1016/j.physletb.2014.05.083} {\bibfield  {journal} {\bibinfo  {journal}
  {Phys. Lett.}\ }\textbf {\bibinfo {volume} {B734}},\ \bibinfo {pages}
  {338}}\BibitemShut {NoStop}%
\bibitem [{\citenamefont {Ishida}\ \emph {et~al.}(2012)\citenamefont {Ishida}
  \emph {et~al.}}]{Ishida2011ds}%
  \BibitemOpen
  \bibfield  {author} {\bibinfo {author} {\bibnamefont {Ishida}, \bibfnamefont
  {A.}},  \emph {et~al.}} (\bibinfo {year} {2012}),\ \href {\doibase
  10.1007/s10751-011-0455-9} {\bibfield  {journal} {\bibinfo  {journal}
  {Hyperfine Interact.}\ }\textbf {\bibinfo {volume} {212}}~(\bibinfo {number}
  {1-3}),\ \bibinfo {pages} {133}}\BibitemShut {NoStop}%
\bibitem [{\citenamefont {Ito}\ \emph {et~al.}(2005)\citenamefont {Ito},
  \citenamefont {Yu}, \citenamefont {Sato}, \citenamefont {Hirata},\ and\
  \citenamefont {Kobayashi}}]{Ito2005}%
  \BibitemOpen
  \bibfield  {author} {\bibinfo {author} {\bibnamefont {Ito}, \bibfnamefont
  {K.}}, \bibinfo {author} {\bibfnamefont {R.}~\bibnamefont {Yu}}, \bibinfo
  {author} {\bibfnamefont {K.}~\bibnamefont {Sato}}, \bibinfo {author}
  {\bibfnamefont {K.}~\bibnamefont {Hirata}}, \ and\ \bibinfo {author}
  {\bibfnamefont {Y.}~\bibnamefont {Kobayashi}}} (\bibinfo {year} {2005}),\
  \href {\doibase 10.1063/1.2125121} {\bibfield  {journal} {\bibinfo  {journal}
  {J. of Appl. Phys.}\ }\textbf {\bibinfo {volume} {98}},\ \bibinfo {pages}
  {094307}}\BibitemShut {NoStop}%
\bibitem [{\citenamefont {Jasinska}\ \emph {et~al.}(2016)\citenamefont
  {Jasinska} \emph {et~al.}}]{Jasinska2016}%
  \BibitemOpen
  \bibfield  {author} {\bibinfo {author} {\bibnamefont {Jasinska},
  \bibfnamefont {B.}},  \emph {et~al.}} (\bibinfo {year} {2016}),\ \href@noop
  {} {\bibfield  {journal} {\bibinfo  {journal} {Acta Phys. Pol. B}\ }\textbf
  {\bibinfo {volume} {47}},\ \bibinfo {pages} {453}}\BibitemShut {NoStop}%
\bibitem [{\citenamefont {Jasinska}\ \emph {et~al.}(2017)\citenamefont
  {Jasinska} \emph {et~al.}}]{Jasinska2017}%
  \BibitemOpen
  \bibfield  {author} {\bibinfo {author} {\bibnamefont {Jasinska},
  \bibfnamefont {B.}},  \emph {et~al.}} (\bibinfo {year} {2017}),\ \href@noop
  {} {\bibfield  {journal} {\bibinfo  {journal} {Acta Phys. Pol.B}\ }\textbf
  {\bibinfo {volume} {48}},\ \bibinfo {pages} {1737}}\BibitemShut {NoStop}%
\bibitem [{\citenamefont {Jean}(2003)}]{Jean2003}%
  \BibitemOpen
  \bibfield  {author} {\bibinfo {author} {\bibnamefont {Jean}, \bibfnamefont
  {Y., M.~P. S.~D.}}} (\bibinfo {year} {2003}),\ \href@noop {} {\emph {\bibinfo
  {title} {{Principles And Applications Of Positron And Positronium
  Chemistry}}}}\ (\bibinfo  {publisher} {Jean,Y., Mallon, P.E.,Schrader, D.M.,
  World Scientific, New Jersey, London, Singapore, Hong Kong})\BibitemShut
  {NoStop}%
\bibitem [{\citenamefont {Jean}\ and\ \citenamefont {Ache}(1977)}]{Jean1977}%
  \BibitemOpen
  \bibfield  {author} {\bibinfo {author} {\bibnamefont {Jean}, \bibfnamefont
  {Y.~C.}}, \ and\ \bibinfo {author} {\bibfnamefont {H.}~\bibnamefont {Ache}}}
  (\bibinfo {year} {1977}),\ \href {\doibase
  https://doi.org/10.1021/ja00447a056} {\bibfield  {journal} {\bibinfo
  {journal} {J. Am. Chem. Soc.}\ }\textbf {\bibinfo {volume} {99}},\ \bibinfo
  {pages} {1623.}}\BibitemShut {Stop}%
\bibitem [{\citenamefont {Jean}\ \emph {et~al.}(2007)\citenamefont {Jean},
  \citenamefont {Chen}, \citenamefont {Liu},\ and\ \citenamefont
  {Gadzin}}]{Jean2007}%
  \BibitemOpen
  \bibfield  {author} {\bibinfo {author} {\bibnamefont {Jean}, \bibfnamefont
  {Y.~C.}}, \bibinfo {author} {\bibfnamefont {H.}~\bibnamefont {Chen}},
  \bibinfo {author} {\bibfnamefont {G.}~\bibnamefont {Liu}}, \ and\ \bibinfo
  {author} {\bibfnamefont {J.}~\bibnamefont {Gadzin}}} (\bibinfo {year}
  {2007}),\ \href {\doibase https://doi.org/10.1016/j.radphyschem.2006.03.008}
  {\bibfield  {journal} {\bibinfo  {journal} {Rad. Phys. Chem.}\ }\textbf
  {\bibinfo {volume} {76}},\ \bibinfo {pages} {70}}\BibitemShut {NoStop}%
\bibitem [{\citenamefont {Jean}\ and\ \citenamefont
  {Hancock}(1982)}]{Jean1982}%
  \BibitemOpen
  \bibfield  {author} {\bibinfo {author} {\bibnamefont {Jean}, \bibfnamefont
  {Y.~C.}}, \ and\ \bibinfo {author} {\bibfnamefont {A.}~\bibnamefont
  {Hancock}}} (\bibinfo {year} {1982}),\ \href {\doibase
  http://dx.doi.org.10.1063/1.443743} {\bibfield  {journal} {\bibinfo
  {journal} {J. Chem. Phys.}\ }\textbf {\bibinfo {volume} {77}},\ \bibinfo
  {pages} {5836}}\BibitemShut {NoStop}%
\bibitem [{\citenamefont {Jean}\ \emph {et~al.}(2006)\citenamefont {Jean},
  \citenamefont {Li}, \citenamefont {Liu}, \citenamefont {Chen}, \citenamefont
  {Zhang},\ and\ \citenamefont {Gadzia}}]{Jean2006}%
  \BibitemOpen
  \bibfield  {author} {\bibinfo {author} {\bibnamefont {Jean}, \bibfnamefont
  {Y.~C.}}, \bibinfo {author} {\bibfnamefont {Y.}~\bibnamefont {Li}}, \bibinfo
  {author} {\bibfnamefont {G.}~\bibnamefont {Liu}}, \bibinfo {author}
  {\bibfnamefont {H.}~\bibnamefont {Chen}}, \bibinfo {author} {\bibfnamefont
  {J.}~\bibnamefont {Zhang}}, \ and\ \bibinfo {author} {\bibfnamefont
  {J.}~\bibnamefont {Gadzia}}} (\bibinfo {year} {2006}),\ \href {\doibase
  https://doi.org/10.1016/j.apsusc.2005.08.101} {\bibfield  {journal} {\bibinfo
   {journal} {App. Surf. Sci}\ }\textbf {\bibinfo {volume} {252}},\ \bibinfo
  {pages} {3166}}\BibitemShut {NoStop}%
\bibitem [{\citenamefont {Jensen}\ \emph {et~al.}(2022)\citenamefont {Jensen},
  \citenamefont {Nyemann}, \citenamefont {Muren}, \citenamefont {Julsgaard},
  \citenamefont {Balling},\ and\ \citenamefont {Turtos}}]{Jensen-Turtos2022}%
  \BibitemOpen
  \bibfield  {author} {\bibinfo {author} {\bibnamefont {Jensen}, \bibfnamefont
  {M.}}, \bibinfo {author} {\bibfnamefont {J.}~\bibnamefont {Nyemann}},
  \bibinfo {author} {\bibfnamefont {L.}~\bibnamefont {Muren}}, \bibinfo
  {author} {\bibfnamefont {B.}~\bibnamefont {Julsgaard}}, \bibinfo {author}
  {\bibfnamefont {P.}~\bibnamefont {Balling}}, \ and\ \bibinfo {author}
  {\bibfnamefont {R.}~\bibnamefont {Turtos}}} (\bibinfo {year} {2022}),\
  \href@noop {} {\bibfield  {journal} {\bibinfo  {journal} {Sci. Rep.}\
  }\textbf {\bibinfo {volume} {12}},\ \bibinfo {pages} {8301}}\BibitemShut
  {NoStop}%
\bibitem [{\citenamefont {Kakimoto}\ \emph {et~al.}(1987)\citenamefont
  {Kakimoto}, \citenamefont {Hyodo}, \citenamefont {Chiba}, \citenamefont
  {Akahane},\ and\ \citenamefont {Chang}}]{Kakimoto1987}%
  \BibitemOpen
  \bibfield  {author} {\bibinfo {author} {\bibnamefont {Kakimoto},
  \bibfnamefont {M.}}, \bibinfo {author} {\bibfnamefont {T.}~\bibnamefont
  {Hyodo}}, \bibinfo {author} {\bibfnamefont {T.}~\bibnamefont {Chiba}},
  \bibinfo {author} {\bibfnamefont {T.}~\bibnamefont {Akahane}}, \ and\
  \bibinfo {author} {\bibfnamefont {T.}~\bibnamefont {Chang}}} (\bibinfo {year}
  {1987}),\ \href {\doibase 10.1088/0022-3700/20/3/007} {\bibfield  {journal}
  {\bibinfo  {journal} {J. of Phys. B: Atom. and Mol. Phys.}\ }\textbf
  {\bibinfo {volume} {20}},\ \bibinfo {pages} {L107}}\BibitemShut {NoStop}%
\bibitem [{\citenamefont {Kami\'nska}\ \emph {et~al.}(2016)\citenamefont
  {Kami\'nska} \emph {et~al.}}]{Kaminska2016fsn}%
  \BibitemOpen
  \bibfield  {author} {\bibinfo {author} {\bibnamefont {Kami\'nska},
  \bibfnamefont {D.}},  \emph {et~al.}} (\bibinfo {year} {2016}),\ \href
  {\doibase 10.1140/epjc/s10052-016-4294-3} {\bibfield  {journal} {\bibinfo
  {journal} {Eur. Phys. J. C}\ }\textbf {\bibinfo {volume} {76}},\ \bibinfo
  {pages} {445}}\BibitemShut {NoStop}%
\bibitem [{\citenamefont {Karakatsanis}\ \emph {et~al.}(2022)\citenamefont
  {Karakatsanis} \emph {et~al.}}]{Karakatsanis2022}%
  \BibitemOpen
  \bibfield  {author} {\bibinfo {author} {\bibnamefont {Karakatsanis},
  \bibfnamefont {N.}},  \emph {et~al.}} (\bibinfo {year} {2022}),\ \href@noop
  {} {\bibfield  {journal} {\bibinfo  {journal} {Phys. Med. Biol.}\ }\textbf
  {\bibinfo {volume} {67}},\ \bibinfo {pages} {105010}}\BibitemShut {NoStop}%
\bibitem [{\citenamefont {Karimi}\ \emph {et~al.}(2020)\citenamefont {Karimi},
  \citenamefont {Leszczynski}, \citenamefont {Kolodziej}, \citenamefont
  {Kubicz}, \citenamefont {Przybylo},\ and\ \citenamefont
  {Stepien}}]{Karimi2020}%
  \BibitemOpen
  \bibfield  {author} {\bibinfo {author} {\bibnamefont {Karimi}, \bibfnamefont
  {H.}}, \bibinfo {author} {\bibfnamefont {B.}~\bibnamefont {Leszczynski}},
  \bibinfo {author} {\bibfnamefont {T.}~\bibnamefont {Kolodziej}}, \bibinfo
  {author} {\bibfnamefont {E.}~\bibnamefont {Kubicz}}, \bibinfo {author}
  {\bibfnamefont {M.}~\bibnamefont {Przybylo}}, \ and\ \bibinfo {author}
  {\bibfnamefont {E.}~\bibnamefont {Stepien}}} (\bibinfo {year} {2020}),\ \href
  {\doibase doi.org.10.1016/j.micron.2020.102917} {\bibfield  {journal}
  {\bibinfo  {journal} {Micron}\ }\textbf {\bibinfo {volume} {137}},\ \bibinfo
  {pages} {102917}}\BibitemShut {NoStop}%
\bibitem [{\citenamefont {Karp}\ \emph {et~al.}(2020)\citenamefont {Karp},
  \citenamefont {Viswanath}, \citenamefont {Geagan}, \citenamefont
  {Muehllehner}, \citenamefont {Pantel}, \citenamefont {Parma}, \citenamefont
  {Schmall}, \citenamefont {Werner},\ and\ \citenamefont
  {Daube-Witherspoon}}]{Karp2020}%
  \BibitemOpen
  \bibfield  {author} {\bibinfo {author} {\bibnamefont {Karp}, \bibfnamefont
  {J.}}, \bibinfo {author} {\bibfnamefont {V.}~\bibnamefont {Viswanath}},
  \bibinfo {author} {\bibfnamefont {M.}~\bibnamefont {Geagan}}, \bibinfo
  {author} {\bibfnamefont {G.}~\bibnamefont {Muehllehner}}, \bibinfo {author}
  {\bibfnamefont {A.}~\bibnamefont {Pantel}}, \bibinfo {author} {\bibfnamefont
  {A.}~\bibnamefont {Parma}, \bibfnamefont {M.J.~Perkins}}, \bibinfo {author}
  {\bibfnamefont {J.}~\bibnamefont {Schmall}}, \bibinfo {author} {\bibfnamefont
  {M.}~\bibnamefont {Werner}}, \ and\ \bibinfo {author} {\bibfnamefont
  {M.}~\bibnamefont {Daube-Witherspoon}}} (\bibinfo {year} {2020}),\ \href@noop
  {} {\bibfield  {journal} {\bibinfo  {journal} {J. Nucl. Med.}\ }\textbf
  {\bibinfo {volume} {61}},\ \bibinfo {pages} {136}}\BibitemShut {NoStop}%
\bibitem [{\citenamefont {Karshenboim}(2004)}]{Karshenboim2003vs}%
  \BibitemOpen
  \bibfield  {author} {\bibinfo {author} {\bibnamefont {Karshenboim},
  \bibfnamefont {S.~G.}}} (\bibinfo {year} {2004}),\ \href {\doibase
  10.1142/S0217751X04020142} {\bibfield  {journal} {\bibinfo  {journal} {Int.
  J. Mod. Phys.}\ }\textbf {\bibinfo {volume} {A19}},\ \bibinfo {pages}
  {3879}}\BibitemShut {NoStop}%
%%CITATION = HEP-PH/0310099;%%
\bibitem [{\citenamefont {Karshenboim}(2005)}]{Karshenboim2005iy}%
  \BibitemOpen
  \bibfield  {author} {\bibinfo {author} {\bibnamefont {Karshenboim},
  \bibfnamefont {S.~G.}}} (\bibinfo {year} {2005}),\ \href {\doibase
  10.1016/j.physrep.2005.08.008} {\bibfield  {journal} {\bibinfo  {journal}
  {Phys. Rept.}\ }\textbf {\bibinfo {volume} {422}},\ \bibinfo {pages}
  {1}}\BibitemShut {NoStop}%
\bibitem [{\citenamefont {Kataoka}\ \emph {et~al.}(2009)\citenamefont
  {Kataoka}, \citenamefont {Asai},\ and\ \citenamefont
  {Kobayashi}}]{Kataoka2008hj}%
  \BibitemOpen
  \bibfield  {author} {\bibinfo {author} {\bibnamefont {Kataoka}, \bibfnamefont
  {Y.}}, \bibinfo {author} {\bibfnamefont {S.}~\bibnamefont {Asai}}, \ and\
  \bibinfo {author} {\bibfnamefont {T.}~\bibnamefont {Kobayashi}}} (\bibinfo
  {year} {2009}),\ \href {\doibase 10.1016/j.physletb.2008.12.008} {\bibfield
  {journal} {\bibinfo  {journal} {Phys. Lett. B}\ }\textbf {\bibinfo {volume}
  {671}},\ \bibinfo {pages} {219}}\BibitemShut {NoStop}%
\bibitem [{\citenamefont {Kerr}\ and\ \citenamefont {Hogg}(1962)}]{Kerr1962}%
  \BibitemOpen
  \bibfield  {author} {\bibinfo {author} {\bibnamefont {Kerr}, \bibfnamefont
  {D.~P.}}, \ and\ \bibinfo {author} {\bibfnamefont {B.~G.}\ \bibnamefont
  {Hogg}}} (\bibinfo {year} {1962}),\ \href {\doibase
  https://doi.org/10.1063/1.1732838} {\bibfield  {journal} {\bibinfo  {journal}
  {J. Chem. Phys.}\ }\textbf {\bibinfo {volume} {36}},\ \bibinfo {pages}
  {2190.}}\BibitemShut {Stop}%
\bibitem [{\citenamefont {Kilburn}\ \emph {et~al.}(2006)\citenamefont
  {Kilburn}, \citenamefont {Townrow}, \citenamefont {Meunier}, \citenamefont
  {Richardson}, \citenamefont {Alam},\ and\ \citenamefont
  {Ubbink}}]{Kilburn2006}%
  \BibitemOpen
  \bibfield  {author} {\bibinfo {author} {\bibnamefont {Kilburn}, \bibfnamefont
  {D.}}, \bibinfo {author} {\bibfnamefont {S.}~\bibnamefont {Townrow}},
  \bibinfo {author} {\bibfnamefont {V.}~\bibnamefont {Meunier}}, \bibinfo
  {author} {\bibfnamefont {R.}~\bibnamefont {Richardson}}, \bibinfo {author}
  {\bibfnamefont {A.}~\bibnamefont {Alam}}, \ and\ \bibinfo {author}
  {\bibfnamefont {J.}~\bibnamefont {Ubbink}}} (\bibinfo {year} {2006}),\ \href
  {\doibase http://dx.doi.org/10.1038/nmat1681} {\bibfield  {journal} {\bibinfo
   {journal} {Nat. Mater.}\ }\textbf {\bibinfo {volume} {5}},\ \bibinfo {pages}
  {632.}}\BibitemShut {Stop}%
\bibitem [{\citenamefont {Kim}\ \emph {et~al.}(1986)\citenamefont {Kim},
  \citenamefont {Pratt}, \citenamefont {Seltzer},\ and\ \citenamefont
  {Berger}}]{Kim1986}%
  \BibitemOpen
  \bibfield  {author} {\bibinfo {author} {\bibnamefont {Kim}, \bibfnamefont
  {L.}}, \bibinfo {author} {\bibfnamefont {R.~H.}\ \bibnamefont {Pratt}},
  \bibinfo {author} {\bibfnamefont {S.~M.}\ \bibnamefont {Seltzer}}, \ and\
  \bibinfo {author} {\bibfnamefont {M.~J.}\ \bibnamefont {Berger}}} (\bibinfo
  {year} {1986}),\ \href {\doibase 10.1103/PhysRevA.33.3002} {\bibfield
  {journal} {\bibinfo  {journal} {Phys. Rev. A}\ }\textbf {\bibinfo {volume}
  {33}},\ \bibinfo {pages} {3002}}\BibitemShut {NoStop}%
\bibitem [{\citenamefont {Kinoshita}\ and\ \citenamefont
  {Lepage}(1990)}]{Kinoshita1990ai}%
  \BibitemOpen
  \bibfield  {author} {\bibinfo {author} {\bibnamefont {Kinoshita},
  \bibfnamefont {T.}}, \ and\ \bibinfo {author} {\bibfnamefont {G.~P.}\
  \bibnamefont {Lepage}}} (\bibinfo {year} {1990}),\ \href {\doibase
  10.1142/9789814503273_0004} {\bibfield  {journal} {\bibinfo  {journal} {Adv.
  Ser. Direct. High Energy Phys.}\ }\textbf {\bibinfo {volume} {7}},\ \bibinfo
  {pages} {81}}\BibitemShut {NoStop}%
%%CITATION = 00319,7,81;%%
\bibitem [{\citenamefont {Klein}\ and\ \citenamefont
  {Nishina}(1929)}]{Klein1929}%
  \BibitemOpen
  \bibfield  {author} {\bibinfo {author} {\bibnamefont {Klein}, \bibfnamefont
  {O.~J.}}, \ and\ \bibinfo {author} {\bibfnamefont {Y.}~\bibnamefont
  {Nishina}}} (\bibinfo {year} {1929}),\ \href@noop {} {\bibfield  {journal}
  {\bibinfo  {journal} {Z. Physik}\ }\textbf {\bibinfo {volume} {52}},\
  \bibinfo {pages} {853}}\BibitemShut {NoStop}%
\bibitem [{\citenamefont {Kotera}\ \emph {et~al.}(2005)\citenamefont {Kotera},
  \citenamefont {Saito},\ and\ \citenamefont {Yamanaka}}]{Kotera2005}%
  \BibitemOpen
  \bibfield  {author} {\bibinfo {author} {\bibnamefont {Kotera}, \bibfnamefont
  {K.}}, \bibinfo {author} {\bibfnamefont {T.}~\bibnamefont {Saito}}, \ and\
  \bibinfo {author} {\bibfnamefont {T.}~\bibnamefont {Yamanaka}}} (\bibinfo
  {year} {2005}),\ \href {\doibase
  https://doi.org/10.1016/j.physleta.2005.07.018} {\bibfield  {journal}
  {\bibinfo  {journal} {Phys. Lett. A}\ }\textbf {\bibinfo {volume} {345}},\
  \bibinfo {pages} {184.}}\BibitemShut {Stop}%
\bibitem [{\citenamefont {Krasnicky}\ \emph {et~al.}(2016)\citenamefont
  {Krasnicky}, \citenamefont {Caravita}, \citenamefont {Canali},\ and\
  \citenamefont {Testera}}]{Krasnicky2016}%
  \BibitemOpen
  \bibfield  {author} {\bibinfo {author} {\bibnamefont {Krasnicky},
  \bibfnamefont {D.}}, \bibinfo {author} {\bibfnamefont {R.}~\bibnamefont
  {Caravita}}, \bibinfo {author} {\bibfnamefont {C.}~\bibnamefont {Canali}}, \
  and\ \bibinfo {author} {\bibfnamefont {G.}~\bibnamefont {Testera}}} (\bibinfo
  {year} {2016}),\ \href {\doibase 10.1103/PhysRevA.94.022714} {\bibfield
  {journal} {\bibinfo  {journal} {Phys. Rev. A}\ }\textbf {\bibinfo {volume}
  {94}},\ \bibinfo {pages} {022714}}\BibitemShut {NoStop}%
\bibitem [{\citenamefont {Kr{\'o}licki}\ and\ \citenamefont
  {Kunikowska}(2021)}]{Krolicki2021}%
  \BibitemOpen
  \bibfield  {author} {\bibinfo {author} {\bibnamefont {Kr{\'o}licki},
  \bibfnamefont {L.}}, \ and\ \bibinfo {author} {\bibfnamefont
  {J.}~\bibnamefont {Kunikowska}}} (\bibinfo {year} {2021}),\ \href {\doibase
  https://doi.org/10.1515/bams-2021-0169} {\bibfield  {journal} {\bibinfo
  {journal} {Bio-Algorithms and Med-Systems}\ }\textbf {\bibinfo {volume}
  {17}},\ \bibinfo {pages} {213}}\BibitemShut {NoStop}%
\bibitem [{\citenamefont {Kubica}\ and\ \citenamefont
  {Stewart}(1975)}]{Kubica1975}%
  \BibitemOpen
  \bibfield  {author} {\bibinfo {author} {\bibnamefont {Kubica}, \bibfnamefont
  {P.}}, \ and\ \bibinfo {author} {\bibfnamefont {A.}~\bibnamefont {Stewart}}}
  (\bibinfo {year} {1975}),\ \href {\doibase 10.1103/PhysRevLett.34.852}
  {\bibfield  {journal} {\bibinfo  {journal} {Phys. Rev. Lett.}\ }\textbf
  {\bibinfo {volume} {34}},\ \bibinfo {pages} {852}}\BibitemShut {NoStop}%
\bibitem [{\citenamefont {Kubicz}\ \emph {et~al.}(2015)\citenamefont {Kubicz}
  \emph {et~al.}}]{Kubicz2015}%
  \BibitemOpen
  \bibfield  {author} {\bibinfo {author} {\bibnamefont {Kubicz}, \bibfnamefont
  {E.}},  \emph {et~al.}} (\bibinfo {year} {2015}),\ \href {\doibase
  https://doi.org/10.1515/nuka-2015-0135} {\bibfield  {journal} {\bibinfo
  {journal} {Nukleoinika}\ }\textbf {\bibinfo {volume} {60}},\ \bibinfo {pages}
  {749}}\BibitemShut {NoStop}%
\bibitem [{\citenamefont {Kuhl}\ and\ \citenamefont
  {Edwards}(1963)}]{Kuhl1963}%
  \BibitemOpen
  \bibfield  {author} {\bibinfo {author} {\bibnamefont {Kuhl}, \bibfnamefont
  {D.~E.}}, \ and\ \bibinfo {author} {\bibfnamefont {R.}~\bibnamefont
  {Edwards}}} (\bibinfo {year} {1963}),\ \href {\doibase
  https://doi.org/10.1148/80.4.653} {\bibfield  {journal} {\bibinfo  {journal}
  {Radiology.}\ }\textbf {\bibinfo {volume} {80}},\ \bibinfo {pages}
  {653}}\BibitemShut {NoStop}%
\bibitem [{\citenamefont {Kwon}\ \emph {et~al.}(2021)\citenamefont {Kwon} \emph
  {et~al.}}]{Kwon2021}%
  \BibitemOpen
  \bibfield  {author} {\bibinfo {author} {\bibnamefont {Kwon}, \bibfnamefont
  {S.}},  \emph {et~al.}} (\bibinfo {year} {2021}),\ \href@noop {} {\bibfield
  {journal} {\bibinfo  {journal} {Nat. Photon.}\ }\textbf {\bibinfo {volume}
  {15}},\ \bibinfo {pages} {914}}\BibitemShut {NoStop}%
\bibitem [{\citenamefont {Labelle}(1992)}]{Labelle1992hd}%
  \BibitemOpen
  \bibfield  {author} {\bibinfo {author} {\bibnamefont {Labelle}, \bibfnamefont
  {P.}}} (\bibinfo {year} {1992}),\ \Eprint
  {http://arxiv.org/abs/hep-ph/9209266} {hep-ph/9209266} \BibitemShut {NoStop}%
\bibitem [{\citenamefont {Lartigau}\ \emph {et~al.}(1997)\citenamefont
  {Lartigau}, \citenamefont {Randrianarivelo}, \citenamefont {Avril},
  \citenamefont {Margulis}, \citenamefont {Spatz}, \citenamefont {Eschwège},\
  and\ \citenamefont {M.}}]{Lartigau1997}%
  \BibitemOpen
  \bibfield  {author} {\bibinfo {author} {\bibnamefont {Lartigau},
  \bibfnamefont {E.}}, \bibinfo {author} {\bibfnamefont {H.}~\bibnamefont
  {Randrianarivelo}}, \bibinfo {author} {\bibfnamefont {M.}~\bibnamefont
  {Avril}}, \bibinfo {author} {\bibfnamefont {A.}~\bibnamefont {Margulis}},
  \bibinfo {author} {\bibfnamefont {A.}~\bibnamefont {Spatz}}, \bibinfo
  {author} {\bibfnamefont {F.}~\bibnamefont {Eschwège}}, \ and\ \bibinfo
  {author} {\bibfnamefont {G.}~\bibnamefont {M.}}} (\bibinfo {year} {1997}),\
  \href {\doibase 10.1097/00008390-199710000-00006} {\bibfield  {journal}
  {\bibinfo  {journal} {Mel. Res.}\ }\textbf {\bibinfo {volume} {7}},\ \bibinfo
  {pages} {400}}\BibitemShut {NoStop}%
\bibitem [{\citenamefont {Lawrentschuk}\ \emph {et~al.}(2005)\citenamefont
  {Lawrentschuk}, \citenamefont {Poon}, \citenamefont {Foo}, \citenamefont
  {Putra}, \citenamefont {Murone}, \citenamefont {Davis}, \citenamefont
  {Bolton},\ and\ \citenamefont {Scott}}]{Lawrentschuk2005}%
  \BibitemOpen
  \bibfield  {author} {\bibinfo {author} {\bibnamefont {Lawrentschuk},
  \bibfnamefont {N.}}, \bibinfo {author} {\bibfnamefont {A.~M.}\ \bibnamefont
  {Poon}}, \bibinfo {author} {\bibfnamefont {S.}~\bibnamefont {Foo}}, \bibinfo
  {author} {\bibfnamefont {L.}~\bibnamefont {Putra}}, \bibinfo {author}
  {\bibfnamefont {C.}~\bibnamefont {Murone}}, \bibinfo {author} {\bibfnamefont
  {I.}~\bibnamefont {Davis}}, \bibinfo {author} {\bibfnamefont
  {D.}~\bibnamefont {Bolton}}, \ and\ \bibinfo {author} {\bibfnamefont
  {A.}~\bibnamefont {Scott}}} (\bibinfo {year} {2005}),\ \href {\doibase
  10.1111/j.1464-410X.2005.05681.x} {\bibfield  {journal} {\bibinfo  {journal}
  {IBJU Int.}\ }\textbf {\bibinfo {volume} {96}},\ \bibinfo {pages}
  {540}}\BibitemShut {NoStop}%
\bibitem [{\citenamefont {Lecoq}(2022)}]{Lecoq2022}%
  \BibitemOpen
  \bibfield  {author} {\bibinfo {author} {\bibnamefont {Lecoq}, \bibfnamefont
  {P.}}} (\bibinfo {year} {2022}),\ \href {\doibase
  10.1140/epjp/s13360-022-03159-8} {\bibfield  {journal} {\bibinfo  {journal}
  {Eur. Phys. J. Plus}\ }\textbf {\bibinfo {volume} {137}},\ \bibinfo {pages}
  {964}}\BibitemShut {NoStop}%
\bibitem [{\citenamefont {Lecoq}\ \emph {et~al.}(2022)\citenamefont {Lecoq}
  \emph {et~al.}}]{Lecoq2022IEEE}%
  \BibitemOpen
  \bibfield  {author} {\bibinfo {author} {\bibnamefont {Lecoq}, \bibfnamefont
  {P.}},  \emph {et~al.}} (\bibinfo {year} {2022}),\ \href@noop {} {\bibfield
  {journal} {\bibinfo  {journal} {IEEE Trans. Radiat. Plasma Med. Sci.}\
  }\textbf {\bibinfo {volume} {6}},\ \bibinfo {pages} {510}}\BibitemShut
  {NoStop}%
\bibitem [{\citenamefont {Lin}\ and\ \citenamefont {Alavi}(2019)}]{Alavi2019}%
  \BibitemOpen
  \bibfield  {author} {\bibinfo {author} {\bibnamefont {Lin}, \bibfnamefont
  {E.~C.}}, \ and\ \bibinfo {author} {\bibfnamefont {A.}~\bibnamefont {Alavi}}}
  (\bibinfo {year} {2019}),\ \href@noop {} {\emph {\bibinfo {title} {{PET and
  PET/CT: A Clinical Guide, Third Edition}}}},\ 3\ (\bibinfo  {publisher}
  {Thieme Medical Publishers},\ \bibinfo {address} {New York})\BibitemShut
  {NoStop}%
\bibitem [{\citenamefont {Liszkay}\ \emph {et~al.}(2012)\citenamefont
  {Liszkay}, \citenamefont {Guillemot}, \citenamefont {Corbel}, \citenamefont
  {Boilot}, \citenamefont {Gacoin}, \citenamefont {Barthel}, \citenamefont
  {Pérez}, \citenamefont {Barthe}, \citenamefont {Desgardin}, \citenamefont
  {Crivelli}, \citenamefont {Gendotti},\ and\ \citenamefont
  {Rubbia}}]{Liszkay2012}%
  \BibitemOpen
  \bibfield  {author} {\bibinfo {author} {\bibnamefont {Liszkay}, \bibfnamefont
  {L.}}, \bibinfo {author} {\bibfnamefont {F.}~\bibnamefont {Guillemot}},
  \bibinfo {author} {\bibfnamefont {C.}~\bibnamefont {Corbel}}, \bibinfo
  {author} {\bibfnamefont {J.}~\bibnamefont {Boilot}}, \bibinfo {author}
  {\bibfnamefont {T.}~\bibnamefont {Gacoin}}, \bibinfo {author} {\bibfnamefont
  {E.}~\bibnamefont {Barthel}}, \bibinfo {author} {\bibfnamefont
  {P.}~\bibnamefont {Pérez}}, \bibinfo {author} {\bibfnamefont
  {M.}~\bibnamefont {Barthe}}, \bibinfo {author} {\bibfnamefont
  {P.}~\bibnamefont {Desgardin}}, \bibinfo {author} {\bibfnamefont
  {P.}~\bibnamefont {Crivelli}}, \bibinfo {author} {\bibfnamefont
  {U.}~\bibnamefont {Gendotti}}, \ and\ \bibinfo {author} {\bibfnamefont
  {A.}~\bibnamefont {Rubbia}}} (\bibinfo {year} {2012}),\ \href {\doibase
  10.1088/1367-2630/14/6/065009} {\bibfield  {journal} {\bibinfo  {journal}
  {New J. of Phys.}\ }\textbf {\bibinfo {volume} {14}},\ \bibinfo {pages}
  {065009}}\BibitemShut {NoStop}%
\bibitem [{\citenamefont {Liu}\ \emph {et~al.}(2008)\citenamefont {Liu},
  \citenamefont {Chen}, \citenamefont {Chakka}, \citenamefont {Cheng},
  \citenamefont {Gadzia}, \citenamefont {Suzuki}, \citenamefont {Ohdaira},
  \citenamefont {Oshima},\ and\ \citenamefont {Jean}}]{Liu2008}%
  \BibitemOpen
  \bibfield  {author} {\bibinfo {author} {\bibnamefont {Liu}, \bibfnamefont
  {G.}}, \bibinfo {author} {\bibfnamefont {H.}~\bibnamefont {Chen}}, \bibinfo
  {author} {\bibfnamefont {L.}~\bibnamefont {Chakka}}, \bibinfo {author}
  {\bibfnamefont {M.-L.}\ \bibnamefont {Cheng}}, \bibinfo {author}
  {\bibfnamefont {J.}~\bibnamefont {Gadzia}}, \bibinfo {author} {\bibfnamefont
  {R.}~\bibnamefont {Suzuki}}, \bibinfo {author} {\bibfnamefont
  {T.}~\bibnamefont {Ohdaira}}, \bibinfo {author} {\bibfnamefont
  {N.}~\bibnamefont {Oshima}}, \ and\ \bibinfo {author} {\bibfnamefont
  {Y.}~\bibnamefont {Jean}}} (\bibinfo {year} {2008}),\ \href {\doibase
  https://doi.org/10.1016/j.apsusc.2008.05.241} {\bibfield  {journal} {\bibinfo
   {journal} {App. Surf. Sci.}\ }\textbf {\bibinfo {volume} {255}},\ \bibinfo
  {pages} {115}}\BibitemShut {NoStop}%
\bibitem [{\citenamefont {Liu}\ \emph {et~al.}(2007)\citenamefont {Liu},
  \citenamefont {Chen}, \citenamefont {Chakka}, \citenamefont {Gadzia},\ and\
  \citenamefont {Jean}}]{Liu2007}%
  \BibitemOpen
  \bibfield  {author} {\bibinfo {author} {\bibnamefont {Liu}, \bibfnamefont
  {G.}}, \bibinfo {author} {\bibfnamefont {H.}~\bibnamefont {Chen}}, \bibinfo
  {author} {\bibfnamefont {L.}~\bibnamefont {Chakka}}, \bibinfo {author}
  {\bibfnamefont {J.}~\bibnamefont {Gadzia}}, \ and\ \bibinfo {author}
  {\bibfnamefont {Y.}~\bibnamefont {Jean}}} (\bibinfo {year} {2007}),\ \href
  {\doibase https://doi.org/10.1002/pssc.200675736} {\bibfield  {journal}
  {\bibinfo  {journal} {Phys. Stat. Sol. C}\ }\textbf {\bibinfo {volume} {4}},\
  \bibinfo {pages} {3912}}\BibitemShut {NoStop}%
\bibitem [{\citenamefont {Lynn}\ and\ \citenamefont {Welch}(1980)}]{Lynn1980}%
  \BibitemOpen
  \bibfield  {author} {\bibinfo {author} {\bibnamefont {Lynn}, \bibfnamefont
  {K.}}, \ and\ \bibinfo {author} {\bibfnamefont {D.}~\bibnamefont {Welch}}}
  (\bibinfo {year} {1980}),\ \href {\doibase 10.1103/PhysRevB.22.99} {\bibfield
   {journal} {\bibinfo  {journal} {Phys. Rev. B}\ }\textbf {\bibinfo {volume}
  {22}},\ \bibinfo {pages} {99}}\BibitemShut {NoStop}%
\bibitem [{\citenamefont {Mariazzi}\ \emph
  {et~al.}(2010{\natexlab{a}})\citenamefont {Mariazzi}, \citenamefont
  {Bettotti},\ and\ \citenamefont {Brusa}}]{Mariazzi2010}%
  \BibitemOpen
  \bibfield  {author} {\bibinfo {author} {\bibnamefont {Mariazzi},
  \bibfnamefont {S.}}, \bibinfo {author} {\bibfnamefont {P.}~\bibnamefont
  {Bettotti}}, \ and\ \bibinfo {author} {\bibfnamefont {R.}~\bibnamefont
  {Brusa}}} (\bibinfo {year} {2010}{\natexlab{a}}),\ \href {\doibase
  10.1103/PhysRevLett.104.243401} {\bibfield  {journal} {\bibinfo  {journal}
  {Phys. Rev. Lett.}\ }\textbf {\bibinfo {volume} {104}},\ \bibinfo {pages}
  {243401}}\BibitemShut {NoStop}%
\bibitem [{\citenamefont {Mariazzi}\ \emph
  {et~al.}(2010{\natexlab{b}})\citenamefont {Mariazzi}, \citenamefont
  {Bettotti}, \citenamefont {Larcheri}, \citenamefont {Toniutti},\ and\
  \citenamefont {Brusa}}]{Mariazzi2010b}%
  \BibitemOpen
  \bibfield  {author} {\bibinfo {author} {\bibnamefont {Mariazzi},
  \bibfnamefont {S.}}, \bibinfo {author} {\bibfnamefont {P.}~\bibnamefont
  {Bettotti}}, \bibinfo {author} {\bibfnamefont {S.}~\bibnamefont {Larcheri}},
  \bibinfo {author} {\bibfnamefont {L.}~\bibnamefont {Toniutti}}, \ and\
  \bibinfo {author} {\bibfnamefont {R.}~\bibnamefont {Brusa}}} (\bibinfo {year}
  {2010}{\natexlab{b}}),\ \href {\doibase 10.1103/PhysRevB.81.235418}
  {\bibfield  {journal} {\bibinfo  {journal} {Phys. Rev. B}\ }\textbf {\bibinfo
  {volume} {81}},\ \bibinfo {pages} {235418}}\BibitemShut {NoStop}%
\bibitem [{\citenamefont {Mariazzi}\ \emph {et~al.}(2020)\citenamefont
  {Mariazzi}, \citenamefont {Caravita}, \citenamefont {Doser}, \citenamefont
  {Nebbia},\ and\ \citenamefont {Brusa}}]{Mariazzi2020bgc}%
  \BibitemOpen
  \bibfield  {author} {\bibinfo {author} {\bibnamefont {Mariazzi},
  \bibfnamefont {S.}}, \bibinfo {author} {\bibfnamefont {R.}~\bibnamefont
  {Caravita}}, \bibinfo {author} {\bibfnamefont {M.}~\bibnamefont {Doser}},
  \bibinfo {author} {\bibfnamefont {G.}~\bibnamefont {Nebbia}}, \ and\ \bibinfo
  {author} {\bibfnamefont {R.~S.}\ \bibnamefont {Brusa}}} (\bibinfo {year}
  {2020}),\ \href {\doibase 10.1140/epjd/e2020-100585-8} {\bibfield  {journal}
  {\bibinfo  {journal} {Eur. Phys. J.}\ }\textbf {\bibinfo {volume}
  {D74}}~(\bibinfo {number} {4}),\ \bibinfo {pages} {79}}\BibitemShut {NoStop}%
%%CITATION = EPHJA,D74,79;%%
\bibitem [{\citenamefont {Mariazzi}\ \emph {et~al.}(2022)\citenamefont
  {Mariazzi}, \citenamefont {Rieneacker}, \citenamefont {Magrin~Maffei},
  \citenamefont {Povolo}, \citenamefont {Sharma}, \citenamefont {Caravita},
  \citenamefont {Penasa}, \citenamefont {Bettotti}, \citenamefont {Doser},\
  and\ \citenamefont {Brusa}}]{Mariazzi2022}%
  \BibitemOpen
  \bibfield  {author} {\bibinfo {author} {\bibnamefont {Mariazzi},
  \bibfnamefont {S.}}, \bibinfo {author} {\bibfnamefont {B.}~\bibnamefont
  {Rieneacker}}, \bibinfo {author} {\bibfnamefont {R.}~\bibnamefont
  {Magrin~Maffei}}, \bibinfo {author} {\bibfnamefont {L.}~\bibnamefont
  {Povolo}}, \bibinfo {author} {\bibfnamefont {S.}~\bibnamefont {Sharma}},
  \bibinfo {author} {\bibfnamefont {R.}~\bibnamefont {Caravita}}, \bibinfo
  {author} {\bibfnamefont {L.}~\bibnamefont {Penasa}}, \bibinfo {author}
  {\bibfnamefont {P.}~\bibnamefont {Bettotti}}, \bibinfo {author}
  {\bibfnamefont {M.}~\bibnamefont {Doser}}, \ and\ \bibinfo {author}
  {\bibfnamefont {R.}~\bibnamefont {Brusa}}} (\bibinfo {year} {2022}),\ \href
  {\doibase 10.1103/PhysRevB.105.115422} {\bibfield  {journal} {\bibinfo
  {journal} {Phys. Rev. B}\ }\textbf {\bibinfo {volume} {105}},\ \bibinfo
  {pages} {115422}}\BibitemShut {NoStop}%
\bibitem [{\citenamefont {Mariazzi}\ \emph {et~al.}(2008)\citenamefont
  {Mariazzi}, \citenamefont {Salemi},\ and\ \citenamefont
  {Brusa}}]{Mariazzi2008}%
  \BibitemOpen
  \bibfield  {author} {\bibinfo {author} {\bibnamefont {Mariazzi},
  \bibfnamefont {S.}}, \bibinfo {author} {\bibfnamefont {A.}~\bibnamefont
  {Salemi}}, \ and\ \bibinfo {author} {\bibfnamefont {R.}~\bibnamefont
  {Brusa}}} (\bibinfo {year} {2008}),\ \href {\doibase
  10.1103/PhysRevB.78.085428} {\bibfield  {journal} {\bibinfo  {journal} {Phys.
  Rev. B}\ }\textbf {\bibinfo {volume} {78}},\ \bibinfo {pages}
  {085428}}\BibitemShut {NoStop}%
\bibitem [{\citenamefont {Mariazzi}\ \emph {et~al.}(2021)\citenamefont
  {Mariazzi} \emph {et~al.}}]{Mariazzi2021}%
  \BibitemOpen
  \bibfield  {author} {\bibinfo {author} {\bibnamefont {Mariazzi},
  \bibfnamefont {S.}},  \emph {et~al.}} (\bibinfo {year} {2021}),\ \href@noop
  {} {\bibfield  {journal} {\bibinfo  {journal} {J. Phys. B: At. Mol. Opt.
  Phys.}\ }\textbf {\bibinfo {volume} {54}},\ \bibinfo {pages}
  {085004}}\BibitemShut {NoStop}%
\bibitem [{\citenamefont {Marom}\ \emph {et~al.}(2001)\citenamefont {Marom},
  \citenamefont {Aloia}, \citenamefont {Hara}, \citenamefont {Herndon},
  \citenamefont {Harpole}, \citenamefont {Goodman},\ and\ \citenamefont
  {Patz}}]{Marom2001}%
  \BibitemOpen
  \bibfield  {author} {\bibinfo {author} {\bibnamefont {Marom}, \bibfnamefont
  {E.~M.}}, \bibinfo {author} {\bibfnamefont {M.}~\bibnamefont {Aloia},
  \bibfnamefont {T.A.and~Moore}}, \bibinfo {author} {\bibfnamefont
  {M.}~\bibnamefont {Hara}}, \bibinfo {author} {\bibfnamefont {J.~n.}\
  \bibnamefont {Herndon}}, \bibinfo {author} {\bibfnamefont {D.~J.}\
  \bibnamefont {Harpole}}, \bibinfo {author} {\bibfnamefont {P.}~\bibnamefont
  {Goodman}}, \ and\ \bibinfo {author} {\bibfnamefont {E.~J.}\ \bibnamefont
  {Patz}}} (\bibinfo {year} {2001}),\ \href {\doibase
  https://doi.org/10.1016/s0169-5002(00)00250-6} {\bibfield  {journal}
  {\bibinfo  {journal} {Lung Cancer}\ }\textbf {\bibinfo {volume} {33}},\
  \bibinfo {pages} {99.}}\BibitemShut {Stop}%
\bibitem [{\citenamefont {Matulewicz}(2021)}]{Matulewicz2021}%
  \BibitemOpen
  \bibfield  {author} {\bibinfo {author} {\bibnamefont {Matulewicz},
  \bibfnamefont {T.}}} (\bibinfo {year} {2021}),\ \href {\doibase
  https://doi.org/10.1515/bams-2021-0142} {\bibfield  {journal} {\bibinfo
  {journal} {Bio-Algorithms and Med-Systems}\ }\textbf {\bibinfo {volume}
  {17}},\ \bibinfo {pages} {235}}\BibitemShut {NoStop}%
\bibitem [{\citenamefont {McKeown}(2014)}]{McKeown2014}%
  \BibitemOpen
  \bibfield  {author} {\bibinfo {author} {\bibnamefont {McKeown}, \bibfnamefont
  {S.~R.}}} (\bibinfo {year} {2014}),\ \href {\doibase 10.1259/bjr.20130676}
  {\bibfield  {journal} {\bibinfo  {journal} {Br. J. Radiol.}\ }\textbf
  {\bibinfo {volume} {87}},\ \bibinfo {pages} {20130676}}\BibitemShut {NoStop}%
\bibitem [{\citenamefont {McNamara}\ \emph {et~al.}(2014)\citenamefont
  {McNamara}, \citenamefont {Toghyani}, \citenamefont {Gillam}, \citenamefont
  {Wu},\ and\ \citenamefont {Kuncic}}]{McNamara2014}%
  \BibitemOpen
  \bibfield  {author} {\bibinfo {author} {\bibnamefont {McNamara},
  \bibfnamefont {A.~L.}}, \bibinfo {author} {\bibfnamefont {M.}~\bibnamefont
  {Toghyani}}, \bibinfo {author} {\bibfnamefont {J.~E.}\ \bibnamefont
  {Gillam}}, \bibinfo {author} {\bibfnamefont {K.}~\bibnamefont {Wu}}, \ and\
  \bibinfo {author} {\bibfnamefont {Z.}~\bibnamefont {Kuncic}}} (\bibinfo
  {year} {2014}),\ \href@noop {} {\bibfield  {journal} {\bibinfo  {journal}
  {Phys. Med. Biol.}\ }\textbf {\bibinfo {volume} {59}},\ \bibinfo {pages}
  {7587}}\BibitemShut {NoStop}%
\bibitem [{\citenamefont {Mikolajczyk}\ \emph {et~al.}(2021)\citenamefont
  {Mikolajczyk} \emph {et~al.}}]{Mikolajczak2021}%
  \BibitemOpen
  \bibfield  {author} {\bibinfo {author} {\bibnamefont {Mikolajczyk},
  \bibfnamefont {R.}},  \emph {et~al.}} (\bibinfo {year} {2021}),\ \href@noop
  {} {\bibfield  {journal} {\bibinfo  {journal} {EJNMMI Radiopharm. Chem.}\
  }\textbf {\bibinfo {volume} {6}},\ \bibinfo {pages} {19}}\BibitemShut
  {NoStop}%
\bibitem [{\citenamefont {Mills}\ and\ \citenamefont
  {Gullikson}(1986)}]{Mills1986}%
  \BibitemOpen
  \bibfield  {author} {\bibinfo {author} {\bibnamefont {Mills}, \bibfnamefont
  {A.}}, \ and\ \bibinfo {author} {\bibfnamefont {E.}~\bibnamefont
  {Gullikson}}} (\bibinfo {year} {1986}),\ \href {\doibase 10.1063/1.97441}
  {\bibfield  {journal} {\bibinfo  {journal} {Appl. Phys. Lett.}\ }\textbf
  {\bibinfo {volume} {49}},\ \bibinfo {pages} {1121}}\BibitemShut {NoStop}%
\bibitem [{\citenamefont {Mills}\ and\ \citenamefont
  {Leventhal}(2002)}]{Mills2002}%
  \BibitemOpen
  \bibfield  {author} {\bibinfo {author} {\bibnamefont {Mills}, \bibfnamefont
  {A.}}, \ and\ \bibinfo {author} {\bibfnamefont {M.}~\bibnamefont
  {Leventhal}}} (\bibinfo {year} {2002}),\ \href {\doibase
  10.1016/S0168-583X(02)00789-9} {\bibfield  {journal} {\bibinfo  {journal}
  {Nucl. Inst. and Meth. B}\ }\textbf {\bibinfo {volume} {192}},\ \bibinfo
  {pages} {102}}\BibitemShut {NoStop}%
\bibitem [{\citenamefont {Mills}\ \emph {et~al.}(1978)\citenamefont {Mills},
  \citenamefont {Platzman},\ and\ \citenamefont {Brown}}]{Mills1978}%
  \BibitemOpen
  \bibfield  {author} {\bibinfo {author} {\bibnamefont {Mills}, \bibfnamefont
  {A.}}, \bibinfo {author} {\bibfnamefont {P.}~\bibnamefont {Platzman}}, \ and\
  \bibinfo {author} {\bibfnamefont {B.}~\bibnamefont {Brown}}} (\bibinfo {year}
  {1978}),\ \href {\doibase 10.1103/PhysRevLett.41.1076} {\bibfield  {journal}
  {\bibinfo  {journal} {Phys. Rev. Lett.}\ }\textbf {\bibinfo {volume} {41}},\
  \bibinfo {pages} {1076}}\BibitemShut {NoStop}%
\bibitem [{\citenamefont {Mills}(1983)}]{Mills1983zzd}%
  \BibitemOpen
  \bibfield  {author} {\bibinfo {author} {\bibnamefont {Mills}, \bibfnamefont
  {A.~P.}}} (\bibinfo {year} {1983}),\ \href {\doibase 10.1103/PhysRevA.27.262}
  {\bibfield  {journal} {\bibinfo  {journal} {Phys. Rev.}\ }\textbf {\bibinfo
  {volume} {A27}},\ \bibinfo {pages} {262}}\BibitemShut {NoStop}%
\bibitem [{\citenamefont {Mills}(2019)}]{PhysRevA.100.063615}%
  \BibitemOpen
  \bibfield  {author} {\bibinfo {author} {\bibnamefont {Mills}, \bibfnamefont
  {A.~P.}}} (\bibinfo {year} {2019}),\ \href {\doibase
  10.1103/PhysRevA.100.063615} {\bibfield  {journal} {\bibinfo  {journal}
  {Phys. Rev. A}\ }\textbf {\bibinfo {volume} {100}},\ \bibinfo {pages}
  {063615}}\BibitemShut {NoStop}%
\bibitem [{\citenamefont {Mills~Jr.}\ \emph {et~al.}(1994)\citenamefont
  {Mills~Jr.}, \citenamefont {Voris~Jr.},\ and\ \citenamefont
  {Andrew}}]{Mills1994}%
  \BibitemOpen
  \bibfield  {author} {\bibinfo {author} {\bibnamefont {Mills~Jr.},
  \bibfnamefont {A.}}, \bibinfo {author} {\bibfnamefont {S.}~\bibnamefont
  {Voris~Jr.}}, \ and\ \bibinfo {author} {\bibfnamefont {T.}~\bibnamefont
  {Andrew}}} (\bibinfo {year} {1994}),\ \href {\doibase 10.1063/1.357550}
  {\bibfield  {journal} {\bibinfo  {journal} {J. of App. Phys.}\ }\textbf
  {\bibinfo {volume} {76}},\ \bibinfo {pages} {2556}}\BibitemShut {NoStop}%
\bibitem [{\citenamefont {Miyazaki}\ \emph {et~al.}(2015)\citenamefont
  {Miyazaki}, \citenamefont {Yamazaki}, \citenamefont {Suehara}, \citenamefont
  {Namba}, \citenamefont {Asai}, \citenamefont {Kobayashi}, \citenamefont
  {Saito}, \citenamefont {Tatematsu}, \citenamefont {Ogawa},\ and\
  \citenamefont {Idehara}}]{Miyazaki2014bla}%
  \BibitemOpen
  \bibfield  {author} {\bibinfo {author} {\bibnamefont {Miyazaki},
  \bibfnamefont {A.}}, \bibinfo {author} {\bibfnamefont {T.}~\bibnamefont
  {Yamazaki}}, \bibinfo {author} {\bibfnamefont {T.}~\bibnamefont {Suehara}},
  \bibinfo {author} {\bibfnamefont {T.}~\bibnamefont {Namba}}, \bibinfo
  {author} {\bibfnamefont {S.}~\bibnamefont {Asai}}, \bibinfo {author}
  {\bibfnamefont {T.}~\bibnamefont {Kobayashi}}, \bibinfo {author}
  {\bibfnamefont {H.}~\bibnamefont {Saito}}, \bibinfo {author} {\bibfnamefont
  {Y.}~\bibnamefont {Tatematsu}}, \bibinfo {author} {\bibfnamefont
  {I.}~\bibnamefont {Ogawa}}, \ and\ \bibinfo {author} {\bibfnamefont
  {T.}~\bibnamefont {Idehara}}} (\bibinfo {year} {2015}),\ \href {\doibase
  10.1093/ptep/ptu181} {\bibfield  {journal} {\bibinfo  {journal} {PTEP}\
  }\textbf {\bibinfo {volume} {2015}}~(\bibinfo {number} {1}),\ \bibinfo
  {pages} {011C01}}\BibitemShut {NoStop}%
\bibitem [{\citenamefont {Mogensen}(1974)}]{Mogensen1974}%
  \BibitemOpen
  \bibfield  {author} {\bibinfo {author} {\bibnamefont {Mogensen},
  \bibfnamefont {O.}}} (\bibinfo {year} {1974}),\ \href {\doibase
  10.1063/1.1681180} {\bibfield  {journal} {\bibinfo  {journal} {J. Chem.
  Phys.}\ }\textbf {\bibinfo {volume} {60}},\ \bibinfo {pages}
  {998}}\BibitemShut {NoStop}%
\bibitem [{\citenamefont {Mohorovicic}(1934)}]{Mohorovicic1934aa}%
  \BibitemOpen
  \bibfield  {author} {\bibinfo {author} {\bibnamefont {Mohorovicic},
  \bibfnamefont {S.}}} (\bibinfo {year} {1934}),\ \href@noop {} {\bibfield
  {journal} {\bibinfo  {journal} {Astron. Nachrichten}\ }\textbf {\bibinfo
  {volume} {253}},\ \bibinfo {pages} {93}}\BibitemShut {NoStop}%
\bibitem [{\citenamefont {Morel}\ \emph {et~al.}(2020)\citenamefont {Morel},
  \citenamefont {Yao}, \citenamefont {Clad\'e},\ and\ \citenamefont
  {Guellati-Kh\'elifa}}]{Morel2020dww}%
  \BibitemOpen
  \bibfield  {author} {\bibinfo {author} {\bibnamefont {Morel}, \bibfnamefont
  {L.}}, \bibinfo {author} {\bibfnamefont {Z.}~\bibnamefont {Yao}}, \bibinfo
  {author} {\bibfnamefont {P.}~\bibnamefont {Clad\'e}}, \ and\ \bibinfo
  {author} {\bibfnamefont {S.}~\bibnamefont {Guellati-Kh\'elifa}}} (\bibinfo
  {year} {2020}),\ \href@noop {} {\bibfield  {journal} {\bibinfo  {journal}
  {Nature}\ }\textbf {\bibinfo {volume} {588}}~(\bibinfo {number} {7836}),\
  \bibinfo {pages} {61}}\BibitemShut {NoStop}%
\bibitem [{\citenamefont {Moskal}(2018)}]{MoskalIEEE2018}%
  \BibitemOpen
  \bibfield  {author} {\bibinfo {author} {\bibnamefont {Moskal}, \bibfnamefont
  {P.}}} (\bibinfo {year} {2018}),\ \href {\doibase
  10.1109/NSSMIC.2018.8824622} {\bibinfo  {journal} {IEEE Xplore, 2018 IEEE
  Nuclear Science Symposium and Medical Imaging Conference}\ ,\ \bibinfo
  {pages} {1}}\BibitemShut {NoStop}%
\bibitem [{\citenamefont {Moskal}(2019)}]{MoskalIEEE2019}%
  \BibitemOpen
\bibfield  {journal} {  }\bibfield  {author} {\bibinfo {author} {\bibnamefont
  {Moskal}, \bibfnamefont {P.}}} (\bibinfo {year} {2019}),\ \href {\doibase
  10.1109/NSS/MIC42101.2019.9059856} {\bibinfo  {journal} {IEEE Xplore, 2019
  IEEE Nuclear Science Symposium and Medical Imaging Conference}\ ,\ \bibinfo
  {pages} {1}}\BibitemShut {NoStop}%
\bibitem [{\citenamefont {Moskal}\ \emph
  {et~al.}(2019{\natexlab{a}})\citenamefont {Moskal}, \citenamefont {Jasinska},
  \citenamefont {Stepien},\ and\ \citenamefont {Bass}}]{NRP2019}%
  \BibitemOpen
\bibfield  {journal} {  }\bibfield  {author} {\bibinfo {author} {\bibnamefont
  {Moskal}, \bibfnamefont {P.}}, \bibinfo {author} {\bibfnamefont
  {B.}~\bibnamefont {Jasinska}}, \bibinfo {author} {\bibfnamefont
  {E.}~\bibnamefont {Stepien}}, \ and\ \bibinfo {author} {\bibfnamefont
  {S.~D.}\ \bibnamefont {Bass}}} (\bibinfo {year} {2019}{\natexlab{a}}),\
  \href@noop {} {\bibfield  {journal} {\bibinfo  {journal} {Nat. Rev. Phys.}\
  }\textbf {\bibinfo {volume} {1}},\ \bibinfo {pages} {527}}\BibitemShut
  {NoStop}%
\bibitem [{\citenamefont {Moskal}\ \emph
  {et~al.}(2021{\natexlab{a}})\citenamefont {Moskal}, \citenamefont {Kubicz},
  \citenamefont {Grudzien}, \citenamefont {Czerwinski}, \citenamefont {Dulski},
  \citenamefont {Leszczynski}, \citenamefont {Niedzwiecki},\ and\ \citenamefont
  {Stepien}}]{Moskal2021}%
  \BibitemOpen
  \bibfield  {author} {\bibinfo {author} {\bibnamefont {Moskal}, \bibfnamefont
  {P.}}, \bibinfo {author} {\bibfnamefont {E.}~\bibnamefont {Kubicz}}, \bibinfo
  {author} {\bibfnamefont {G.}~\bibnamefont {Grudzien}}, \bibinfo {author}
  {\bibfnamefont {E.}~\bibnamefont {Czerwinski}}, \bibinfo {author}
  {\bibfnamefont {K.}~\bibnamefont {Dulski}}, \bibinfo {author} {\bibfnamefont
  {B.}~\bibnamefont {Leszczynski}}, \bibinfo {author} {\bibfnamefont
  {S.}~\bibnamefont {Niedzwiecki}}, \ and\ \bibinfo {author} {\bibfnamefont
  {E.}~\bibnamefont {Stepien}}} (\bibinfo {year} {2021}{\natexlab{a}}),\ \href
  {\doibase doi.org/10.1101/2021.08.05.455285} {\bibfield  {journal} {\bibinfo
  {journal} {bioRxiv}\ }\textbf {\bibinfo {volume} {08.05}},\ \bibinfo {pages}
  {455285}}\BibitemShut {NoStop}%
\bibitem [{\citenamefont {Moskal}\ and\ \citenamefont
  {Stepien}(2021)}]{MoskalStepien2021hypoxia}%
  \BibitemOpen
  \bibfield  {author} {\bibinfo {author} {\bibnamefont {Moskal}, \bibfnamefont
  {P.}}, \ and\ \bibinfo {author} {\bibfnamefont {E.}~\bibnamefont {Stepien}}}
  (\bibinfo {year} {2021}),\ \href {\doibase
  https://doi.org/10.1515/bams-2021-0189} {\bibfield  {journal} {\bibinfo
  {journal} {Bio-Algorithms and Med-Systems}\ }\textbf {\bibinfo {volume}
  {17}},\ \bibinfo {pages} {311}}\BibitemShut {NoStop}%
\bibitem [{\citenamefont {Moskal}\ and\ \citenamefont
  {Stepien}(2020)}]{Moskal2020petclin}%
  \BibitemOpen
  \bibfield  {author} {\bibinfo {author} {\bibnamefont {Moskal}, \bibfnamefont
  {P.}}, \ and\ \bibinfo {author} {\bibfnamefont {E.~L.}\ \bibnamefont
  {Stepien}}} (\bibinfo {year} {2020}),\ \href@noop {} {\bibfield  {journal}
  {\bibinfo  {journal} {PET Clin.}\ }\textbf {\bibinfo {volume} {15}},\
  \bibinfo {pages} {439}}\BibitemShut {NoStop}%
\bibitem [{\citenamefont {Moskal}\ and\ \citenamefont
  {Stepien}(2022)}]{MoskalStepien2022}%
  \BibitemOpen
  \bibfield  {author} {\bibinfo {author} {\bibnamefont {Moskal}, \bibfnamefont
  {P.}}, \ and\ \bibinfo {author} {\bibfnamefont {E.~L.}\ \bibnamefont
  {Stepien}}} (\bibinfo {year} {2022}),\ \href {\doibase
  10.3389/fphy.2022.969806} {\bibfield  {journal} {\bibinfo  {journal} {Front.
  Phys.}\ }\textbf {\bibinfo {volume} {10}},\ \bibinfo {pages}
  {969806}}\BibitemShut {NoStop}%
\bibitem [{\citenamefont {Moskal}\ \emph {et~al.}(2014)\citenamefont {Moskal}
  \emph {et~al.}}]{Moskal2014nim}%
  \BibitemOpen
  \bibfield  {author} {\bibinfo {author} {\bibnamefont {Moskal}, \bibfnamefont
  {P.}},  \emph {et~al.}} (\bibinfo {year} {2014}),\ \href@noop {} {\bibfield
  {journal} {\bibinfo  {journal} {Nucl. Instrum. Meth. A}\ }\textbf {\bibinfo
  {volume} {764}},\ \bibinfo {pages} {317}}\BibitemShut {NoStop}%
\bibitem [{\citenamefont {Moskal}\ \emph {et~al.}(2016)\citenamefont {Moskal}
  \emph {et~al.}}]{Moskal2016moj}%
  \BibitemOpen
  \bibfield  {author} {\bibinfo {author} {\bibnamefont {Moskal}, \bibfnamefont
  {P.}},  \emph {et~al.}} (\bibinfo {year} {2016}),\ \href@noop {} {\bibfield
  {journal} {\bibinfo  {journal} {Acta Phys. Polon. B}\ }\textbf {\bibinfo
  {volume} {47}},\ \bibinfo {pages} {509}}\BibitemShut {NoStop}%
\bibitem [{\citenamefont {Moskal}\ \emph {et~al.}(2018)\citenamefont {Moskal}
  \emph {et~al.}}]{Moskal2018pus}%
  \BibitemOpen
  \bibfield  {author} {\bibinfo {author} {\bibnamefont {Moskal}, \bibfnamefont
  {P.}},  \emph {et~al.}} (\bibinfo {year} {2018}),\ \href {\doibase
  10.1140/epjc/s10052-018-6461-1} {\bibfield  {journal} {\bibinfo  {journal}
  {Eur. Phys. J. C}\ }\textbf {\bibinfo {volume} {78}}~(\bibinfo {number}
  {11}),\ \bibinfo {pages} {970}}\BibitemShut {NoStop}%
\bibitem [{\citenamefont {Moskal}\ \emph
  {et~al.}(2019{\natexlab{b}})\citenamefont {Moskal} \emph
  {et~al.}}]{Moskal2019pmb}%
  \BibitemOpen
  \bibfield  {author} {\bibinfo {author} {\bibnamefont {Moskal}, \bibfnamefont
  {P.}},  \emph {et~al.}} (\bibinfo {year} {2019}{\natexlab{b}}),\ \href
  {\doibase 10.1088/1361-6560/aafe20} {\bibfield  {journal} {\bibinfo
  {journal} {Phys. Med. Biol.}\ }\textbf {\bibinfo {volume} {64}},\ \bibinfo
  {pages} {055017}}\BibitemShut {NoStop}%
\bibitem [{\citenamefont {Moskal}\ \emph {et~al.}(2020)\citenamefont {Moskal}
  \emph {et~al.}}]{Moskal2020ejnmmi}%
  \BibitemOpen
  \bibfield  {author} {\bibinfo {author} {\bibnamefont {Moskal}, \bibfnamefont
  {P.}},  \emph {et~al.}} (\bibinfo {year} {2020}),\ \href {\doibase
  10.1186/s40658-020-00307-w} {\bibfield  {journal} {\bibinfo  {journal}
  {EJNMMI Phys.}\ }\textbf {\bibinfo {volume} {7}},\ \bibinfo {pages}
  {44}}\BibitemShut {NoStop}%
\bibitem [{\citenamefont {Moskal}\ \emph
  {et~al.}(2021{\natexlab{b}})\citenamefont {Moskal} \emph
  {et~al.}}]{Moskal2021science}%
  \BibitemOpen
  \bibfield  {author} {\bibinfo {author} {\bibnamefont {Moskal}, \bibfnamefont
  {P.}},  \emph {et~al.}} (\bibinfo {year} {2021}{\natexlab{b}}),\ \href
  {\doibase 10.1126/sciadv.abh4394} {\bibfield  {journal} {\bibinfo  {journal}
  {Sci. Adv.}\ }\textbf {\bibinfo {volume} {7}},\ \bibinfo {pages}
  {eabh4394}}\BibitemShut {NoStop}%
\bibitem [{\citenamefont {Moskal}\ \emph
  {et~al.}(2021{\natexlab{c}})\citenamefont {Moskal} \emph
  {et~al.}}]{Moskal2021nature}%
  \BibitemOpen
  \bibfield  {author} {\bibinfo {author} {\bibnamefont {Moskal}, \bibfnamefont
  {P.}},  \emph {et~al.}} (\bibinfo {year} {2021}{\natexlab{c}}),\ \href
  {\doibase https://doi.org/10.1038/s41467-021-25905-9} {\bibfield  {journal}
  {\bibinfo  {journal} {Nat. Commun.}\ }\textbf {\bibinfo {volume} {12}},\
  \bibinfo {pages} {5658}}\BibitemShut {NoStop}%
\bibitem [{\citenamefont {Murphy}\ and\ \citenamefont
  {Surko}(1992)}]{Murphy1992}%
  \BibitemOpen
  \bibfield  {author} {\bibinfo {author} {\bibnamefont {Murphy}, \bibfnamefont
  {T.}}, \ and\ \bibinfo {author} {\bibfnamefont {C.}~\bibnamefont {Surko}}}
  (\bibinfo {year} {1992}),\ \href {\doibase 10.1103/PhysRevA.46.5696}
  {\bibfield  {journal} {\bibinfo  {journal} {Phys. Rev. A}\ }\textbf {\bibinfo
  {volume} {46}},\ \bibinfo {pages} {5696}}\BibitemShut {NoStop}%
\bibitem [{\citenamefont {Nagashima}(2014)}]{Nagashima2014}%
  \BibitemOpen
  \bibfield  {author} {\bibinfo {author} {\bibnamefont {Nagashima},
  \bibfnamefont {Y.}}} (\bibinfo {year} {2014}),\ \href@noop {} {\bibfield
  {journal} {\bibinfo  {journal} {Phys. Rep.}\ }\textbf {\bibinfo {volume}
  {545}},\ \bibinfo {pages} {95}}\BibitemShut {NoStop}%
\bibitem [{\citenamefont {Nagashima}\ \emph {et~al.}(1995)\citenamefont
  {Nagashima}, \citenamefont {Kakimoto}, \citenamefont {Hyodo}, \citenamefont
  {Fujiwara}, \citenamefont {Ichimura}, \citenamefont {Chang}, \citenamefont
  {Deng}, \citenamefont {Akahane}, \citenamefont {Chiba}, \citenamefont
  {Suzuki}, \citenamefont {McKee},\ and\ \citenamefont
  {Stewart}}]{Nagashima1995}%
  \BibitemOpen
  \bibfield  {author} {\bibinfo {author} {\bibnamefont {Nagashima},
  \bibfnamefont {Y.}}, \bibinfo {author} {\bibfnamefont {M.}~\bibnamefont
  {Kakimoto}}, \bibinfo {author} {\bibfnamefont {T.}~\bibnamefont {Hyodo}},
  \bibinfo {author} {\bibfnamefont {K.}~\bibnamefont {Fujiwara}}, \bibinfo
  {author} {\bibfnamefont {A.}~\bibnamefont {Ichimura}}, \bibinfo {author}
  {\bibfnamefont {T.}~\bibnamefont {Chang}}, \bibinfo {author} {\bibfnamefont
  {J.}~\bibnamefont {Deng}}, \bibinfo {author} {\bibfnamefont {T.}~\bibnamefont
  {Akahane}}, \bibinfo {author} {\bibfnamefont {T.}~\bibnamefont {Chiba}},
  \bibinfo {author} {\bibfnamefont {K.}~\bibnamefont {Suzuki}}, \bibinfo
  {author} {\bibfnamefont {B.}~\bibnamefont {McKee}}, \ and\ \bibinfo {author}
  {\bibfnamefont {A.}~\bibnamefont {Stewart}}} (\bibinfo {year} {1995}),\ \href
  {\doibase 10.1103/PhysRevA.52.258} {\bibfield  {journal} {\bibinfo  {journal}
  {Phys. Rev. A}\ }\textbf {\bibinfo {volume} {52}},\ \bibinfo {pages}
  {258}}\BibitemShut {NoStop}%
\bibitem [{\citenamefont {Nagashima}\ \emph {et~al.}(1998)\citenamefont
  {Nagashima}, \citenamefont {Morinaka}, \citenamefont {Kurihara},
  \citenamefont {Nagai}, \citenamefont {Hyodo}, \citenamefont {Shidara},\ and\
  \citenamefont {Nakahara}}]{Nagashima1998}%
  \BibitemOpen
  \bibfield  {author} {\bibinfo {author} {\bibnamefont {Nagashima},
  \bibfnamefont {Y.}}, \bibinfo {author} {\bibfnamefont {Y.}~\bibnamefont
  {Morinaka}}, \bibinfo {author} {\bibfnamefont {T.}~\bibnamefont {Kurihara}},
  \bibinfo {author} {\bibfnamefont {Y.}~\bibnamefont {Nagai}}, \bibinfo
  {author} {\bibfnamefont {T.}~\bibnamefont {Hyodo}}, \bibinfo {author}
  {\bibfnamefont {T.}~\bibnamefont {Shidara}}, \ and\ \bibinfo {author}
  {\bibfnamefont {K.}~\bibnamefont {Nakahara}}} (\bibinfo {year} {1998}),\
  \href {\doibase 10.1103/PhysRevB.58.12676} {\bibfield  {journal} {\bibinfo
  {journal} {Phys. Rev. B}\ }\textbf {\bibinfo {volume} {58}},\ \bibinfo
  {pages} {12676}}\BibitemShut {NoStop}%
\bibitem [{\citenamefont {Niedzwiecki}\ \emph {et~al.}(2017)\citenamefont
  {Niedzwiecki} \emph {et~al.}}]{Niedzwiecki2017}%
  \BibitemOpen
  \bibfield  {author} {\bibinfo {author} {\bibnamefont {Niedzwiecki},
  \bibfnamefont {S.}},  \emph {et~al.}} (\bibinfo {year} {2017}),\ \href@noop
  {} {\bibfield  {journal} {\bibinfo  {journal} {Acta Phys. Polon. B}\ }\textbf
  {\bibinfo {volume} {48}},\ \bibinfo {pages} {1567}}\BibitemShut {NoStop}%
\bibitem [{\citenamefont {Nielsen}\ \emph {et~al.}(1986)\citenamefont
  {Nielsen}, \citenamefont {Lynn},\ and\ \citenamefont {Chen}}]{Nielsen1986}%
  \BibitemOpen
  \bibfield  {author} {\bibinfo {author} {\bibnamefont {Nielsen}, \bibfnamefont
  {B.}}, \bibinfo {author} {\bibfnamefont {K.}~\bibnamefont {Lynn}}, \ and\
  \bibinfo {author} {\bibfnamefont {Y.}~\bibnamefont {Chen}}} (\bibinfo {year}
  {1986}),\ \href {\doibase 10.1103/PhysRevLett.57.1789} {\bibfield  {journal}
  {\bibinfo  {journal} {Phys. Rev. Lett.}\ }\textbf {\bibinfo {volume} {57}},\
  \bibinfo {pages} {1789}}\BibitemShut {NoStop}%
\bibitem [{\citenamefont {Nieminen}\ and\ \citenamefont
  {Oliva}(1980)}]{Nieminen1980}%
  \BibitemOpen
  \bibfield  {author} {\bibinfo {author} {\bibnamefont {Nieminen},
  \bibfnamefont {R.}}, \ and\ \bibinfo {author} {\bibfnamefont
  {J.}~\bibnamefont {Oliva}}} (\bibinfo {year} {1980}),\ \href {\doibase
  10.1103/PhysRevB.22.2226} {\bibfield  {journal} {\bibinfo  {journal} {Phys.
  Rev. B}\ }\textbf {\bibinfo {volume} {22}},\ \bibinfo {pages}
  {2226}}\BibitemShut {NoStop}%
\bibitem [{\citenamefont {Nordsmark}\ \emph {et~al.}(1994)\citenamefont
  {Nordsmark}, \citenamefont {Bentzen},\ and\ \citenamefont
  {Overgaard}}]{Nordsmark1994}%
  \BibitemOpen
  \bibfield  {author} {\bibinfo {author} {\bibnamefont {Nordsmark},
  \bibfnamefont {M.}}, \bibinfo {author} {\bibfnamefont {S.}~\bibnamefont
  {Bentzen}}, \ and\ \bibinfo {author} {\bibfnamefont {J.}~\bibnamefont
  {Overgaard}}} (\bibinfo {year} {1994}),\ \href {\doibase
  10.3109/02841869409098433} {\bibfield  {journal} {\bibinfo  {journal} {Acta.
  Oncol.}\ }\textbf {\bibinfo {volume} {33}},\ \bibinfo {pages}
  {383}}\BibitemShut {NoStop}%
\bibitem [{\citenamefont {Nowakowski}\ and\ \citenamefont
  {Bedoya~Fierro}(2017)}]{Marek2017}%
  \BibitemOpen
  \bibfield  {author} {\bibinfo {author} {\bibnamefont {Nowakowski},
  \bibfnamefont {M.}}, \ and\ \bibinfo {author} {\bibfnamefont
  {D.}~\bibnamefont {Bedoya~Fierro}}} (\bibinfo {year} {2017}),\ \href@noop {}
  {\bibfield  {journal} {\bibinfo  {journal} {Acta Phys. Polon. B}\ }\textbf
  {\bibinfo {volume} {48}},\ \bibinfo {pages} {1955}}\BibitemShut {NoStop}%
\bibitem [{\citenamefont {Oberthaler}(2002)}]{Oberthaler2002}%
  \BibitemOpen
  \bibfield  {author} {\bibinfo {author} {\bibnamefont {Oberthaler},
  \bibfnamefont {M.}}} (\bibinfo {year} {2002}),\ \href {\doibase
  10.1016/S0168-583X(02)00793-0} {\bibfield  {journal} {\bibinfo  {journal}
  {Nucl. Inst. and Meth. B}\ }\textbf {\bibinfo {volume} {192}},\ \bibinfo
  {pages} {129}}\BibitemShut {NoStop}%
\bibitem [{\citenamefont {Ore}(1949)}]{Ore1949}%
  \BibitemOpen
  \bibfield  {author} {\bibinfo {author} {\bibnamefont {Ore}, \bibfnamefont
  {A.}}} (\bibinfo {year} {1949}),\ \href@noop {} {\bibfield  {journal}
  {\bibinfo  {journal} {Univ. Bergen Arbok}\ }\textbf {\bibinfo {volume}
  {9}}}\BibitemShut {NoStop}%
\bibitem [{\citenamefont {Ota}\ \emph {et~al.}(2019)\citenamefont {Ota} \emph
  {et~al.}}]{Ota2019}%
  \BibitemOpen
  \bibfield  {author} {\bibinfo {author} {\bibnamefont {Ota}, \bibfnamefont
  {R.}},  \emph {et~al.}} (\bibinfo {year} {2019}),\ \href@noop {} {\bibfield
  {journal} {\bibinfo  {journal} {Phys. Med. Biol.}\ }\textbf {\bibinfo
  {volume} {64}},\ \bibinfo {pages} {07LT01}}\BibitemShut {NoStop}%
\bibitem [{\citenamefont {Pachucki}\ and\ \citenamefont
  {Karshenboim}(1998)}]{Pachucki1997vm}%
  \BibitemOpen
  \bibfield  {author} {\bibinfo {author} {\bibnamefont {Pachucki},
  \bibfnamefont {K.}}, \ and\ \bibinfo {author} {\bibfnamefont {S.~G.}\
  \bibnamefont {Karshenboim}}} (\bibinfo {year} {1998}),\ \href {\doibase
  10.1103/PhysRevLett.80.2101} {\bibfield  {journal} {\bibinfo  {journal}
  {Phys. Rev. Lett.}\ }\textbf {\bibinfo {volume} {80}},\ \bibinfo {pages}
  {2101}}\BibitemShut {NoStop}%
%%CITATION = HEP-PH/9709387;%%
\bibitem [{\citenamefont {Pages}\ \emph {et~al.}(1972)\citenamefont {Pages},
  \citenamefont {Bertel}, \citenamefont {Joffre},\ and\ \citenamefont
  {Sklavenitis}}]{Pages1972}%
  \BibitemOpen
  \bibfield  {author} {\bibinfo {author} {\bibnamefont {Pages}, \bibfnamefont
  {L.}}, \bibinfo {author} {\bibfnamefont {E.}~\bibnamefont {Bertel}}, \bibinfo
  {author} {\bibfnamefont {H.}~\bibnamefont {Joffre}}, \ and\ \bibinfo {author}
  {\bibfnamefont {L.}~\bibnamefont {Sklavenitis}}} (\bibinfo {year} {1972}),\
  \href {\doibase 10.1016/S0092-640X(72)80002-0} {\bibfield  {journal}
  {\bibinfo  {journal} {Atomic Data and Nuclear Data Tables}\ }\textbf
  {\bibinfo {volume} {4}},\ \bibinfo {pages} {1}}\BibitemShut {NoStop}%
\bibitem [{\citenamefont {Pamula}\ and\ \citenamefont
  {Dryzek}(2008)}]{Pamula2008}%
  \BibitemOpen
  \bibfield  {author} {\bibinfo {author} {\bibnamefont {Pamula}, \bibfnamefont
  {E.}}, \ and\ \bibinfo {author} {\bibfnamefont {E.}~\bibnamefont {Dryzek}}}
  (\bibinfo {year} {2008}),\ \href@noop {} {\bibfield  {journal} {\bibinfo
  {journal} {Acta Phys. Polon. A}\ }\textbf {\bibinfo {volume} {113}},\
  \bibinfo {pages} {1485.}}\BibitemShut {Stop}%
\bibitem [{\citenamefont {Pamula}\ \emph {et~al.}(2006)\citenamefont {Pamula},
  \citenamefont {Dryzek},\ and\ \citenamefont {Dobrzynski}}]{Pamula2006}%
  \BibitemOpen
  \bibfield  {author} {\bibinfo {author} {\bibnamefont {Pamula}, \bibfnamefont
  {E.}}, \bibinfo {author} {\bibfnamefont {E.}~\bibnamefont {Dryzek}}, \ and\
  \bibinfo {author} {\bibfnamefont {P.}~\bibnamefont {Dobrzynski}}} (\bibinfo
  {year} {2006}),\ \href@noop {} {\bibfield  {journal} {\bibinfo  {journal}
  {Acta Phys. Polon. A}\ }\textbf {\bibinfo {volume} {110}},\ \bibinfo {pages}
  {631.}}\BibitemShut {Stop}%
\bibitem [{\citenamefont {Parker}\ \emph {et~al.}(2018)\citenamefont {Parker},
  \citenamefont {Yu}, \citenamefont {Zhong}, \citenamefont {Estey},\ and\
  \citenamefont {Müller}}]{Parker2018vye}%
  \BibitemOpen
  \bibfield  {author} {\bibinfo {author} {\bibnamefont {Parker}, \bibfnamefont
  {R.~H.}}, \bibinfo {author} {\bibfnamefont {C.}~\bibnamefont {Yu}}, \bibinfo
  {author} {\bibfnamefont {W.}~\bibnamefont {Zhong}}, \bibinfo {author}
  {\bibfnamefont {B.}~\bibnamefont {Estey}}, \ and\ \bibinfo {author}
  {\bibfnamefont {H.}~\bibnamefont {Müller}}} (\bibinfo {year} {2018}),\
  \href@noop {} {\bibfield  {journal} {\bibinfo  {journal} {Science}\ }\textbf
  {\bibinfo {volume} {360}},\ \bibinfo {pages} {191}}\BibitemShut {NoStop}%
\bibitem [{\citenamefont {Perez}\ and\ \citenamefont
  {Sacquin}(2012)}]{Perez2012}%
  \BibitemOpen
  \bibfield  {author} {\bibinfo {author} {\bibnamefont {Perez}, \bibfnamefont
  {P.}}, \ and\ \bibinfo {author} {\bibfnamefont {Y.}~\bibnamefont {Sacquin}}}
  (\bibinfo {year} {2012}),\ \href {\doibase 10.1088/0264-9381/29/18/184008}
  {\bibfield  {journal} {\bibinfo  {journal} {Class. Quant. Grav.}\ }\textbf
  {\bibinfo {volume} {29}},\ \bibinfo {pages} {184008}}\BibitemShut {NoStop}%
\bibitem [{\citenamefont {Perkins}\ and\ \citenamefont
  {Carbotte}(1970)}]{Perkins1970}%
  \BibitemOpen
  \bibfield  {author} {\bibinfo {author} {\bibnamefont {Perkins}, \bibfnamefont
  {A.}}, \ and\ \bibinfo {author} {\bibfnamefont {J.}~\bibnamefont {Carbotte}}}
  (\bibinfo {year} {1970}),\ \href {\doibase 10.1103/PhysRevB.1.101} {\bibfield
   {journal} {\bibinfo  {journal} {Phys. Rev. B}\ }\textbf {\bibinfo {volume}
  {1}},\ \bibinfo {pages} {101}}\BibitemShut {NoStop}%
\bibitem [{\citenamefont {Pethrick}(1997)}]{Pethrick1997}%
  \BibitemOpen
  \bibfield  {author} {\bibinfo {author} {\bibnamefont {Pethrick},
  \bibfnamefont {R.~A.}}} (\bibinfo {year} {1997}),\ \href {\doibase
  https://doi.org/10.1016/S0079-6700(96)00023-8} {\bibfield  {journal}
  {\bibinfo  {journal} {Prog. Polym. Sci.}\ }\textbf {\bibinfo {volume} {22}},\
  \bibinfo {pages} {1.}}\BibitemShut {Stop}%
\bibitem [{\citenamefont {Platzman}\ and\ \citenamefont
  {Mills}(1994)}]{PhysRevB.49.454}%
  \BibitemOpen
  \bibfield  {author} {\bibinfo {author} {\bibnamefont {Platzman},
  \bibfnamefont {P.~M.}}, \ and\ \bibinfo {author} {\bibfnamefont {A.~P.}\
  \bibnamefont {Mills}}} (\bibinfo {year} {1994}),\ \href {\doibase
  10.1103/PhysRevB.49.454} {\bibfield  {journal} {\bibinfo  {journal} {Phys.
  Rev. B}\ }\textbf {\bibinfo {volume} {49}},\ \bibinfo {pages}
  {454}}\BibitemShut {NoStop}%
\bibitem [{\citenamefont {Prenosil}\ \emph {et~al.}(2022)\citenamefont
  {Prenosil} \emph {et~al.}}]{Quadra2022}%
  \BibitemOpen
  \bibfield  {author} {\bibinfo {author} {\bibnamefont {Prenosil},
  \bibfnamefont {G.}},  \emph {et~al.}} (\bibinfo {year} {2022}),\ \href
  {\doibase doi:10.2967/jnumed.121.261972} {\bibfield  {journal} {\bibinfo
  {journal} {J. Nucl. Med.}\ }\textbf {\bibinfo {volume} {63}},\ \bibinfo
  {pages} {476}}\BibitemShut {NoStop}%
\bibitem [{\citenamefont {Pruszynski}\ \emph {et~al.}(2010)\citenamefont
  {Pruszynski} \emph {et~al.}}]{Pruszynski2010}%
  \BibitemOpen
  \bibfield  {author} {\bibinfo {author} {\bibnamefont {Pruszynski},
  \bibfnamefont {M.}},  \emph {et~al.}} (\bibinfo {year} {2010}),\ \href@noop
  {} {\bibfield  {journal} {\bibinfo  {journal} {Appl. Radiat. Isot.}\ }\textbf
  {\bibinfo {volume} {68}},\ \bibinfo {pages} {1636}}\BibitemShut {NoStop}%
\bibitem [{\citenamefont {Puska}\ and\ \citenamefont
  {Nieminen}(1994)}]{Puska1994}%
  \BibitemOpen
  \bibfield  {author} {\bibinfo {author} {\bibnamefont {Puska}, \bibfnamefont
  {M.}}, \ and\ \bibinfo {author} {\bibfnamefont {R.}~\bibnamefont {Nieminen}}}
  (\bibinfo {year} {1994}),\ \href {\doibase 10.1103/RevModPhys.66.841}
  {\bibfield  {journal} {\bibinfo  {journal} {Rev. Mod. Phys.}\ }\textbf
  {\bibinfo {volume} {66}},\ \bibinfo {pages} {841}}\BibitemShut {NoStop}%
\bibitem [{\citenamefont {Qi}\ and\ \citenamefont
  {Huang}(2022)}]{Qi-positronium-2022}%
  \BibitemOpen
  \bibfield  {author} {\bibinfo {author} {\bibnamefont {Qi}, \bibfnamefont
  {J.}}, \ and\ \bibinfo {author} {\bibfnamefont {B.}~\bibnamefont {Huang}}}
  (\bibinfo {year} {2022}),\ \href@noop {} {\bibfield  {journal} {\bibinfo
  {journal} {IEEE Trans. Med. Imag.}\ }\textbf {\bibinfo {volume} {41}},\
  \bibinfo {pages} {2848}}\BibitemShut {NoStop}%
\bibitem [{\citenamefont {Reivich}\ \emph {et~al.}(1979)\citenamefont {Reivich}
  \emph {et~al.}}]{Reivich1979}%
  \BibitemOpen
  \bibfield  {author} {\bibinfo {author} {\bibnamefont {Reivich}, \bibfnamefont
  {M.}},  \emph {et~al.}} (\bibinfo {year} {1979}),\ \href {\doibase
  https://doi.org/10.1161/01.res.44.1.127} {\bibfield  {journal} {\bibinfo
  {journal} {Circ. Res.}\ }\textbf {\bibinfo {volume} {44}},\ \bibinfo {pages}
  {127.}}\BibitemShut {Stop}%
\bibitem [{\citenamefont {Ritter}\ \emph {et~al.}(1984)\citenamefont {Ritter},
  \citenamefont {Egan}, \citenamefont {Hughes},\ and\ \citenamefont
  {Woodle}}]{Ritter1984mqy}%
  \BibitemOpen
  \bibfield  {author} {\bibinfo {author} {\bibnamefont {Ritter}, \bibfnamefont
  {M.~W.}}, \bibinfo {author} {\bibfnamefont {P.~O.}\ \bibnamefont {Egan}},
  \bibinfo {author} {\bibfnamefont {V.~W.}\ \bibnamefont {Hughes}}, \ and\
  \bibinfo {author} {\bibfnamefont {K.~A.}\ \bibnamefont {Woodle}}} (\bibinfo
  {year} {1984}),\ \href {\doibase 10.1103/PhysRevA.30.1331} {\bibfield
  {journal} {\bibinfo  {journal} {Phys. Rev.}\ }\textbf {\bibinfo {volume}
  {A30}},\ \bibinfo {pages} {1331}}\BibitemShut {NoStop}%
%%CITATION = PHRVA,A30,1331;%%
\bibitem [{\citenamefont {Roesch}(2012)}]{Roesch2012}%
  \BibitemOpen
  \bibfield  {author} {\bibinfo {author} {\bibnamefont {Roesch}, \bibfnamefont
  {F.}}} (\bibinfo {year} {2012}),\ \href@noop {} {\bibfield  {journal}
  {\bibinfo  {journal} {Curr. Radiopharm.}\ }\textbf {\bibinfo {volume} {5}},\
  \bibinfo {pages} {187}}\BibitemShut {NoStop}%
\bibitem [{\citenamefont {Rosar}\ \emph {et~al.}(2020)\citenamefont {Rosar},
  \citenamefont {Buchholz}, \citenamefont {Michels}, \citenamefont {Hoffmann},
  \citenamefont {Piel}, \citenamefont {Waldmann}, \citenamefont {Rösch},
  \citenamefont {Reuss},\ and\ \citenamefont {Schreckenberger}}]{Rosar2020}%
  \BibitemOpen
  \bibfield  {author} {\bibinfo {author} {\bibnamefont {Rosar}, \bibfnamefont
  {F.}}, \bibinfo {author} {\bibfnamefont {H.}~\bibnamefont {Buchholz}},
  \bibinfo {author} {\bibfnamefont {S.}~\bibnamefont {Michels}}, \bibinfo
  {author} {\bibfnamefont {M.}~\bibnamefont {Hoffmann}}, \bibinfo {author}
  {\bibfnamefont {M.}~\bibnamefont {Piel}}, \bibinfo {author} {\bibfnamefont
  {C.}~\bibnamefont {Waldmann}}, \bibinfo {author} {\bibfnamefont
  {F.}~\bibnamefont {Rösch}}, \bibinfo {author} {\bibfnamefont
  {S.}~\bibnamefont {Reuss}}, \ and\ \bibinfo {author} {\bibfnamefont
  {M.}~\bibnamefont {Schreckenberger}}} (\bibinfo {year} {2020}),\ \href
  {\doibase 10.1186/s40658-020-0286-3} {\bibfield  {journal} {\bibinfo
  {journal} {EJNMMI Phys.}\ }\textbf {\bibinfo {volume} {7}},\ \bibinfo {pages}
  {16}}\BibitemShut {NoStop}%
\bibitem [{\citenamefont {Rubin}\ and\ \citenamefont
  {Yarden}(2001)}]{Rubin2001}%
  \BibitemOpen
  \bibfield  {author} {\bibinfo {author} {\bibnamefont {Rubin}, \bibfnamefont
  {I.}}, \ and\ \bibinfo {author} {\bibfnamefont {Y.}~\bibnamefont {Yarden}}}
  (\bibinfo {year} {2001}),\ \href {\doibase
  doi.org.10.1093/annonc/12.suppl_1.s3} {\bibfield  {journal} {\bibinfo
  {journal} {Ann. Oncol.}\ }\textbf {\bibinfo {volume} {Suppl 1}},\ \bibinfo
  {pages} {S3}}\BibitemShut {NoStop}%
\bibitem [{\citenamefont {Saito}\ and\ \citenamefont
  {Hyodo}(1999)}]{Saito1999}%
  \BibitemOpen
  \bibfield  {author} {\bibinfo {author} {\bibnamefont {Saito}, \bibfnamefont
  {H.}}, \ and\ \bibinfo {author} {\bibfnamefont {T.}~\bibnamefont {Hyodo}}}
  (\bibinfo {year} {1999}),\ \href {\doibase 10.1103/PhysRevB.60.11070}
  {\bibfield  {journal} {\bibinfo  {journal} {Phys. Rev. B}\ }\textbf {\bibinfo
  {volume} {60}},\ \bibinfo {pages} {11070}}\BibitemShut {NoStop}%
\bibitem [{\citenamefont {Sane}\ \emph {et~al.}(2010)\citenamefont {Sane},
  \citenamefont {Wiedmer}, \citenamefont {Nyman}, \citenamefont {Vattulainen},\
  and\ \citenamefont {Holopainen}}]{Sane2010}%
  \BibitemOpen
  \bibfield  {author} {\bibinfo {author} {\bibnamefont {Sane}, \bibfnamefont
  {P.~Tuomisto, F.}}, \bibinfo {author} {\bibfnamefont {S.}~\bibnamefont
  {Wiedmer}}, \bibinfo {author} {\bibfnamefont {T.}~\bibnamefont {Nyman}},
  \bibinfo {author} {\bibfnamefont {I.}~\bibnamefont {Vattulainen}}, \ and\
  \bibinfo {author} {\bibfnamefont {J.}~\bibnamefont {Holopainen}}} (\bibinfo
  {year} {2010}),\ \href {\doibase doi.org.10.1016/j.bbamem.2010.01.011}
  {\bibfield  {journal} {\bibinfo  {journal} {Biochim. Biophys. Acta}\ }\textbf
  {\bibinfo {volume} {1798}},\ \bibinfo {pages} {958}}\BibitemShut {NoStop}%
\bibitem [{\citenamefont {Sane}\ \emph {et~al.}(2009)\citenamefont {Sane},
  \citenamefont {Salonen}, \citenamefont {Falck}, \citenamefont {Repakova},
  \citenamefont {Tuomisto}, \citenamefont {Holopainen},\ and\ \citenamefont
  {Vattulainen}}]{Sane2009}%
  \BibitemOpen
  \bibfield  {author} {\bibinfo {author} {\bibnamefont {Sane}, \bibfnamefont
  {P.}}, \bibinfo {author} {\bibfnamefont {E.}~\bibnamefont {Salonen}},
  \bibinfo {author} {\bibfnamefont {E.}~\bibnamefont {Falck}}, \bibinfo
  {author} {\bibfnamefont {J.}~\bibnamefont {Repakova}}, \bibinfo {author}
  {\bibfnamefont {F.}~\bibnamefont {Tuomisto}}, \bibinfo {author}
  {\bibfnamefont {J.}~\bibnamefont {Holopainen}}, \ and\ \bibinfo {author}
  {\bibfnamefont {I.}~\bibnamefont {Vattulainen}}} (\bibinfo {year} {2009}),\
  \href {\doibase doi.org.10.1021/jp809308j} {\bibfield  {journal} {\bibinfo
  {journal} {J. Phys. Chem. B.}\ }\textbf {\bibinfo {volume} {113}},\ \bibinfo
  {pages} {16810}}\BibitemShut {NoStop}%
\bibitem [{\citenamefont {Sane}\ \emph {et~al.}(2011)\citenamefont {Sane},
  \citenamefont {Toumisto},\ and\ \citenamefont {Holopainen}}]{Sane2011}%
  \BibitemOpen
  \bibfield  {author} {\bibinfo {author} {\bibnamefont {Sane}, \bibfnamefont
  {P.}}, \bibinfo {author} {\bibfnamefont {F.}~\bibnamefont {Toumisto}}, \ and\
  \bibinfo {author} {\bibfnamefont {J.}~\bibnamefont {Holopainen}}} (\bibinfo
  {year} {2011}),\ \href {\doibase doi.org.10.1016/j.clae.2010.06.008}
  {\bibfield  {journal} {\bibinfo  {journal} {Cont. Lens. Anterior. Eye}\
  }\textbf {\bibinfo {volume} {34}},\ \bibinfo {pages} {2}}\BibitemShut
  {NoStop}%
\bibitem [{\citenamefont {Sato}\ \emph {et~al.}(2015)\citenamefont {Sato},
  \citenamefont {Xu}, \citenamefont {Yoshiie}, \citenamefont {Sano},
  \citenamefont {Kawabe}, \citenamefont {Nagai}, \citenamefont {Nagumo},
  \citenamefont {Inoue}, \citenamefont {Toyama}, \citenamefont {Oshima},
  \citenamefont {Kinomura},\ and\ \citenamefont {Shirai}}]{Sato2015}%
  \BibitemOpen
  \bibfield  {author} {\bibinfo {author} {\bibnamefont {Sato}, \bibfnamefont
  {K.}}, \bibinfo {author} {\bibfnamefont {Q.}~\bibnamefont {Xu}}, \bibinfo
  {author} {\bibfnamefont {T.}~\bibnamefont {Yoshiie}}, \bibinfo {author}
  {\bibfnamefont {T.}~\bibnamefont {Sano}}, \bibinfo {author} {\bibfnamefont
  {H.}~\bibnamefont {Kawabe}}, \bibinfo {author} {\bibfnamefont
  {Y.}~\bibnamefont {Nagai}}, \bibinfo {author} {\bibfnamefont
  {K.}~\bibnamefont {Nagumo}}, \bibinfo {author} {\bibfnamefont
  {K.}~\bibnamefont {Inoue}}, \bibinfo {author} {\bibfnamefont
  {T.}~\bibnamefont {Toyama}}, \bibinfo {author} {\bibfnamefont
  {N.}~\bibnamefont {Oshima}}, \bibinfo {author} {\bibfnamefont
  {A.}~\bibnamefont {Kinomura}}, \ and\ \bibinfo {author} {\bibfnamefont
  {Y.}~\bibnamefont {Shirai}}} (\bibinfo {year} {2015}),\ \href {\doibase
  10.1016/j.nimb.2014.09.022} {\bibfield  {journal} {\bibinfo  {journal} {Nucl.
  Inst. and Meth.B}\ }\textbf {\bibinfo {volume} {342}},\ \bibinfo {pages}
  {104}}\BibitemShut {NoStop}%
\bibitem [{\citenamefont {Sawyers}(2009)}]{Sawyers2009}%
  \BibitemOpen
  \bibfield  {author} {\bibinfo {author} {\bibnamefont {Sawyers}, \bibfnamefont
  {C.~L.}}} (\bibinfo {year} {2009}),\ \href {\doibase
  doi.org.10.1016/j.cell.2019.08.027} {\bibfield  {journal} {\bibinfo
  {journal} {Cell}\ }\textbf {\bibinfo {volume} {179}},\ \bibinfo {pages}
  {8}}\BibitemShut {NoStop}%
\bibitem [{\citenamefont {Schultz}\ and\ \citenamefont
  {Campbell}(1985)}]{Schultz1985}%
  \BibitemOpen
  \bibfield  {author} {\bibinfo {author} {\bibnamefont {Schultz}, \bibfnamefont
  {P.}}, \ and\ \bibinfo {author} {\bibfnamefont {J.}~\bibnamefont {Campbell}}}
  (\bibinfo {year} {1985}),\ \href {\doibase 10.1016/0375-9601(85)90349-4}
  {\bibfield  {journal} {\bibinfo  {journal} {Appl. Phys. A}\ }\textbf
  {\bibinfo {volume} {112}},\ \bibinfo {pages} {316}}\BibitemShut {NoStop}%
\bibitem [{\citenamefont {Schultz}\ and\ \citenamefont
  {Lynn}(1988)}]{Schultz1988}%
  \BibitemOpen
  \bibfield  {author} {\bibinfo {author} {\bibnamefont {Schultz}, \bibfnamefont
  {P.}}, \ and\ \bibinfo {author} {\bibfnamefont {K.}~\bibnamefont {Lynn}}}
  (\bibinfo {year} {1988}),\ \href {\doibase 10.1103/RevModPhys.60.701}
  {\bibfield  {journal} {\bibinfo  {journal} {Rev. Mod. Phys.}\ }\textbf
  {\bibinfo {volume} {60}},\ \bibinfo {pages} {701}}\BibitemShut {NoStop}%
\bibitem [{\citenamefont {Schut}\ \emph {et~al.}(2004)\citenamefont {Schut},
  \citenamefont {van Veen}, \citenamefont {de~Roode},\ and\ \citenamefont
  {Labohm}}]{Schut2004}%
  \BibitemOpen
  \bibfield  {author} {\bibinfo {author} {\bibnamefont {Schut}, \bibfnamefont
  {H.}}, \bibinfo {author} {\bibfnamefont {A.}~\bibnamefont {van Veen}},
  \bibinfo {author} {\bibfnamefont {J.}~\bibnamefont {de~Roode}}, \ and\
  \bibinfo {author} {\bibfnamefont {F.}~\bibnamefont {Labohm}}} (\bibinfo
  {year} {2004}),\ \href {\doibase 10.4028/www.scientific.net/MSF.445-446.507}
  {\bibfield  {journal} {\bibinfo  {journal} {Mater.Sci.Forum}\ }\textbf
  {\bibinfo {volume} {444}},\ \bibinfo {pages} {507}}\BibitemShut {NoStop}%
\bibitem [{\citenamefont {Sferlazzo}\ \emph {et~al.}(1985)\citenamefont
  {Sferlazzo}, \citenamefont {Berko},\ and\ \citenamefont
  {Canter}}]{Sferlazzo1985}%
  \BibitemOpen
  \bibfield  {author} {\bibinfo {author} {\bibnamefont {Sferlazzo},
  \bibfnamefont {P.}}, \bibinfo {author} {\bibfnamefont {S.}~\bibnamefont
  {Berko}}, \ and\ \bibinfo {author} {\bibfnamefont {K.}~\bibnamefont
  {Canter}}} (\bibinfo {year} {1985}),\ \href {\doibase
  10.1103/PhysRevB.32.6067} {\bibfield  {journal} {\bibinfo  {journal} {Phys.
  Rev. B}\ }\textbf {\bibinfo {volume} {32}},\ \bibinfo {pages}
  {6067}}\BibitemShut {NoStop}%
\bibitem [{\citenamefont {Sharma}\ \emph {et~al.}(2022)\citenamefont {Sharma},
  \citenamefont {Kumar},\ and\ \citenamefont {Moskal}}]{Sharma2022puzzle}%
  \BibitemOpen
  \bibfield  {author} {\bibinfo {author} {\bibnamefont {Sharma}, \bibfnamefont
  {S.}}, \bibinfo {author} {\bibfnamefont {D.}~\bibnamefont {Kumar}}, \ and\
  \bibinfo {author} {\bibfnamefont {P.}~\bibnamefont {Moskal}}} (\bibinfo
  {year} {2022}),\ \href@noop {} {\bibfield  {journal} {\bibinfo  {journal}
  {Acta Phys. Polon. A}\ }\textbf {\bibinfo {volume} {142}},\ \bibinfo {pages}
  {428}}\BibitemShut {NoStop}%
\bibitem [{\citenamefont {Shibuya}\ \emph {et~al.}(2020)\citenamefont
  {Shibuya}, \citenamefont {Saito}, \citenamefont {Nishikido}, \citenamefont
  {Takahashi},\ and\ \citenamefont {T.~Yamaya}}]{Shibuya2020}%
  \BibitemOpen
  \bibfield  {author} {\bibinfo {author} {\bibnamefont {Shibuya}, \bibfnamefont
  {K.}}, \bibinfo {author} {\bibfnamefont {H.}~\bibnamefont {Saito}}, \bibinfo
  {author} {\bibfnamefont {F.}~\bibnamefont {Nishikido}}, \bibinfo {author}
  {\bibfnamefont {M.}~\bibnamefont {Takahashi}}, \ and\ \bibinfo {author}
  {\bibfnamefont {T.}~\bibnamefont {T.~Yamaya}}} (\bibinfo {year} {2020}),\
  \href {\doibase https://doi.org/10.1038/s42005-020-00440-z} {\bibfield
  {journal} {\bibinfo  {journal} {Commun. Phys.}\ }\textbf {\bibinfo {volume}
  {3}},\ \bibinfo {pages} {173}}\BibitemShut {NoStop}%
\bibitem [{\citenamefont {Shibuya}\ \emph {et~al.}(2022)\citenamefont
  {Shibuya}, \citenamefont {Saito},\ and\ \citenamefont
  {Tashima}}]{Shibua-positronium-2022}%
  \BibitemOpen
  \bibfield  {author} {\bibinfo {author} {\bibnamefont {Shibuya}, \bibfnamefont
  {K.}}, \bibinfo {author} {\bibfnamefont {H.}~\bibnamefont {Saito}}, \ and\
  \bibinfo {author} {\bibfnamefont {T.}~\bibnamefont {Tashima}, \bibfnamefont
  {H.~amd~Yamaya}}} (\bibinfo {year} {2022}),\ \href@noop {} {\bibfield
  {journal} {\bibinfo  {journal} {Phys. Med. Biol.}\ }\textbf {\bibinfo
  {volume} {67}},\ \bibinfo {pages} {025009}}\BibitemShut {NoStop}%
\bibitem [{\citenamefont {Shinohara}\ \emph {et~al.}(2001)\citenamefont
  {Shinohara}, \citenamefont {Suzuki}, \citenamefont {Chang},\ and\
  \citenamefont {Hyodo}}]{Shinohara2001}%
  \BibitemOpen
  \bibfield  {author} {\bibinfo {author} {\bibnamefont {Shinohara},
  \bibfnamefont {N.}}, \bibinfo {author} {\bibfnamefont {N.}~\bibnamefont
  {Suzuki}}, \bibinfo {author} {\bibfnamefont {T.}~\bibnamefont {Chang}}, \
  and\ \bibinfo {author} {\bibfnamefont {T.}~\bibnamefont {Hyodo}}} (\bibinfo
  {year} {2001}),\ \href {\doibase 10.1103/PhysRevA.64.042702} {\bibfield
  {journal} {\bibinfo  {journal} {Phys. Rev. A}\ }\textbf {\bibinfo {volume}
  {64}},\ \bibinfo {pages} {042702}}\BibitemShut {NoStop}%
\bibitem [{\citenamefont {Stepanov}\ \emph {et~al.}(2020)\citenamefont
  {Stepanov}, \citenamefont {Selim}, \citenamefont {Stepanov}, \citenamefont
  {Bokov}, \citenamefont {Ilyukhina}, \citenamefont {Duplatre},\ and\
  \citenamefont {Byakov}}]{Stepanov2020}%
  \BibitemOpen
  \bibfield  {author} {\bibinfo {author} {\bibnamefont {Stepanov},
  \bibfnamefont {P.~S.}}, \bibinfo {author} {\bibfnamefont {F.}~\bibnamefont
  {Selim}}, \bibinfo {author} {\bibfnamefont {S.}~\bibnamefont {Stepanov}},
  \bibinfo {author} {\bibfnamefont {A.}~\bibnamefont {Bokov}}, \bibinfo
  {author} {\bibfnamefont {O.}~\bibnamefont {Ilyukhina}}, \bibinfo {author}
  {\bibfnamefont {G.}~\bibnamefont {Duplatre}}, \ and\ \bibinfo {author}
  {\bibfnamefont {V.}~\bibnamefont {Byakov}}} (\bibinfo {year} {2020}),\
  \href@noop {} {\bibfield  {journal} {\bibinfo  {journal} {Phys. Chem. Chem.
  Phys.}\ }\textbf {\bibinfo {volume} {22}},\ \bibinfo {pages}
  {5123}}\BibitemShut {NoStop}%
\bibitem [{\citenamefont {Stepanov}\ and\ \citenamefont
  {Byakov}(2002)}]{Stepanov2002}%
  \BibitemOpen
  \bibfield  {author} {\bibinfo {author} {\bibnamefont {Stepanov},
  \bibfnamefont {S.}}, \ and\ \bibinfo {author} {\bibfnamefont
  {V.}~\bibnamefont {Byakov}}} (\bibinfo {year} {2002}),\ \href {\doibase
  10.1063/1.1451244} {\bibfield  {journal} {\bibinfo  {journal} {J. Chem.
  Phys.}\ }\textbf {\bibinfo {volume} {116}},\ \bibinfo {pages}
  {6178}}\BibitemShut {NoStop}%
\bibitem [{\citenamefont {Stepanov}\ \emph {et~al.}(2009)\citenamefont
  {Stepanov}, \citenamefont {Byakov}, \citenamefont {Duplâtre}, \citenamefont
  {Zvezhinskiy},\ and\ \citenamefont {Lomachuk}}]{Stepanov2009}%
  \BibitemOpen
  \bibfield  {author} {\bibinfo {author} {\bibnamefont {Stepanov},
  \bibfnamefont {S.}}, \bibinfo {author} {\bibfnamefont {V.}~\bibnamefont
  {Byakov}}, \bibinfo {author} {\bibfnamefont {G.}~\bibnamefont {Duplâtre}},
  \bibinfo {author} {\bibfnamefont {D.}~\bibnamefont {Zvezhinskiy}}, \ and\
  \bibinfo {author} {\bibfnamefont {Y.}~\bibnamefont {Lomachuk}}} (\bibinfo
  {year} {2009}),\ \href {\doibase 10.1002/pssc.200982059} {\bibfield
  {journal} {\bibinfo  {journal} {hys. Status Solidi (c)}\ }\textbf {\bibinfo
  {volume} {6}},\ \bibinfo {pages} {2476}}\BibitemShut {NoStop}%
\bibitem [{\citenamefont {Stepanov}\ \emph {et~al.}(2007)\citenamefont
  {Stepanov}, \citenamefont {Byakov},\ and\ \citenamefont
  {Hirade}}]{Stepanov2007}%
  \BibitemOpen
  \bibfield  {author} {\bibinfo {author} {\bibnamefont {Stepanov},
  \bibfnamefont {S.}}, \bibinfo {author} {\bibfnamefont {V.}~\bibnamefont
  {Byakov}}, \ and\ \bibinfo {author} {\bibfnamefont {T.}~\bibnamefont
  {Hirade}}} (\bibinfo {year} {2007}),\ \href {\doibase
  10.1016/j.radphyschem.2006.03.012} {\bibfield  {journal} {\bibinfo  {journal}
  {Rad. Phys. and Chem.}\ }\textbf {\bibinfo {volume} {76}},\ \bibinfo {pages}
  {90}}\BibitemShut {NoStop}%
\bibitem [{\citenamefont {Stepanov}\ \emph {et~al.}(2005)\citenamefont
  {Stepanov}, \citenamefont {Byakov},\ and\ \citenamefont
  {Kobayashi}}]{Stepanov2005}%
  \BibitemOpen
  \bibfield  {author} {\bibinfo {author} {\bibnamefont {Stepanov},
  \bibfnamefont {S.}}, \bibinfo {author} {\bibfnamefont {V.}~\bibnamefont
  {Byakov}}, \ and\ \bibinfo {author} {\bibfnamefont {Y.}~\bibnamefont
  {Kobayashi}}} (\bibinfo {year} {2005}),\ \href {\doibase
  10.1103/PhysRevB.72.054205} {\bibfield  {journal} {\bibinfo  {journal} {Phys.
  Rev. B}\ }\textbf {\bibinfo {volume} {72}},\ \bibinfo {pages}
  {054205}}\BibitemShut {NoStop}%
\bibitem [{\citenamefont {Stepanov}\ \emph {et~al.}(2021)\citenamefont
  {Stepanov}, \citenamefont {Byakova},\ and\ \citenamefont
  {Stepanov}}]{Stepanov2021}%
  \BibitemOpen
  \bibfield  {author} {\bibinfo {author} {\bibnamefont {Stepanov},
  \bibfnamefont {S.~V.}}, \bibinfo {author} {\bibfnamefont {V.}~\bibnamefont
  {Byakova}}, \ and\ \bibinfo {author} {\bibfnamefont {P.}~\bibnamefont
  {Stepanov}}} (\bibinfo {year} {2021}),\ \href@noop {} {\bibfield  {journal}
  {\bibinfo  {journal} {Physics of Wave Phenomena}\ }\textbf {\bibinfo {volume}
  {29}},\ \bibinfo {pages} {174}}\BibitemShut {NoStop}%
\bibitem [{\citenamefont {Stinson}\ \emph {et~al.}(1980)\citenamefont
  {Stinson}, \citenamefont {King}, \citenamefont {Marsh}, \citenamefont
  {Fabian},\ and\ \citenamefont {MacKenzie}}]{Stinson1980}%
  \BibitemOpen
  \bibfield  {author} {\bibinfo {author} {\bibnamefont {Stinson}, \bibfnamefont
  {R.~H.}}, \bibinfo {author} {\bibfnamefont {D.}~\bibnamefont {King}},
  \bibinfo {author} {\bibfnamefont {J.}~\bibnamefont {Marsh}}, \bibinfo
  {author} {\bibfnamefont {J.}~\bibnamefont {Fabian}}, \ and\ \bibinfo {author}
  {\bibfnamefont {I.}~\bibnamefont {MacKenzie}}} (\bibinfo {year} {1980}),\
  \href {\doibase https://doi.org/10.1016/0375-9601(80)90128-0} {\bibfield
  {journal} {\bibinfo  {journal} {Phys. Lett. A}\ }\textbf {\bibinfo {volume}
  {74}},\ \bibinfo {pages} {246.}}\BibitemShut {Stop}%
\bibitem [{\citenamefont {Surko}\ \emph {et~al.}(1989)\citenamefont {Surko},
  \citenamefont {Leventhal},\ and\ \citenamefont {Passner}}]{Surko1989}%
  \BibitemOpen
  \bibfield  {author} {\bibinfo {author} {\bibnamefont {Surko}, \bibfnamefont
  {C.}}, \bibinfo {author} {\bibfnamefont {M.}~\bibnamefont {Leventhal}}, \
  and\ \bibinfo {author} {\bibfnamefont {A.}~\bibnamefont {Passner}}} (\bibinfo
  {year} {1989}),\ \href {\doibase 10.1103/PhysRevLett.62.901} {\bibfield
  {journal} {\bibinfo  {journal} {Phys. Rev. Lett.}\ }\textbf {\bibinfo
  {volume} {62}},\ \bibinfo {pages} {901}}\BibitemShut {NoStop}%
\bibitem [{\citenamefont {Surti}\ \emph {et~al.}(2020)\citenamefont {Surti},
  \citenamefont {Pantel},\ and\ \citenamefont {JS.}}]{Surti-review2020}%
  \BibitemOpen
  \bibfield  {author} {\bibinfo {author} {\bibnamefont {Surti}, \bibfnamefont
  {S.}}, \bibinfo {author} {\bibfnamefont {A.}~\bibnamefont {Pantel}}, \ and\
  \bibinfo {author} {\bibfnamefont {K.}~\bibnamefont {JS.}}} (\bibinfo {year}
  {2020}),\ \href {\doibase 10.1109/TRPMS.2020.2985403} {\bibfield  {journal}
  {\bibinfo  {journal} {IEEE Trans. Plasma Sci.}\ }\textbf {\bibinfo {volume}
  {4}},\ \bibinfo {pages} {283}}\BibitemShut {NoStop}%
\bibitem [{\citenamefont {Swartz}\ \emph {et~al.}(2020)\citenamefont {Swartz},
  \citenamefont {Flood}, \citenamefont {Schaner}, \citenamefont {Halpern},
  \citenamefont {Williams}, \citenamefont {Pogue}, \citenamefont {Gallez},\
  and\ \citenamefont {Vaupel}}]{Swartz2020}%
  \BibitemOpen
  \bibfield  {author} {\bibinfo {author} {\bibnamefont {Swartz}, \bibfnamefont
  {H.~M.}}, \bibinfo {author} {\bibfnamefont {A.}~\bibnamefont {Flood}},
  \bibinfo {author} {\bibfnamefont {P.}~\bibnamefont {Schaner}}, \bibinfo
  {author} {\bibfnamefont {H.}~\bibnamefont {Halpern}}, \bibinfo {author}
  {\bibfnamefont {B.}~\bibnamefont {Williams}}, \bibinfo {author}
  {\bibfnamefont {B.}~\bibnamefont {Pogue}}, \bibinfo {author} {\bibfnamefont
  {B.}~\bibnamefont {Gallez}}, \ and\ \bibinfo {author} {\bibfnamefont
  {P.}~\bibnamefont {Vaupel}}} (\bibinfo {year} {2020}),\ \href {\doibase
  https://doi.org/10.14814/phy2.14541} {\bibfield  {journal} {\bibinfo
  {journal} {Physiol. Rep.}\ }\textbf {\bibinfo {volume} {8}},\ \bibinfo
  {pages} {e14541}}\BibitemShut {NoStop}%
\bibitem [{\citenamefont {Tanaka}\ \emph {et~al.}(2006)\citenamefont {Tanaka},
  \citenamefont {Kurihara},\ and\ \citenamefont {Mills}}]{Tanaka2006}%
  \BibitemOpen
  \bibfield  {author} {\bibinfo {author} {\bibnamefont {Tanaka}, \bibfnamefont
  {H.}}, \bibinfo {author} {\bibfnamefont {T.}~\bibnamefont {Kurihara}}, \ and\
  \bibinfo {author} {\bibfnamefont {A.}~\bibnamefont {Mills}}} (\bibinfo {year}
  {2006}),\ \href {\doibase 10.1351/pac199466081739} {\bibfield  {journal}
  {\bibinfo  {journal} {J. Phys.: Condens. Matter}\ }\textbf {\bibinfo {volume}
  {18}},\ \bibinfo {pages} {8581}}\BibitemShut {NoStop}%
\bibitem [{\citenamefont {Tao}\ \emph {et~al.}(2021)\citenamefont {Tao},
  \citenamefont {Coffee}, \citenamefont {Jeong},\ and\ \citenamefont
  {Levin}}]{Tao-Levin2021}%
  \BibitemOpen
  \bibfield  {author} {\bibinfo {author} {\bibnamefont {Tao}, \bibfnamefont
  {L.}}, \bibinfo {author} {\bibfnamefont {R.}~\bibnamefont {Coffee}}, \bibinfo
  {author} {\bibfnamefont {D.}~\bibnamefont {Jeong}}, \ and\ \bibinfo {author}
  {\bibfnamefont {C.}~\bibnamefont {Levin}}} (\bibinfo {year} {2021}),\ \href
  {\doibase 10.1088/1361-6560/abd951} {\bibfield  {journal} {\bibinfo
  {journal} {Phys. Med. Biol}\ }\textbf {\bibinfo {volume} {66}},\ \bibinfo
  {pages} {045032}}\BibitemShut {NoStop}%
\bibitem [{\citenamefont {Thirolf}\ \emph {et~al.}(2015)\citenamefont
  {Thirolf}, \citenamefont {Lang},\ and\ \citenamefont {Parodi}}]{Thirolf2015}%
  \BibitemOpen
  \bibfield  {author} {\bibinfo {author} {\bibnamefont {Thirolf}, \bibfnamefont
  {P.~G.}}, \bibinfo {author} {\bibfnamefont {C.}~\bibnamefont {Lang}}, \ and\
  \bibinfo {author} {\bibfnamefont {K.}~\bibnamefont {Parodi}}} (\bibinfo
  {year} {2015}),\ \href@noop {} {\bibfield  {journal} {\bibinfo  {journal}
  {Acta Phys. Polon. A}\ }\textbf {\bibinfo {volume} {127}},\ \bibinfo {pages}
  {1441}}\BibitemShut {NoStop}%
\bibitem [{\citenamefont {Toghyani}\ \emph {et~al.}(2016)\citenamefont
  {Toghyani}, \citenamefont {Gillam}, \citenamefont {McNamara},\ and\
  \citenamefont {Kuncic}}]{Toghyani2016}%
  \BibitemOpen
  \bibfield  {author} {\bibinfo {author} {\bibnamefont {Toghyani},
  \bibfnamefont {M.}}, \bibinfo {author} {\bibfnamefont {J.}~\bibnamefont
  {Gillam}}, \bibinfo {author} {\bibfnamefont {A.}~\bibnamefont {McNamara}}, \
  and\ \bibinfo {author} {\bibfnamefont {Z.}~\bibnamefont {Kuncic}}} (\bibinfo
  {year} {2016}),\ \href@noop {} {\bibfield  {journal} {\bibinfo  {journal}
  {Phys. Med. Biol.}\ }\textbf {\bibinfo {volume} {61}},\ \bibinfo {pages}
  {5803}}\BibitemShut {NoStop}%
\bibitem [{\citenamefont {Tuomisaari}\ \emph {et~al.}(1989)\citenamefont
  {Tuomisaari}, \citenamefont {Howell},\ and\ \citenamefont
  {McMullen}}]{Tuomisaari1989}%
  \BibitemOpen
  \bibfield  {author} {\bibinfo {author} {\bibnamefont {Tuomisaari},
  \bibfnamefont {M.}}, \bibinfo {author} {\bibfnamefont {R.}~\bibnamefont
  {Howell}}, \ and\ \bibinfo {author} {\bibfnamefont {T.}~\bibnamefont
  {McMullen}}} (\bibinfo {year} {1989}),\ \href {\doibase
  10.1103/PhysRevB.40.2060} {\bibfield  {journal} {\bibinfo  {journal} {Phys.
  Rev. B}\ }\textbf {\bibinfo {volume} {40}},\ \bibinfo {pages}
  {2060}}\BibitemShut {NoStop}%
\bibitem [{\citenamefont {Valkealahti}\ and\ \citenamefont
  {Nieminen}(1983)}]{Valkealahti1983}%
  \BibitemOpen
  \bibfield  {author} {\bibinfo {author} {\bibnamefont {Valkealahti},
  \bibfnamefont {S.}}, \ and\ \bibinfo {author} {\bibfnamefont
  {R.}~\bibnamefont {Nieminen}}} (\bibinfo {year} {1983}),\ \href {\doibase
  10.1007/BF00617834} {\bibfield  {journal} {\bibinfo  {journal} {Appl. Phys.
  A}\ }\textbf {\bibinfo {volume} {32}},\ \bibinfo {pages} {95}}\BibitemShut
  {NoStop}%
\bibitem [{\citenamefont {Valkealahti}\ and\ \citenamefont
  {Nieminen}(1984)}]{Valkealahti1984}%
  \BibitemOpen
  \bibfield  {author} {\bibinfo {author} {\bibnamefont {Valkealahti},
  \bibfnamefont {S.}}, \ and\ \bibinfo {author} {\bibfnamefont
  {R.}~\bibnamefont {Nieminen}}} (\bibinfo {year} {1984}),\ \href {\doibase
  10.1007/BF00620300} {\bibfield  {journal} {\bibinfo  {journal} {Appl. Phys.
  A}\ }\textbf {\bibinfo {volume} {35}},\ \bibinfo {pages} {51}}\BibitemShut
  {NoStop}%
\bibitem [{\citenamefont {Vallery}\ \emph {et~al.}(2003)\citenamefont
  {Vallery}, \citenamefont {Zitzewitz},\ and\ \citenamefont
  {Gidley}}]{Vallery2003iz}%
  \BibitemOpen
  \bibfield  {author} {\bibinfo {author} {\bibnamefont {Vallery}, \bibfnamefont
  {R.~S.}}, \bibinfo {author} {\bibfnamefont {P.}~\bibnamefont {Zitzewitz}}, \
  and\ \bibinfo {author} {\bibfnamefont {D.}~\bibnamefont {Gidley}}} (\bibinfo
  {year} {2003}),\ \href {\doibase 10.1103/PhysRevLett.90.203402} {\bibfield
  {journal} {\bibinfo  {journal} {Phys. Rev. Lett.}\ }\textbf {\bibinfo
  {volume} {90}},\ \bibinfo {pages} {203402}}\BibitemShut {NoStop}%
\bibitem [{\citenamefont {Van~Dyck}\ \emph {et~al.}(1987)\citenamefont
  {Van~Dyck}, \citenamefont {Schwinberg},\ and\ \citenamefont
  {Dehmelt}}]{VanDyck1987ay}%
  \BibitemOpen
  \bibfield  {author} {\bibinfo {author} {\bibnamefont {Van~Dyck},
  \bibfnamefont {R.~S.}}, \bibinfo {author} {\bibfnamefont {P.}~\bibnamefont
  {Schwinberg}}, \ and\ \bibinfo {author} {\bibfnamefont {H.}~\bibnamefont
  {Dehmelt}}} (\bibinfo {year} {1987}),\ \href {\doibase
  10.1103/PhysRevLett.59.26} {\bibfield  {journal} {\bibinfo  {journal} {Phys.
  Rev. Lett.}\ }\textbf {\bibinfo {volume} {59}},\ \bibinfo {pages}
  {26}}\BibitemShut {NoStop}%
\bibitem [{\citenamefont {Van~House}\ \emph {et~al.}(1984)\citenamefont
  {Van~House}, \citenamefont {Rich},\ and\ \citenamefont
  {Zitzewitz}}]{VanHouse1984}%
  \BibitemOpen
  \bibfield  {author} {\bibinfo {author} {\bibnamefont {Van~House},
  \bibfnamefont {J.}}, \bibinfo {author} {\bibfnamefont {A.}~\bibnamefont
  {Rich}}, \ and\ \bibinfo {author} {\bibfnamefont {P.~W.}\ \bibnamefont
  {Zitzewitz}}} (\bibinfo {year} {1984}),\ \href {\doibase
  10.1103/PhysRevLett.53.953} {\bibfield  {journal} {\bibinfo  {journal} {Phys.
  Rev. Lett.}\ }\textbf {\bibinfo {volume} {53}},\ \bibinfo {pages}
  {953}}\BibitemShut {NoStop}%
\bibitem [{\citenamefont {Van~Petegem}\ \emph {et~al.}(2004)\citenamefont
  {Van~Petegem}, \citenamefont {Dauwe}, \citenamefont {Van~Hoecke},
  \citenamefont {De~Baerdemaeker},\ and\ \citenamefont
  {Segers}}]{VanPetegem2004}%
  \BibitemOpen
  \bibfield  {author} {\bibinfo {author} {\bibnamefont {Van~Petegem},
  \bibfnamefont {S.}}, \bibinfo {author} {\bibfnamefont {C.}~\bibnamefont
  {Dauwe}}, \bibinfo {author} {\bibfnamefont {T.}~\bibnamefont {Van~Hoecke}},
  \bibinfo {author} {\bibfnamefont {J.}~\bibnamefont {De~Baerdemaeker}}, \ and\
  \bibinfo {author} {\bibfnamefont {D.}~\bibnamefont {Segers}}} (\bibinfo
  {year} {2004}),\ \href {\doibase 10.1103/PhysRevB.70.115410} {\bibfield
  {journal} {\bibinfo  {journal} {Phys. Rev. B}\ }\textbf {\bibinfo {volume}
  {70}},\ \bibinfo {pages} {115410}}\BibitemShut {NoStop}%
\bibitem [{\citenamefont {Vanderberghe}\ \emph {et~al.}(2020)\citenamefont
  {Vanderberghe}, \citenamefont {Moskal},\ and\ \citenamefont
  {Karp}}]{Vanderberghe2020}%
  \BibitemOpen
  \bibfield  {author} {\bibinfo {author} {\bibnamefont {Vanderberghe},
  \bibfnamefont {S.}}, \bibinfo {author} {\bibfnamefont {P.}~\bibnamefont
  {Moskal}}, \ and\ \bibinfo {author} {\bibfnamefont {J.~S.}\ \bibnamefont
  {Karp}}} (\bibinfo {year} {2020}),\ \href@noop {} {\bibfield  {journal}
  {\bibinfo  {journal} {EJNMMI Phys.}\ }\textbf {\bibinfo {volume} {7}},\
  \bibinfo {pages} {35}}\BibitemShut {NoStop}%
\bibitem [{\citenamefont {Vaupel}\ \emph {et~al.}(2021)\citenamefont {Vaupel},
  \citenamefont {Flood},\ and\ \citenamefont {Swartz}}]{Vaupel2021}%
  \BibitemOpen
  \bibfield  {author} {\bibinfo {author} {\bibnamefont {Vaupel}, \bibfnamefont
  {P.}}, \bibinfo {author} {\bibfnamefont {A.}~\bibnamefont {Flood}}, \ and\
  \bibinfo {author} {\bibfnamefont {H.}~\bibnamefont {Swartz}}} (\bibinfo
  {year} {2021}),\ \href {\doibase https://doi.org/10.1007/s00723-021-01383-6}
  {\bibfield  {journal} {\bibinfo  {journal} {Appl. Magn. Reson.}\ }\textbf
  {\bibinfo {volume} {52}},\ \bibinfo {pages} {451}}\BibitemShut {NoStop}%
\bibitem [{\citenamefont {Vaupel}\ \emph {et~al.}(2007)\citenamefont {Vaupel},
  \citenamefont {Hockel},\ and\ \citenamefont {Maye}}]{Vaupel2007}%
  \BibitemOpen
  \bibfield  {author} {\bibinfo {author} {\bibnamefont {Vaupel}, \bibfnamefont
  {P.}}, \bibinfo {author} {\bibfnamefont {M.}~\bibnamefont {Hockel}}, \ and\
  \bibinfo {author} {\bibfnamefont {A.}~\bibnamefont {Maye}}} (\bibinfo {year}
  {2007}),\ \href {\doibase 10.1089/ars.2007.1628} {\bibfield  {journal}
  {\bibinfo  {journal} {Antioxid. Redox. Signal.}\ }\textbf {\bibinfo {volume}
  {9}},\ \bibinfo {pages} {1221}}\BibitemShut {NoStop}%
\bibitem [{\citenamefont {Venter}\ \emph {et~al.}(1987)\citenamefont {Venter},
  \citenamefont {Tuzi}, \citenamefont {Kumar},\ and\ \citenamefont
  {Gullick}}]{Venter1987}%
  \BibitemOpen
  \bibfield  {author} {\bibinfo {author} {\bibnamefont {Venter}, \bibfnamefont
  {D.~J.}}, \bibinfo {author} {\bibfnamefont {N.}~\bibnamefont {Tuzi}},
  \bibinfo {author} {\bibfnamefont {S.}~\bibnamefont {Kumar}}, \ and\ \bibinfo
  {author} {\bibfnamefont {W.}~\bibnamefont {Gullick}}} (\bibinfo {year}
  {1987}),\ \href {\doibase doi.org.10.1016/s0140-6736(87)92736-x} {\bibfield
  {journal} {\bibinfo  {journal} {Lancet}\ }\textbf {\bibinfo {volume} {2}},\
  \bibinfo {pages} {69}}\BibitemShut {NoStop}%
\bibitem [{\citenamefont {Vigo}\ \emph {et~al.}(2020)\citenamefont {Vigo},
  \citenamefont {Gerchow}, \citenamefont {Radics}, \citenamefont {Raaijmakers},
  \citenamefont {Rubbia},\ and\ \citenamefont {Crivelli}}]{Raaijmakers2019hqj}%
  \BibitemOpen
  \bibfield  {author} {\bibinfo {author} {\bibnamefont {Vigo}, \bibfnamefont
  {C.}}, \bibinfo {author} {\bibfnamefont {L.}~\bibnamefont {Gerchow}},
  \bibinfo {author} {\bibfnamefont {B.}~\bibnamefont {Radics}}, \bibinfo
  {author} {\bibfnamefont {M.}~\bibnamefont {Raaijmakers}}, \bibinfo {author}
  {\bibfnamefont {A.}~\bibnamefont {Rubbia}}, \ and\ \bibinfo {author}
  {\bibfnamefont {P.}~\bibnamefont {Crivelli}}} (\bibinfo {year} {2020}),\
  \href {\doibase 10.1103/PhysRevLett.124.101803} {\bibfield  {journal}
  {\bibinfo  {journal} {Phys. Rev. Lett.}\ }\textbf {\bibinfo {volume} {124}},\
  \bibinfo {pages} {101803}}\BibitemShut {NoStop}%
%%CITATION = ARXIV:1905.09128;%%
\bibitem [{\citenamefont {Wada}\ \emph {et~al.}(2012)\citenamefont {Wada},
  \citenamefont {Hyodo}, \citenamefont {Yagishita}, \citenamefont {Ikeda},
  \citenamefont {Ohsawa}, \citenamefont {Shidara}, \citenamefont {Michishio},
  \citenamefont {Tachibana}, \citenamefont {Nagashima}, \citenamefont {Fukaya},
  \citenamefont {Maekawa},\ and\ \citenamefont {Kawasuso}}]{Wada2012}%
  \BibitemOpen
  \bibfield  {author} {\bibinfo {author} {\bibnamefont {Wada}, \bibfnamefont
  {K.}}, \bibinfo {author} {\bibfnamefont {T.}~\bibnamefont {Hyodo}}, \bibinfo
  {author} {\bibfnamefont {A.}~\bibnamefont {Yagishita}}, \bibinfo {author}
  {\bibfnamefont {M.}~\bibnamefont {Ikeda}}, \bibinfo {author} {\bibfnamefont
  {S.}~\bibnamefont {Ohsawa}}, \bibinfo {author} {\bibfnamefont
  {T.}~\bibnamefont {Shidara}}, \bibinfo {author} {\bibfnamefont
  {K.}~\bibnamefont {Michishio}}, \bibinfo {author} {\bibfnamefont
  {T.}~\bibnamefont {Tachibana}}, \bibinfo {author} {\bibfnamefont
  {Y.}~\bibnamefont {Nagashima}}, \bibinfo {author} {\bibfnamefont
  {Y.}~\bibnamefont {Fukaya}}, \bibinfo {author} {\bibfnamefont
  {M.}~\bibnamefont {Maekawa}}, \ and\ \bibinfo {author} {\bibfnamefont
  {A.}~\bibnamefont {Kawasuso}}} (\bibinfo {year} {2012}),\ \href {\doibase
  10.1140/epjd/e2012-20641-4} {\bibfield  {journal} {\bibinfo  {journal} {Eur.
  Phys. J. D}\ }\textbf {\bibinfo {volume} {66}},\ \bibinfo {pages}
  {37}}\BibitemShut {NoStop}%
\bibitem [{\citenamefont {Wall}\ \emph {et~al.}(2015)\citenamefont {Wall},
  \citenamefont {Alonso}, \citenamefont {Cooper}, \citenamefont {Deller},
  \citenamefont {Hogan},\ and\ \citenamefont {Cassidy}}]{Wall2015}%
  \BibitemOpen
  \bibfield  {author} {\bibinfo {author} {\bibnamefont {Wall}, \bibfnamefont
  {T.}}, \bibinfo {author} {\bibfnamefont {A.}~\bibnamefont {Alonso}}, \bibinfo
  {author} {\bibfnamefont {B.}~\bibnamefont {Cooper}}, \bibinfo {author}
  {\bibfnamefont {A.}~\bibnamefont {Deller}}, \bibinfo {author} {\bibfnamefont
  {S.}~\bibnamefont {Hogan}}, \ and\ \bibinfo {author} {\bibfnamefont
  {D.}~\bibnamefont {Cassidy}}} (\bibinfo {year} {2015}),\ \href {\doibase
  10.1103/PhysRevLett.114.173001} {\bibfield  {journal} {\bibinfo  {journal}
  {Phys. Rev. Lett.}\ }\textbf {\bibinfo {volume} {114}},\ \bibinfo {pages}
  {173001}}\BibitemShut {NoStop}%
\bibitem [{\citenamefont {Wang}\ \emph {et~al.}(1998)\citenamefont {Wang},
  \citenamefont {Hirata}, \citenamefont {Kawahara},\ and\ \citenamefont
  {Kobayashi}}]{Wang1998}%
  \BibitemOpen
  \bibfield  {author} {\bibinfo {author} {\bibnamefont {Wang}, \bibfnamefont
  {C.}}, \bibinfo {author} {\bibfnamefont {K.}~\bibnamefont {Hirata}}, \bibinfo
  {author} {\bibfnamefont {J.}~\bibnamefont {Kawahara}}, \ and\ \bibinfo
  {author} {\bibfnamefont {Y.}~\bibnamefont {Kobayashi}}} (\bibinfo {year}
  {1998}),\ \href {\doibase 10.1103/PhysRevB.58.14864} {\bibfield  {journal}
  {\bibinfo  {journal} {Phys. Rev. B}\ }\textbf {\bibinfo {volume} {58}},\
  \bibinfo {pages} {14864}}\BibitemShut {NoStop}%
\bibitem [{\citenamefont {Wang}\ \emph {et~al.}(2022)\citenamefont {Wang} \emph
  {et~al.}}]{Wang2022}%
  \BibitemOpen
  \bibfield  {author} {\bibinfo {author} {\bibnamefont {Wang}, \bibfnamefont
  {G.}},  \emph {et~al.}} (\bibinfo {year} {2022}),\ \href {\doibase
  10.2967/jnumed.121.262668} {\bibfield  {journal} {\bibinfo  {journal} {J.
  Nucl. Med}\ }\textbf {\bibinfo {volume} {63}},\ \bibinfo {pages}
  {1274}}\BibitemShut {NoStop}%
\bibitem [{\citenamefont {Watts}\ \emph {et~al.}(2021)\citenamefont {Watts},
  \citenamefont {Bordes}, \citenamefont {Brown}, \citenamefont {Cherlin},
  \citenamefont {Newton}, \citenamefont {Allison}, \citenamefont {Bashkanov},
  \citenamefont {Efthimiou},\ and\ \citenamefont {Zachariou}}]{Watts2021}%
  \BibitemOpen
  \bibfield  {author} {\bibinfo {author} {\bibnamefont {Watts}, \bibfnamefont
  {D.~P.}}, \bibinfo {author} {\bibfnamefont {J.}~\bibnamefont {Bordes}},
  \bibinfo {author} {\bibfnamefont {J.~R.}\ \bibnamefont {Brown}}, \bibinfo
  {author} {\bibfnamefont {A.}~\bibnamefont {Cherlin}}, \bibinfo {author}
  {\bibfnamefont {R.}~\bibnamefont {Newton}}, \bibinfo {author} {\bibfnamefont
  {J.}~\bibnamefont {Allison}}, \bibinfo {author} {\bibfnamefont
  {M.}~\bibnamefont {Bashkanov}}, \bibinfo {author} {\bibfnamefont
  {N.}~\bibnamefont {Efthimiou}}, \ and\ \bibinfo {author} {\bibfnamefont
  {N.}~\bibnamefont {Zachariou}}} (\bibinfo {year} {2021}),\ \href {\doibase
  10.1038/s41467-021-22907-5} {\bibfield  {journal} {\bibinfo  {journal} {Nat.
  Commun.}\ }\textbf {\bibinfo {volume} {12}},\ \bibinfo {pages}
  {2646}}\BibitemShut {NoStop}%
\bibitem [{\citenamefont {Weber}\ \emph {et~al.}(1999)\citenamefont {Weber},
  \citenamefont {Hunt}, \citenamefont {Golovchenko},\ and\ \citenamefont
  {Lynn}}]{Weber1999}%
  \BibitemOpen
  \bibfield  {author} {\bibinfo {author} {\bibnamefont {Weber}, \bibfnamefont
  {M.}}, \bibinfo {author} {\bibfnamefont {A.}~\bibnamefont {Hunt}}, \bibinfo
  {author} {\bibfnamefont {J.}~\bibnamefont {Golovchenko}}, \ and\ \bibinfo
  {author} {\bibfnamefont {K.}~\bibnamefont {Lynn}}} (\bibinfo {year} {1999}),\
  \href {\doibase 10.1103/PhysRevLett.83.4658} {\bibfield  {journal} {\bibinfo
  {journal} {Phys. Rev. Lett.}\ }\textbf {\bibinfo {volume} {83}},\ \bibinfo
  {pages} {4658}}\BibitemShut {NoStop}%
\bibitem [{\citenamefont {Yamazaki}\ \emph {et~al.}(2010)\citenamefont
  {Yamazaki}, \citenamefont {Namba}, \citenamefont {Asai},\ and\ \citenamefont
  {Kobayashi}}]{Yamazaki2009hp}%
  \BibitemOpen
  \bibfield  {author} {\bibinfo {author} {\bibnamefont {Yamazaki},
  \bibfnamefont {T.}}, \bibinfo {author} {\bibfnamefont {T.}~\bibnamefont
  {Namba}}, \bibinfo {author} {\bibfnamefont {S.}~\bibnamefont {Asai}}, \ and\
  \bibinfo {author} {\bibfnamefont {T.}~\bibnamefont {Kobayashi}}} (\bibinfo
  {year} {2010}),\ \href {\doibase 10.1103/PhysRevLett.104.083401} {\bibfield
  {journal} {\bibinfo  {journal} {Phys. Rev. Lett.}\ }\textbf {\bibinfo
  {volume} {104}},\ \bibinfo {pages} {083401}},\ \bibinfo {note} {[Erratum:
  Phys.Rev.Lett. 120, 239902 (2018)]}\BibitemShut {NoStop}%
\bibitem [{\citenamefont {Yan}\ \emph {et~al.}(2015)\citenamefont {Yan},
  \citenamefont {Schwaederle}, \citenamefont {Arguello}, \citenamefont
  {Millis}, \citenamefont {Gatalica},\ and\ \citenamefont
  {Kurzrock}}]{Yan2015}%
  \BibitemOpen
  \bibfield  {author} {\bibinfo {author} {\bibnamefont {Yan}, \bibfnamefont
  {M.}}, \bibinfo {author} {\bibfnamefont {M.}~\bibnamefont {Schwaederle}},
  \bibinfo {author} {\bibfnamefont {D.}~\bibnamefont {Arguello}}, \bibinfo
  {author} {\bibfnamefont {S.}~\bibnamefont {Millis}}, \bibinfo {author}
  {\bibfnamefont {Z.}~\bibnamefont {Gatalica}}, \ and\ \bibinfo {author}
  {\bibfnamefont {R.}~\bibnamefont {Kurzrock}}} (\bibinfo {year} {2015}),\
  \href {\doibase doi.org.10.1007/s10555-015-9552-6} {\bibfield  {journal}
  {\bibinfo  {journal} {Cancer. Metastasis. Rev.}\ }\textbf {\bibinfo {volume}
  {34}},\ \bibinfo {pages} {157}}\BibitemShut {NoStop}%
\bibitem [{\citenamefont {Zabaglo}\ \emph {et~al.}(2013)\citenamefont {Zabaglo}
  \emph {et~al.}}]{Zabaglo2013}%
  \BibitemOpen
  \bibfield  {author} {\bibinfo {author} {\bibnamefont {Zabaglo}, \bibfnamefont
  {L.}},  \emph {et~al.}} (\bibinfo {year} {2013}),\ \href {\doibase
  doi.org.10.1093/annonc/mdt275} {\bibfield  {journal} {\bibinfo  {journal}
  {Ann. Oncol.}\ }\textbf {\bibinfo {volume} {24}},\ \bibinfo {pages}
  {2761}}\BibitemShut {NoStop}%
\bibitem [{\citenamefont {Zare}\ \emph {et~al.}(2022)\citenamefont {Zare},
  \citenamefont {Ghasemi}, \citenamefont {Kakuee},\ and\ \citenamefont
  {Biganeh}}]{Zare-Biganeh-2022}%
  \BibitemOpen
  \bibfield  {author} {\bibinfo {author} {\bibnamefont {Zare}, \bibfnamefont
  {M.}}, \bibinfo {author} {\bibfnamefont {B.}~\bibnamefont {Ghasemi}},
  \bibinfo {author} {\bibfnamefont {O.}~\bibnamefont {Kakuee}}, \ and\ \bibinfo
  {author} {\bibfnamefont {A.}~\bibnamefont {Biganeh}}} (\bibinfo {year}
  {2022}),\ \href {\doibase 10.22034/rpe.2022.336848.1072} {\bibfield
  {journal} {\bibinfo  {journal} {Rad. Phys. Eng.}\ }\textbf {\bibinfo {volume}
  {3}},\ \bibinfo {pages} {1}}\BibitemShut {NoStop}%
\bibitem [{\citenamefont {Zgardzinska}\ \emph {et~al.}(2020)\citenamefont
  {Zgardzinska}, \citenamefont {Cholubek}, \citenamefont {Jarosz},
  \citenamefont {Wysoglad}, \citenamefont {Gorgol}, \citenamefont {Gozdziuk},
  \citenamefont {Cholubek},\ and\ \citenamefont {Jasinska}}]{Zgardzinska2020}%
  \BibitemOpen
  \bibfield  {author} {\bibinfo {author} {\bibnamefont {Zgardzinska},
  \bibfnamefont {B.}}, \bibinfo {author} {\bibfnamefont {G.}~\bibnamefont
  {Cholubek}}, \bibinfo {author} {\bibfnamefont {B.}~\bibnamefont {Jarosz}},
  \bibinfo {author} {\bibfnamefont {K.}~\bibnamefont {Wysoglad}}, \bibinfo
  {author} {\bibfnamefont {M.}~\bibnamefont {Gorgol}}, \bibinfo {author}
  {\bibfnamefont {M.}~\bibnamefont {Gozdziuk}}, \bibinfo {author}
  {\bibfnamefont {M.}~\bibnamefont {Cholubek}}, \ and\ \bibinfo {author}
  {\bibfnamefont {B.}~\bibnamefont {Jasinska}}} (\bibinfo {year} {2020}),\
  \href {\doibase https://doi.org/10.1038/s41598-020-68727-3} {\bibfield
  {journal} {\bibinfo  {journal} {Sci. Rep.}\ }\textbf {\bibinfo {volume}
  {10}},\ \bibinfo {pages} {111890}}\BibitemShut {NoStop}%
\bibitem [{\citenamefont {Zhang}\ \emph {et~al.}(2020)\citenamefont {Zhang}
  \emph {et~al.}}]{zhang2020}%
  \BibitemOpen
  \bibfield  {author} {\bibinfo {author} {\bibnamefont {Zhang}, \bibfnamefont
  {X.}},  \emph {et~al.}} (\bibinfo {year} {2020}),\ \href@noop {} {\bibfield
  {journal} {\bibinfo  {journal} {J. Nucl. Med}\ }\textbf {\bibinfo {volume}
  {61}},\ \bibinfo {pages} {285}}\BibitemShut {NoStop}%
\bibitem [{\citenamefont {Zhou}\ \emph {et~al.}(2015)\citenamefont {Zhou},
  \citenamefont {Mao}, \citenamefont {Li}, \citenamefont {Wang},\ and\
  \citenamefont {He}}]{Zhou2015}%
  \BibitemOpen
  \bibfield  {author} {\bibinfo {author} {\bibnamefont {Zhou}, \bibfnamefont
  {Y.}}, \bibinfo {author} {\bibfnamefont {W.}~\bibnamefont {Mao}}, \bibinfo
  {author} {\bibfnamefont {Q.}~\bibnamefont {Li}}, \bibinfo {author}
  {\bibfnamefont {J.}~\bibnamefont {Wang}}, \ and\ \bibinfo {author}
  {\bibfnamefont {C.}~\bibnamefont {He}}} (\bibinfo {year} {2015}),\ \href
  {\doibase 10.1016/j.chemphys.2015.07.030} {\bibfield  {journal} {\bibinfo
  {journal} {Chem. Phys.}\ }\textbf {\bibinfo {volume} {459}},\ \bibinfo
  {pages} {81}}\BibitemShut {NoStop}%
\bibitem [{\citenamefont {Zhu}\ \emph {et~al.}(2022)\citenamefont {Zhu},
  \citenamefont {Harrison},\ and\ \citenamefont {C-M}}]{Zhu-positronium-2022}%
  \BibitemOpen
  \bibfield  {author} {\bibinfo {author} {\bibnamefont {Zhu}, \bibfnamefont
  {Z.}}, \bibinfo {author} {\bibfnamefont {C.}~\bibnamefont {Harrison}}, \ and\
  \bibinfo {author} {\bibfnamefont {K.}~\bibnamefont {C-M}}} (\bibinfo {year}
  {2022}),\ \href {\doibase https://doi.org/10.48550/arXiv.2206.06463}
  {\bibinfo  {journal} {arXiv:2206.06463v1}\ ,\ \bibinfo {pages} {in
  print}}\BibitemShut {NoStop}%
\bibitem [{\citenamefont {Čížek}\ \emph {et~al.}(2012)\citenamefont
  {Čížek}, \citenamefont {Vlček},\ and\ \citenamefont
  {Procházk}}]{Cizek2012}%
  \BibitemOpen
\bibfield  {journal} {  }\bibfield  {author} {\bibinfo {author} {\bibnamefont
  {Čížek}, \bibfnamefont {J.}}, \bibinfo {author} {\bibfnamefont
  {M.}~\bibnamefont {Vlček}}, \ and\ \bibinfo {author} {\bibfnamefont
  {P.}~\bibnamefont {Procházk}}} (\bibinfo {year} {2012}),\ \href {\doibase
  10.1088/1367-2630/14/3/035005} {\bibfield  {journal} {\bibinfo  {journal}
  {New J. Phys.}\ }\textbf {\bibinfo {volume} {14}},\ \bibinfo {pages}
  {035005}}\BibitemShut {NoStop}%
\end{thebibliography}%

\end{document}